\documentclass[sigconf, nonacm]{acmart}

\usepackage{color}
\usepackage{graphicx}
\usepackage{subfigure}
\usepackage{threeparttable}
\usepackage{setspace}
\usepackage{booktabs}
\usepackage{diagbox}
\usepackage{multirow}
\usepackage{amsmath}

\usepackage[font=small]{caption}
\usepackage{ragged2e}
\usepackage{listings}
\usepackage[ruled,vlined]{algorithm2e}

\makeatletter
  \newcommand\figcaption{\def\@captype{figure}\caption}
  \newcommand\tabcaption{\def\@captype{table}\caption}
\makeatother

\newcommand\vldbdoi{XX.XX/XXX.XX}
\newcommand\vldbpages{XXX-XXX}
\newcommand\vldbvolume{14}
\newcommand\vldbissue{1}
\newcommand\vldbyear{2021}

\newcommand\vldbavailabilityurl{https://github.com/Lsyhprum/WEAVESS}
\newcommand\vldbpagestyle{plain} 

\begin{document}
\title{A Comprehensive Survey and Experimental Comparison of Graph-Based Approximate Nearest Neighbor Search}

\author{Mengzhao Wang$^1$, Xiaoliang Xu$^1$, Qiang Yue$^1$, Yuxiang Wang$^{1,*}$}
\affiliation{%
 \institution{$^1$Hangzhou Dianzi University, China}
}
\email{{mzwang,xxl,yq,lsswyx}@hdu.edu.cn}

\begin{abstract}
  Approximate nearest neighbor search (ANNS) constitutes an important operation in a multitude of applications, including recommendation systems, information retrieval, and pattern recognition. In the past decade, graph-based ANNS algorithms have been the leading paradigm in this domain, with dozens of graph-based ANNS algorithms proposed. Such algorithms aim to provide effective, efficient solutions for retrieving the nearest neighbors for a given query. Nevertheless, these efforts focus on developing and optimizing algorithms with different approaches, so there is a real need for a comprehensive survey about the approaches' relative performance, strengths, and pitfalls. Thus here we provide a thorough comparative analysis and experimental evaluation of 13 representative graph-based ANNS algorithms via a new taxonomy and fine-grained pipeline. We compared each algorithm in a uniform test environment on eight real-world datasets and 12 synthetic datasets with varying sizes and characteristics. Our study yields novel discoveries, offerings several useful principles to improve algorithms, thus designing an optimized method that outperforms the state-of-the-art algorithms. This effort also helped us pinpoint algorithms’ working portions, along with rule-of-thumb recommendations about promising research directions and suitable algorithms for practitioners in different fields.
\end{abstract}

\maketitle

\pagestyle{\vldbpagestyle}
\begingroup\small\noindent\raggedright\textbf{PVLDB Reference Format:}\\
Mengzhao Wang, Xiaoliang Xu, Qiang Yue, Yuxiang Wang. A Comprehensive Survey and Experimental Comparison of Graph-Based Approximate Nearest Neighbor Search. PVLDB, \vldbvolume(\vldbissue): \vldbpages, \vldbyear.\\
\href{https://doi.org/\vldbdoi}{doi:\vldbdoi}
\endgroup
\begingroup
\renewcommand\thefootnote{}\footnote{\noindent
$^*$Corresponding author.\\
This work is licensed under the Creative Commons BY-NC-ND 4.0 International License. Visit \url{https://creativecommons.org/licenses/by-nc-nd/4.0/} to view a copy of this license. For any use beyond those covered by this license, obtain permission by emailing \href{mailto:info@vldb.org}{info@vldb.org}. Copyright is held by the owner/author(s). Publication rights licensed to the VLDB Endowment. \\
\raggedright Proceedings of the VLDB Endowment, Vol. \vldbvolume, No. \vldbissue\ %
ISSN 2150-8097. \\
\href{https://doi.org/\vldbdoi}{doi:\vldbdoi} \\
}\addtocounter{footnote}{-1}\endgroup

\ifdefempty{\vldbavailabilityurl}{}{
\vspace{.3cm}
\begingroup\small\noindent\raggedright\textbf{PVLDB Artifact Availability:}\\
The source code, data, and/or other artifacts have been made available at \url{\vldbavailabilityurl}.
\endgroup
}

\section{Introduction} \label{sec1}

Nearest Neighbor Search (NNS) is a fundamental building block in various application domains~\cite{milvus,NSG,HNSW,riegger2010literature,arora2018hd,aoyama2011fast,zhang2018zoom,zhou2013large}, such as information retrieval~\cite{flickner1995query,zhu2019accelerating}, pattern recognition~\cite{kosuge2019object,cover1967nearest}, data mining~\cite{yahoo2,huang2017query}, machine learning~\cite{cost1993weighted,cao2017binary}, and recommendation systems~\cite{sarwar2001item,meng2020pmd}. With the explosive growth of datasets’ scale and the inevitable \textit{curse of dimensionality}, accurate NNS cannot meet actual requirements for efficiency and cost~\cite{DPG}. Thus, much of the literature has focused on efforts to research approximate NNS (ANNS) and find an algorithm that improves efficiency substantially while mildly relaxing accuracy constraints (an accuracy-versus-efficiency tradeoff~\cite{li2020improving}).

ANNS is a task that finds the approximate nearest neighbors among a high-dimensional dataset for a query via a well-designed index. According to the index adopted, the existing ANNS algorithms can be divided into four major types: hashing-based~\cite{gong2020idec,huang2015query}; tree-based~\cite{silpa2008optimised,arora2018hd}; quantization-based~\cite{jegou2010product,pan2020product}; and graph-based~\cite{HNSW,NSG} algorithms. Recently, graph-based algorithms have emerged as a highly effective option for ANNS~\cite{ANN-benchmarks,aoyama2013graph,hacid2010neighborhood,KNNG1}. Thanks to graph-based ANNS algorithms’ extraordinary ability to express neighbor relationships~\cite{NSG,weber1998quantitative}, they only need to evaluate fewer points of dataset to receive more accurate results~\cite{HNSW,NSG,DPG,BDG,HCNNG}.

\begin{figure}
\centering
  \setlength{\abovecaptionskip}{0cm}
  \setlength{\belowcaptionskip}{-0.6cm}
  \includegraphics[width=\linewidth]{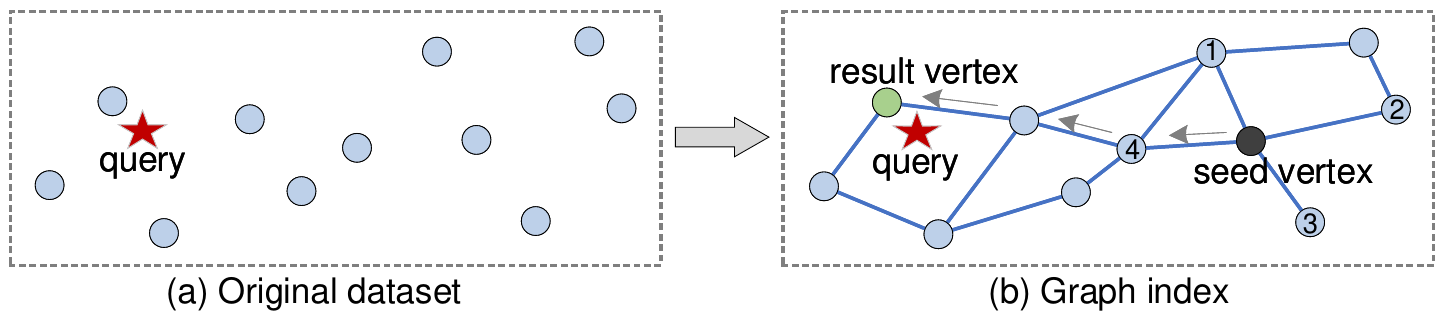}
  \caption{A toy example for the graph-based ANNS algorithm.}
  \label{fig0}
\end{figure}

As \autoref{fig0} shows, graph-based ANNS algorithms build a graph index (\autoref{fig0}(b)) on the original dataset (\autoref{fig0}(a)), the vertices in the graph correspond to the points of the original dataset, and neighboring vertices (marked as $x$, $y$) are associated with an edge by evaluating their distance $\delta (x,y)$, where $\delta$ is a distance function. In \autoref{fig0}(b), the four vertices (numbered 1–4) connected to the black vertex are its neighbors, and the black vertex can visit its neighbors along these edges. Given this graph index and a query $q$ (the red star), ANNS aims to get a set of vertices that are close to $q$. We take the case of returning $q$’s nearest neighbor as an example to show ANNS’ general procedure: Initially, a seed vertex (the black vertex, it can be randomly sampled or obtained by additional approaches~\cite{HNSW,yahoo2}) is selected as the result vertex $r$, and we can conduct ANNS from this seed vertex. Specifically, if $\delta (n,q) < \delta (r,q)$, where $n$ is one of the neighbors of $r$, $r$ will be replaced by ${n}$. We repeat this process until the termination condition (e.g., $\forall {n}, \delta (n,q) \geq \delta (r,q)$) is met, and the final ${r}$ (the green vertex) is $q$’s nearest neighbor. Compared with other index structures, graph-based algorithms are a proven superior tradeoff in terms of accuracy versus efficiency ~\cite{ANN-benchmarks,NSW,HNSW,DPG,NSG}, which is probably why they enjoy widespread use among high-tech companies nowadays (e.g., Microsoft~\cite{wang2012query,wang2012scalable}, Alibaba~\cite{BDG,NSG}, and Yahoo~\cite{yahoo1,yahoo2,yahoo3}).

\vspace{-0.2cm}
\subsection{Motivation} \label{sec1_1}

The problem of graph-based ANNS on high-dimensional and large-scale data has been studied intensively across the literature~\cite{NSG}. Dozens of algorithms have been proposed to solve this problem from different optimizations~\cite{NSW,HNSW,FANNG,NSSG,HCNNG,IEH,EFANNA}. For these algorithms, existing surveys \cite{ANN-benchmarks,DPG,graph_survey} provide some meaningful explorations. However, they are limited to a small subset about algorithms, datasets, and metrics, as well as studying algorithms from a macro perspective, and the analysis and evaluation of intra-algorithm components are ignored. For example, \cite{DPG} includes a few graph-based algorithms (only three), \cite{ANN-benchmarks} focuses on efficiency vs accuracy tradeoff, \cite{graph_survey} only considers several classic graphs. This motivates us to carry out a thorough comparative analysis and experimental evaluation of existing graph-based algorithms via a new taxonomy and micro perspective (i.e., some fine-grained components). We detail the issues of existing work that ensued.



\noindent\textbf{I1: Lack of a reasonable taxonomy and comparative analysis of inter-algorithms.} Many studies in other fields show that an insightful taxonomy can serve as a guideline for promising research in this domain\cite{wang2017survey,wang2015learning,wu2020comprehensive}. Thus, a reasonable taxonomy needs to be established, to point to the different directions of graph-based algorithms (\textbf{\S \ref{sec3}}). The index of existing graph-based ANNS algorithms are generally derivatives of four classic base graphs from different perspectives, i.e., Delaunay Graph (DG)~\cite{DG1}, Relative Neighborhood Graph (RNG)~\cite{RNG1}, K-Nearest Neighbor Graph (KNNG)~\cite{KNNG1}, and Minimum Spanning Tree (MST)~\cite{MST1}. Some representative ANNS algorithms, such as KGraph~\cite{KGraph}, HNSW~\cite{HNSW}, DPG~\cite{DPG}, SPTAG~\cite{SPTAG} can be categorized into KNNG-based (KGraph and SPTAG) and RNG-based (DPG and HNSW) groups working off the base graphs upon which they rely. Under this classification, we can pinpoint differences between algorithms of the same category or different categories, to provide a comprehensive inter-algorithm analysis.

\noindent\textbf{I2: Omission in analysis and evaluation for intra-algorithm fine-grained components.} Many studies only compare and analyze graph-based ANNS algorithms from two coarse-grained components, i.e., construction and search~\cite{graph_survey,rachkovskij2018index}, which hinders insight into the key components. Construction and search, however, can be divided into many fine-grained components such as \textit{candidate neighbor acquisition}, \textit{neighbor selection}~\cite{NSG,NSSG}, \textit{seed acquisition}~\cite{EFANNA,arya1998optimal}, and \textit{routing}~\cite{vargas2019genetic,learn_to_route} (we discuss the details of these components in \textbf{\S \ref{sec4}}). Evaluating these fine-grained components (\textbf{\S \ref{sec5}}) led to some interesting phenomena. For example, some algorithms' performance improvements are not so remarkable for their claimed major contribution (optimization on one component) in the paper, but instead by another small optimization for another component (e.g., NSSG~\cite{NSSG}). Additionally, the key performance of completely different algorithms may be dominated by the same fine-grained component (e.g., the \textit{neighbor selection} of NSG~\cite{NSG} and HNSW~\cite{HNSW}). Such unusual but key discoveries occur by analyzing the components in detail to clarify which part of an algorithm mainly works in practice, thereby assisting researchers’ further optimization goals.


\noindent\textbf{I3: Richer metrics are required for evaluating graph-based ANNS algorithms’ overall performance.} Many evaluations of graph-based algorithms focus on the tradeoff of accuracy vs efficiency ~\cite{NSW,FANNG,magliani2019efficient}, which primarily reflects related algorithms’ search performance ~\cite{li2020improving}. With the explosion of data scale and increasingly frequent requirements to update, the index construction efficiency and algorithm’s index size have received more and more attention~\cite{BDG}. Related metrics such as graph quality (it can be measured by the percentage of vertices that are linked to their nearest neighbor on the graph)~\cite{boutet2016being}, average out-degree, and so on indirectly affect the index construction efficiency and index size, so they are vital for comprehensive analysis of the index performance. From our abundance of experiments (see \textbf{\S \ref{sec5}} for details), we gain a novel discovery: higher graph quality does not necessarily achieve better search performance. For instance, HNSW~\cite{HNSW} and DPG~\cite{DPG} yield similar search performances on the GIST1M dataset~\cite{texmex}. However, in terms of graph quality, HNSW (63.3\%) is significantly lower than DPG (99.2\%) (\textbf{\S \ref{sec5}}). Note that DPG spends a lot of time improving graph quality during index construction, but it is unnecessary; this is not uncommon, as we also see it in \cite{EFANNA,NSSG,NSG,KGraph}.



\noindent\textbf{I4: Diversified datasets are essential for graph-based ANNS algorithms’ scalability evaluation.} Some graph-based ANNS algorithms are evaluated only on a small number of datasets, which limits analysis on how well they scale on different datasets. Looking at the evaluation results on various datasets (see \textbf{\S \ref{sec5}} for details), we find that many algorithms have significant discrepancies in terms of performance on different datasets. That is, the advantages of an algorithm on some datasets may be difficult to extend to other datasets. For example, when the search accuracy reaches 0.99, NSG’s speedup is 125× more than that of HNSW for each query on Msong~\cite{Msong}. However, on Crawl~\cite{Crawl}, NSG’s speedup is 80× lower than that of HNSW when it achieves the same search accuracy of 0.99. This shows that an algorithm’s superiority is contingent on the dataset rather than being fixed in its performance. Evaluating and analyzing different scenarios’ datasets leads to understanding performance differences better for graph-based ANNS algorithms in diverse scenarios, which provides a basis for practitioners in different fields to choose the most suitable algorithm.


\vspace{-0.4cm}
\subsection{Our Contributions} \label{sec1_2}

Driven by the aforementioned issues, we provide a comprehensive comparative analysis and experimental evaluation of representative graph-based algorithms on carefully selected datasets of varying characteristics. It is worth noting that we try our best to reimplement all algorithms using the same design pattern, programming language and tricks, and experimental setup, which makes the comparison fairer. Our key contributions are summarized as follows.

\noindent\textbf{(1) We provide a new taxonomy of the graph-based ANNS algorithms based on four base graphs.} For \textbf{I1}, we classify graph-based algorithms based on four base graphs (\textbf{\S \ref{sec3}}), which brings a new perspective to understanding existing work. On this basis, we compare and analyze the features of inter-algorithms, make connections if different algorithms use similar techniques, and elaborate upon the inheritance and improvement of relevant algorithms, thus exhibiting diversified development roadmaps (\hyperref[tab2]{Table 2} and \autoref{fig: roadmap}). 

\noindent\textbf{(2) We present a unified pipeline with seven fine-grained components for analyzing graph-based ANNS algorithms.} As for \textbf{I2}, we break all graph-based ANNS algorithms down to seven fine-grained components in an unified pipeline (\autoref{fig: components}): We divide \textit{construction} into \textit{initialization}, \textit{candidate neighbor acquisition}, \textit{neighbor selection}, \textit{connectivity}, and \textit{seed preprocessing} components, and divide \textit{search} into \textit{seed acquisition} and \textit{routing} components (\textbf{\S \ref{sec4}}). This not only allows us to have a deeper understanding of the algorithm, but also to achieve a fair evaluation of a component by controlling other components’ consistency in the pipeline (\textbf{\S \ref{sec5}}).

\noindent\textbf{(3) We conduct a comprehensive evaluation for representative graph-based ANNS algorithms with more metrics and diverse datasets.} In terms of \textbf{I3}, we perform a thorough evaluation of algorithms and components in \textbf{\S \ref{sec5}}, with abundant metrics involved in index construction and search. For \textbf{I4}, we investigate different algorithms’ scalability over different datasets (eight real-world and 12 synthetic datasets), covering multimedia data such as video, voice, image, and text.

\noindent\textbf{(4) We discuss the recommendations, guidelines, improvement, tendencies, and challenges about graph-based ANNS algorithms.} Based on our investigation, we provide some rule-of-thumb recommendations about the most suitable scenario for each single algorithm, along with useful guidelines to optimize algorithms, thus designing an algorithm obtains the state-of-the-art performance. Then we analyze graph-based ANNS algorithms’ promising research directions and outstanding challenges (\textbf{\S \ref{sec6}}).


\vspace{-0.15cm}
\section{Preliminaries} \label{sec2}

\noindent\underline{\textbf{Notations.}} Unless otherwise specified, relative notations appear in this paper by default as described in \autoref{tab1}.

\setlength{\textfloatsep}{0cm}
\setlength{\floatsep}{0cm}
\begin{table}[!tb]
  \centering
  \setlength{\abovecaptionskip}{0.05cm}
  \setstretch{0.9}
  \fontsize{8pt}{3.3mm}\selectfont
  \caption{Notations used in this paper}
  \label{tab1}
  \begin{tabular}{p{32pt}|p{185pt}}
    \hline
    \textbf{Notations} & \textbf{Descriptions}\\
    \hline
    \hline
    $E^d$ & The Euclidean space with dimension $d$ \\
    \hline
    $|\cdot|$ & The cardinality of a set \\
    \hline
    $S$ & A limited dataset in $E^d$, where every element is a vector \\
    \hline
    ${q}$ & The query point in $E^d$; it is represented by a vector \\
    \hline
    $\delta (,)$ & The Euclidean distance between points \\
    \hline
    $G(V,E)$ & A graph index $G$ where the set of vertices and edges are $V$ and $E$, respectively \\
    \hline
    $N({v})$ & The neighbors of the vertex ${v}$ in a graph \\
    \hline
  \end{tabular}
\end{table}

\noindent\underline{\textbf{Modeling.}} For a dataset $S=\left \{ s_{0}, s_{1}, \cdots, s_{n-1} \right \}$ of $n$ points, each element $s_{i}$ (denoted as $x$) in $S$ is represented by a vector $\textbf{x}=[ x_0, x_1, \cdots, x_{d-1} ]$ with dimension $d$. Using a similarity calculation of vectors with a similarity function on $S$, we can realize the analysis and retrieval of the corresponding data~\cite{chavez2001searching,milvus}.

\noindent\underline{\textbf{Similarity function.}} For the two points ${x},{y}$ on dataset $S$, a variety of applications employ a distance function to calculate the similarity between the two points ${x}$ and ${y}$~\cite{zezula2006similarity}. The most commonly used distance function is the Euclidean distance $\delta({x},{y})$ ($l_2$ norm)~\cite{graph_survey}, which is given in \autoref{eq1}.
\begin{equation}
  \label{eq1}
  \delta({x},{y})= \sqrt {\sum_{i=0}^{d-1} (x_i-y_i)^2},
\end{equation}
where $x$ and $y$ correspond to the vectors $\textbf{x} = [ x_0, x_1, \cdots, x_{d-1} ]$, and $\textbf{y} = [ y_0, y_1, \cdots, y_{d-1}]$, respectively, here $d$ represents the vectors’ dimension. The larger the $\delta({x},{y})$, the more dissimilar ${x}$ and ${y}$ are, and the closer to zero, the more similar they are~\cite{zezula2006similarity}.

\subsection{Problem Definition}

Before formally describing ANNS, we first define NNS.

\vspace{-0.5em}
\begin{definition}
  \label{def1}
  \textbf{NNS.} Given a finite dataset $S$ in Euclidean space $E^d$ and a query ${q}$, NNS obtains $k$ nearest neighbors $\mathcal{R}$ of ${q}$ by evaluating $\delta ({x},{q})$, where ${x} \in S$. $\mathcal{R}$ is described as follows:
  \begin{equation}
    \label{eq2}
    \mathcal{R}=\arg \min_{\mathcal{R} \subset S, \vert R \vert = k} \sum_{{x} \in \mathcal{R}}\delta({x},{q}).
  \end{equation}
\end{definition}
\vspace{-0.5em}

As the volume of data grows, $\vert S \vert$ becomes exceedingly large (ranging from millions to billions in scale), which makes it impractical to perform NNS on large-scale data because of the high computational cost~\cite{zhou2020learning}. Instead of NNS, a large amount of practical techniques have been proposed for ANNS, which relaxes the guarantee of accuracy for efficiency by evaluating a small subset of $S$~\cite{wang2013trinary}. The ANNS problem is defined as follows:

\vspace{-0.5em}
\begin{definition}
  \label{def2}
  \textbf{ANNS.} Given a finite dataset $S$ in Euclidean space $E^d$, and a query ${q}$, ANNS builds an index $\mathcal{I}$ on $S$. It then gets a subset $\mathcal{C}$ of $S$ by $\mathcal{I}$, and evaluates $\delta ({x},{q})$ to obtain the approximate $k$ nearest neighbors $\tilde{\mathcal{R}}$ of ${q}$, where ${x} \in \mathcal{C}$.
\end{definition}
\vspace{-0.5em}

Generally, we use recall rate $Recall@k = \frac{\vert \mathcal{R} \cap \tilde{\mathcal{R}} \vert}{k}$ to evaluate the search results’ accuracy. ANNS algorithms aim to maximize $Recall@k$ while making $\mathcal{C}$ as small as possible (e.g., $\vert \mathcal{C} \vert$ is only a few thousand when $\vert S \vert$ is millions on the SIFT1M~\cite{texmex} dataset). As mentioned earlier, ANNS algorithms based on graphs have risen in prominence because of their advantages in accuracy versus efficiency. We define graph-based ANNS as follows.

\vspace{-0.5em}
\begin{definition}
  \label{def3}
  \textbf{Graph-based ANNS.} Given a finite dataset $S$ in Euclidean space $E^d$, $G(V,E)$ denotes a graph (the index $\mathcal{I}$ in \hyperref[def2]{Definition 2.2}) constructed on $S$, $\forall {v} \in V$ that uniquely corresponds to a point ${x}$ in $S$. Here $\forall ({u}, {v}) \in E$ represents the neighbor relationship between ${u}$ and ${v}$, and ${u},{v} \in V$. Given a query ${q}$, seeds $\widehat{S}$, routing strategy, and termination condition, the graph-based ANNS initializes approximate $k$ nearest neighbors $\tilde{\mathcal{R}}$ of $q$ with $\widehat{S}$, then conducts a search from $\widehat{S}$ and updates $\tilde{\mathcal{R}}$ via a routing strategy. Finally, it returns the query result $\tilde{\mathcal{R}}$ once the termination condition is met.
\end{definition}
\vspace{-0.5em}

\vspace{-0.15cm}
\subsection{Scope Illustration}

To make our survey and comparison focused yet comprehensive, we employ some necessary constraints.

\noindent\underline{\textbf{Graph-based ANNS.}} We only consider algorithms whose index structures are based on graphs for ANNS. Although some effective algorithms based on other structures exist, these methods’ search performance is far inferior to that of graph-based algorithms. Over time, graph-based algorithms have become mainstream for research and practice in academia and industry.

\noindent\underline{\textbf{Dataset.}} ANNS techniques have been used in various multimedia fields. To comprehensively evaluate the performance of comparative algorithms, we select a variety of multimedia data, including video, image, voice, and text (for details, see \autoref{tab: Dataset} in \S \ref{sec5}). The base data and query data comprise high-dimensional feature vectors extracted by deep learning technology (such as VGG ~\cite{DBLP:journals/corr/SimonyanZ14a} for image), and the ground-truth data comprise the query’s 20 or 100 nearest neighbors calculated in $E^d$ by linear scanning.

\noindent\underline{\textbf{Core algorithms.}} This paper mainly focuses on in-memory core algorithms. For some hardware (e.g., GPU~\cite{SONG} and SSD~\cite{DiskANN}), heterogeneous (e.g., distributed deployment~\cite{deng2019pyramid}), and machine learning (ML)-based optimizations~\cite{learn_to_route, prokhorenkova2020graph, li2020improving} (see \S \ref{machine_learning_optimization} for the evaluation of a few ML-based optimizations), we do not discuss these in detail, keeping in mind that core algorithms are the basis of these optimizations. In future work, we will focus on comparing graph-based ANNS algorithms with GPU, SSD, ML and so on.

\section{Overview of Graph-Based ANNS} \label{sec3}

In this section, we present a taxonomy and overall analysis of graph-based ANNS algorithms from a new perspective. To this end, we first dissect several classic base graphs~\cite{bose2012proximity,toussaint2002proximity}, including Delaunay Graph~\cite{DG1,DG2}, Relative Neighborhood Graph~\cite{RNG1,RNG2}, K-Nearest Neighbor Graph~\cite{KNNG1,arya1998optimal} and Minimum Spanning Tree~\cite{MST1,MST2}. After that, we review 13 representative graph-based ANNS algorithms working off different optimizations to these base graphs.

\vspace{-0.75em}
\subsection{Base Graphs for ANNS} \label{sec3_1}

\begin{figure}[!t]
  \centering
  \setlength{\abovecaptionskip}{0cm}
  \includegraphics[width=.972\linewidth]{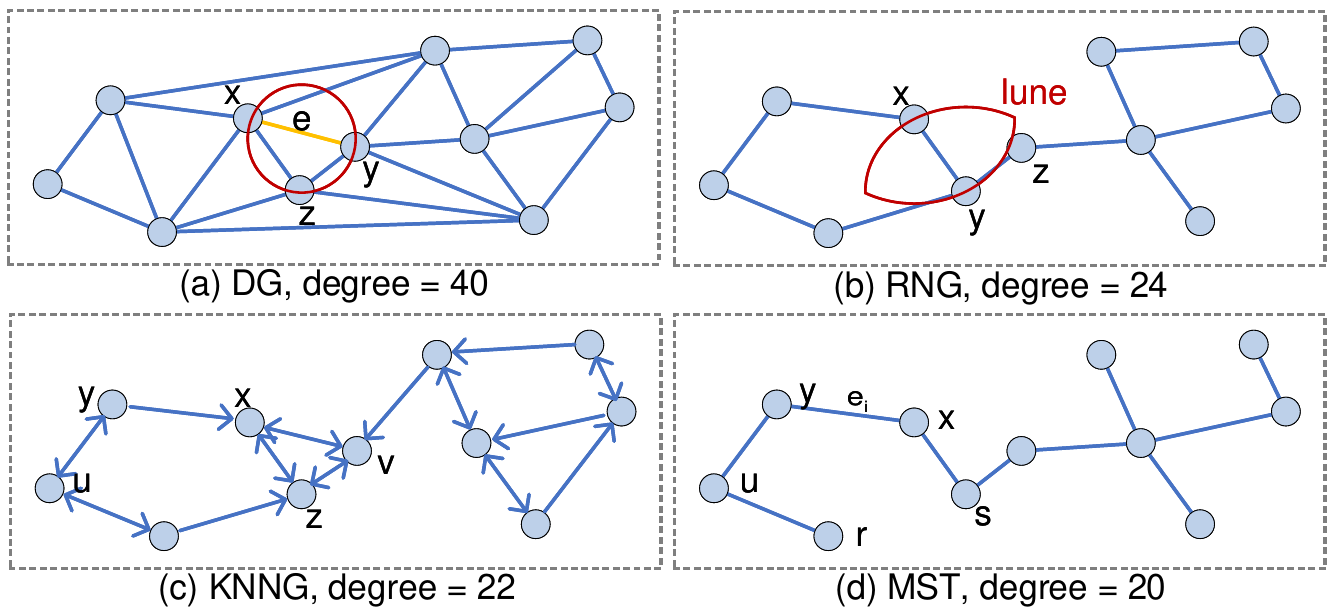}
  \caption{Schematic diagram of different base graphs’ construction results on the same dataset with dimension $d=2$.}
  \label{fig2}
\end{figure}

The four base graphs that graph-based ANNS algorithms depend on are the foundation for analyzing these algorithms. Next, we will give a formal description of each base graph, and visually show their differences through a toy example in \autoref{fig2}.

\noindent\underline{\textbf{Delaunay Graph (DG).}} In Euclidean space $E^d$, the DG $G(V,E)$ constructed on dataset $S$ satisfies the following conditions: For $\forall e \in E$ (e.g., the yellow line in \autoref{fig2}(a)), where its corresponding two vertices are ${x}$, ${y}$, there exists a circle (the red circle in \autoref{fig2}(a)) passing through ${x}$, ${y}$, and no other vertices inside the circle, and there are at most three vertices (i.e., ${x}, {y}, {z}$) on the circle at the same time (see \cite{DG1} for DG’s standard definition). DG ensures that the ANNS always return precise results~\cite{HNSW}, but the disadvantage is that DG is almost fully connected when the dimension $d$ is extremely high, which leads to a large search space~\cite{FANNG,NSG}.

\noindent\underline{\textbf{Relative Neighborhood Graph (RNG).}} In Euclidean space $E^d$, the RNG $G(V,E)$ built on dataset $S$ has the following property: For ${x}, {y} \in V$, if ${x}$ and ${y}$ are connected by edge $e\in E$, then $\forall {z} \in V$, with $\delta ({x},{y}) <\delta ({x}, {z})$, or $\delta ({x}, {y}) <\delta ({z}, {y})$. In other words, ${z}$ is not in the red lune in \autoref{fig2}(b) (for RNG’s standard definition, refer to \cite{RNG1}). Compared with DG, RNG cuts off some redundant neighbors (close to each other) that violate its aforementioned property, and makes the remaining neighbors distribute omnidirectionally, thereby reducing ANNS’ distance calculations~\cite{HNSW}. However, the time complexity of constructing RNG on $S$ is $O(\vert S \vert ^3)$~\cite{jaromczyk1991constructing}.

\noindent\underline{\textbf{K-Nearest Neighbor Graph (KNNG).}} Each point in dataset $S$ is connected to its nearest $K$ points to form a KNNG $G(V, E)$ in Euclidean space $E^d$. As \autoref{fig2}(c) ($K=2$) shows, for ${x}, {y} \in V$, ${x} \in N({y})=\left \{{x},{u}\right \}$, but ${y} \not \in N({x})=\left \{ {z}, {v} \right \}$, where $N({y}), N({x})$ are the neighbor sets of ${y}$ and ${x}$, respectively. Therefore, the edge between $y$ and $x$ is a directed edge, so KNNG is a directed graph. KNNG limits the number of neighbors of each vertex to $K$ at most, thus avoiding the surge of neighbors, which works well in scenarios with limited memory and high demand for efficiency. It can be seen that KNNG does not guarantee global connectivity in \autoref{fig2}(c), which is unfavorable for ANNS.

\noindent\underline{\textbf{Minimum Spanning Tree (MST).}} In Euclidean space $E^d$, MST is the $G(V, E)$ with the smallest $\sum_{i=1}^{\vert E \vert} w(e_i)$ on dataset $S$, where the two vertices associated with $e_i \in E$ are ${x}$ and ${y}$, $w(e_i)=\delta ({x},{y})$. If $\exists e_i,e_j \in E$, $w(e_i)=w(e_j)$, then MST is not unique~\cite{PG}. Although MST has not been adopted by most current graph-based ANNS algorithms, HCNNG~\cite{HCNNG} confirms MST’s effectiveness as a neighbor selection strategy for ANNS. The main advantage for using MST as a base graph relies on the fact that MST uses the least edges to ensure the graph’s global connectivity, so that keeping vertices with low degrees and any two vertices are reachable. However, because of a lack of shortcuts, it may detour when searching on MST~\cite{NSW,NSG}. For example, in \autoref{fig2}(d), when search goes from $s$ to $r$, it must detour with ${s} \rightarrow {x} \rightarrow {y} \rightarrow {u} \rightarrow {r}$. This can be avoided if there is an edge between $s$ and $r$.



\vspace{-0.18em}
\subsection{Graph-Based ANNS Algorithms} \label{sec3_2}

\setlength{\textfloatsep}{0cm}
\setlength{\floatsep}{0cm}
\begin{figure*}
\setlength{\abovecaptionskip}{0cm}
\setstretch{0.9}
\fontsize{6.5pt}{3.3mm}\selectfont
\begin{minipage}{0.59\textwidth}
    \centering
    \tabcaption{Summary of important representative graph-based ANNS algorithms}
    \label{tab2}
    \begin{threeparttable}
    \begin{tabular}{p{35pt}|p{44pt}|p{27pt}|p{80pt}|p{56pt}}
    \hline
    \textbf{Algorithm} & \textbf{Base Graph} & \textbf{Edge} & \textbf{Build Complexity} & \textbf{Search Complexity}\\
    \hline
    \hline
    KGraph~\cite{KGraph} & KNNG & directed & $O(\vert S \vert ^{1.14})$ & $O(\vert S \vert ^{0.54})$\tnote{$\ddagger$} \\
    \hline
    NGT~\cite{NGT} & KNNG+DG+RNG & directed & $O(\vert S \vert ^{1.14})$\tnote{$\ddagger$} & $O(\vert S \vert ^{0.59})$\tnote{$\ddagger$}\\
    \hline
    SPTAG~\cite{SPTAG} & KNNG+RNG & directed & $O(\vert S \vert \cdot \log(\vert S \vert ^{c} + t^t))$\tnote{$\dagger$} & $O(\vert S \vert ^{0.68})$\tnote{$\ddagger$}\\
    \hline
    NSW~\cite{NSW} & DG & undirected & $O(\vert S \vert \cdot \log^{2}(\vert S \vert))$\tnote{$\ddagger$} & $O(\log^{2}(\vert S \vert))^{\dagger}$\\
    \hline
    IEH~\cite{IEH} & KNNG & directed & $O(\vert S \vert ^2 \cdot \log(\vert S \vert) + \vert S \vert ^{2})$\tnote{$\ddagger$} & $O(\vert S \vert ^{0.52})$\tnote{$\ddagger$}\\
    \hline
    FANNG~\cite{FANNG} & RNG & directed & $O(\vert S \vert ^{2} \cdot \log(\vert S \vert))$ & $O(\vert S \vert ^{0.2})$\\
    \hline
    HNSW~\cite{HNSW} & DG+RNG & directed & $O(\vert S \vert \cdot \log(\vert S \vert))$ & $O(\log(\vert S \vert))$\\
    \hline
    EFANNA~\cite{EFANNA} & KNNG & directed & $O(\vert S \vert ^{1.13})$\tnote{$\ddagger$} & $O(\vert S \vert ^{0.55})$\tnote{$\ddagger$}\\
    \hline
    DPG~\cite{DPG} & KNNG+RNG & undirected & $O(\vert S \vert ^{1.14} + \vert S \vert)$\tnote{$\ddagger$} & $O(\vert S \vert ^{0.28})$\tnote{$\ddagger$}\\
    \hline
    NSG~\cite{NSG} & KNNG+RNG & directed & $O(\vert S \vert ^{\frac{1+c}{c}} \cdot \log(\vert S \vert) + \vert S \vert ^{1.14})$\tnote{$\dagger$} & $O(\log(\vert S \vert))$\\
    \hline
    HCNNG~\cite{HCNNG} & MST & directed & $O(\vert S \vert \cdot \log(\vert S \vert))$ & $O(\vert S \vert ^{0.4})$\tnote{$\ddagger$}\\
    \hline
    Vamana~\cite{DiskANN} & RNG & directed & $O(\vert S \vert ^{1.16}$)\tnote{$\ddagger$} & $O(\vert S \vert ^{0.75}$)\tnote{$\ddagger$}\\
    \hline
    NSSG~\cite{NSSG} & KNNG+RNG & directed & $O(\vert S \vert + \vert S \vert ^{1.14})$ & $O(\log(\vert S \vert))$ \\
    \hline
    \end{tabular}
      \footnotesize
      {$^\dagger$ $c$, $t$ are the constants. $^\ddagger$ Complexity is not informed by the authors; we derive it based on the related papers’ descriptions and experimental estimates. See \hyperref[Appendix_D]{Appendix D} for deatils.}
    \end{threeparttable}
\end{minipage}
\begin{minipage}{0.405\textwidth}
  \centering
  \includegraphics[width=0.7\linewidth]{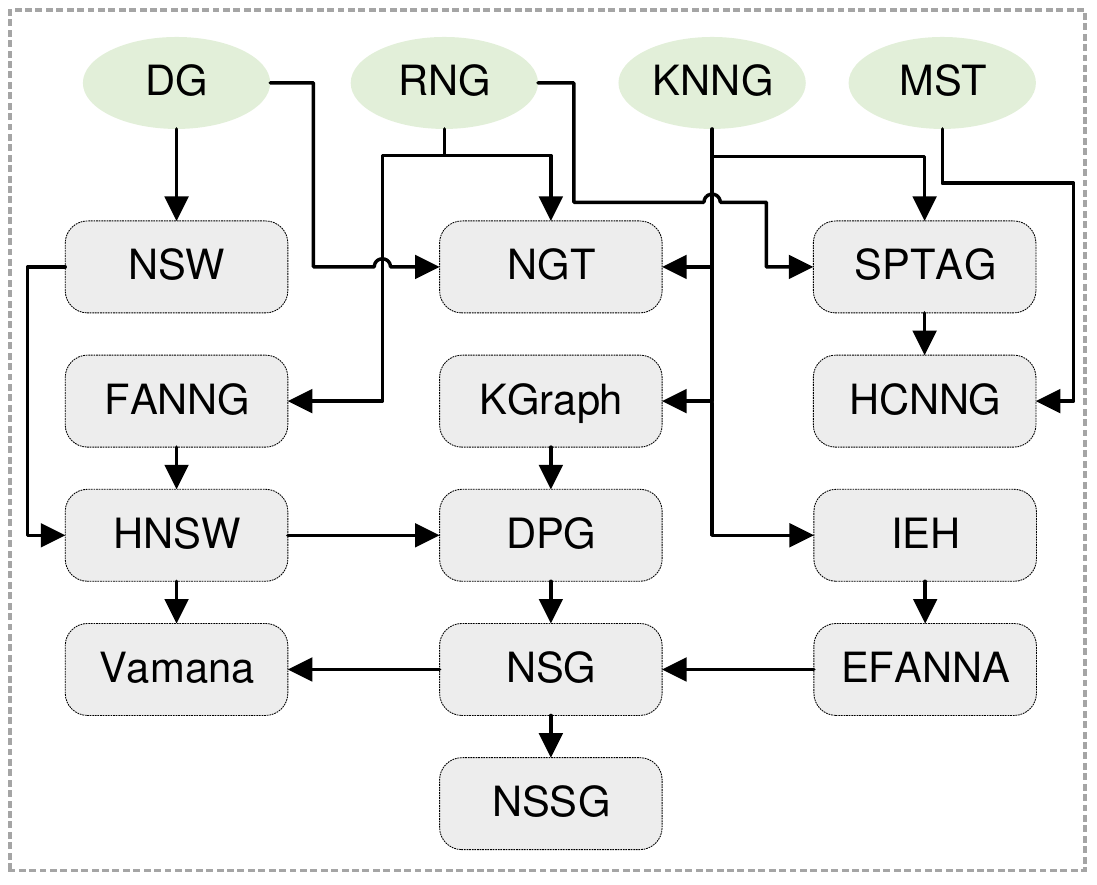}
  \figcaption{Roadmaps of graph-based ANNS algorithms. The arrows from a base graph (green shading) to an algorithm (gray shading) and from one algorithm to another indicate the dependence and development relationships.}
  \label{fig: roadmap}
\end{minipage}
\vspace{-0.55cm}

\end{figure*}

Although the formal definition of base graphs facilitates theoretical analysis, it is impractical for them to be applied directly to ANNS~\cite{NSG}. Obviously, their high construction complexity is difficult to scale to large-scale datasets. This has become even truer with the advent of frequently updated databases~\cite{li2018design}. In addition, it is difficult for base graphs to achieve high search efficiency in high-dimensional scenarios~\cite{FANNG,NSG,ponomarenko2014comparative}. Thus, a number of graph-based ANNS algorithms tackle improving the base graphs from one or several aspects. Next, we outline 13 representative graph-based ANNS algorithms (\textbf{A1–A13}) based on the aforementioned four base graphs and their development roadmaps (\autoref{fig: roadmap}). \hyperref[tab2]{Table 2} summarizes some important properties about algorithms.

\vspace{0.5em}

\noindent\textbf{DG-based and RNG-based ANNS algorithms (NSW, HNSW, FANNG, NGT).} To address the high degree of DG in high dimension, some slight improvements have been proposed~\cite{kleinberg2000small,beaumont2007voronet,beaumont2007peer}. However, they rely heavily on DG’s quality and exist the \textit{curse of dimensionality}~\cite{NSW}. Therefore, some algorithms add an RNG approximation on DG to diversify the distribution of neighbors~\cite{HNSW}.

\noindent\underline{\textbf{A1: Navigable Small World graph (NSW).}} NSW~\cite{NSW} constructs an undirected graph through continuous insertion of elements and ensures global connectivity (approximate DG). The intuition is that the result of a greedy traversal (random seeds) is always the nearest neighbor on DG~\cite{HNSW}. The long edges formed in the beginning of construction have small-world navigation performance to ensure search efficiency, and the vertices inserted later form short-range edges, which ensure search accuracy. NSW also achieved excellent results in the maximum inner product search~\cite{morozov2018non,liu2020understanding}. However, according to the evaluation of \cite{ponomarenko2014comparative}, NSW provides limited best tradeoffs between efficiency and effectiveness compared to non-graph-based indexes, because its search complexity is poly-logarithmic~\cite{NaidanBN15}. In addition, NSW uses undirected edges to connect vertices, which results in vertices in dense areas acting as the “traffic hubs” (high out-degrees), thus damaging search efficiency.

\noindent\underline{\textbf{A2: Hierarchical Navigable Small World graphs (HNSW).}} An improvement direction is put forth by \cite{malkov2016growing,boguna2009navigability} to overcome NSW’s poly-logarithmic search complexity. Motivated by this, HNSW~\cite{HNSW} generates a hierarchical graph and fixes the upper bound of each vertex’s number of neighbors, thereby allowing a logarithmic complexity scaling of search. Its basic idea is to separate neighbors to different levels according to the distance scale, and the search is an iterative process from top to bottom. For an inserted point, HNSW not only selects its nearest neighbors (approximate DG), but also considers the distribution of neighbors (approximate RNG). HNSW has been deployed in various applications \cite{boytsov2016off,kosuge2019object,bee2020content} because of its unprecedented superiority. However, its multilayer structure significantly increases the memory usage and makes it difficult to scale to larger datasets~\cite{NSG}. Meawhile, \cite{lin2019graph} experimentally verifies that the hierarchy’s advantage fades away as intrinsic dimension goes up (>32). Hence, recent works try to optimize HNSW by hardware or heterogeneous implementation to alleviate these problems~\cite{zhang2019grip,deng2019pyramid}.

\noindent\underline{\textbf{A3: Fast Approximate Nearest Neighbor Graph (FANNG).}} An occlusion rule is proposed by FANNG~\cite{FANNG} to cut off redundant neighbors (approximate RNG). Unlike HNSW's approximation to RNG (HNSW only considers a small number of vertices returned by greedy search), FANNG's occlusion rule is applied to all other points on the dataset except the target point, which leads to high construction complexity. Thus, two intuitive optimizations of candidate neighbor acquisition are proposed to alleviate this problem~\cite{FANNG}. To improve the accuracy, FANNG uses backtrack to the second-closest vertex and considers its edges that have not been explored yet.


\noindent\underline{\textbf{A4: Neighborhood Graph and Tree (NGT).}} NGT~\cite{NGT} is a library for performing high-speed ANNS released by Yahoo Japan Corporation. It contains two construction methods. One is to transform KNNG into Bi-directed KNNG (BKNNG), which adds reverse edges to each directed edge on KNNG~\cite{yahoo2}. The other is constructed incrementally like NSW (approximate to DG)~\cite{yahoo2}. The difference from NSW is range search (a variant of greedy search) used during construction. Both of the aforementioned methods make certain hub vertices have a high out-degree, which will seriously affect search efficiency. Therefore, NGT uses three degree-adjustment methods to alleviate this problem, and within the more effective path adjustment is an approximation to RNG (see \hyperref[Appendix_B]{Appendix B} for proof)~\cite{yahoo3}. This reduces memory overhead and improves search efficiency. NGT obtains the seed vertex through the VP-tree~\cite{yahoo3}, and then uses the range search to perform routing. Interestingly, the NGT-like path adjustment and range search are also used by the k-DR algorithm in~\cite{aoyama2011fast} (see \hyperref[Appendix_N]{Appendix N} for details).

\vspace{0.5em}
\noindent\textbf{KNNG-based ANNS algorithms (SPTAG, KGraph, EFANNA, IEH).} A naive construction for KNNG is exhaustively comparing all pairs of points, which is prohibitively slow and unsuitable for large dataset $S$. Some early solutions construct an additional index (such as tree~\cite{paredes2006practical} or hash~\cite{uno2009efficient,zhang2013fast}), and then find the neighbors of each point through ANNS. However, such methods generally suffer from high index construction complexity~\cite{chen2009fast}. This is because they ignore this fact: the queries belong to $S$ in the graph construction process, but the queries of ANNS generally do not~\cite{wang2012scalable}. Thus, it is unnecessary to ensure a good result is given for general queries on additional index~\cite{wang2012scalable}. There are two types of representative solutions, which only focus on graph construction.

\noindent\underline{\textbf{A5: Space Partition Tree and Graph (SPTAG).}} One is based on divide and conquer, and its representative is SPTAG~\cite{SPTAG}, a library released by Microsoft. SPTAG hierarchically divides dataset $S$ into subsets (through Trinary-Projection Trees~\cite{wang2013trinary}) and builds an exact KNNG over each subset. This process repeats multiple times to produce a more accurate KNNG on $S$. Moreover, SPTAG further improves KNNG’s accuracy by performing neighborhood propagation~\cite{wang2012scalable}. The early version of SPTAG added multiple KD-trees on $S$ to iteratively obtain the seeds closer to the query~\cite{wang2012query}. However, on extremely high-dimensional $S$, the KD-trees will produce an inaccurate distance bound estimation. In response, the balanced k-means trees are constructed to replace the KD-trees~\cite{SPTAG}. Inspired by the universal superiority brought about by RNG, SPTAG has recently added the option of approximating RNG in the project~\cite{SPTAG}.

\noindent\underline{\textbf{A6: KGraph.}} The other is based on NN-Descent~\cite{NNDescent}; its basic idea is \textit{neighbors are more likely to be neighbors of each other}~\cite{EFANNA}. KGraph~\cite{KGraph} first adopts this idea to reduce KNNG’s construction complexity to $O(\vert S \vert ^{1.14})$ on dataset $S$. It achieves better search performance than NSW~\cite{Benchmarking}. Therefore, some NN-Descent-based derivatives are developed to explore its potential~\cite{bratic2018nn,zhao2019merge,zhao2018k}.

\noindent\underline{\textbf{A7: EFANNA and A8: IEH.}} Instead of random initialization during construction (such as KGraph), Extremely Fast Approximate Nearest Neighbor Search Algorithm (\textbf{EFANNA})~\cite{EFANNA} first builds multiple KD-trees on $S$, and better initializes the neighbors of each vertex through ANNS on these KD-trees, then executes NN-Descent. At the search stage, EFANNA also uses these KD-trees to obtain seeds that are closer to the query. The idea of initializing seeds through additional structures is inspired by Iterative Expanding Hashing (\textbf{IEH})~\cite{IEH}, which uses hash buckets to obtain better seeds. However, IEH's KNNG is constructed by brute force in \cite{IEH}.

\vspace{0.5em}
\noindent\textbf{KNNG-based and RNG-based ANNS algorithms (DPG, NSG, NSSG, Vamana).} The early optimization of KGraph was limited to improving graph quality~\cite{EFANNA,zhao2018k}. Their intuition is that higher graph quality leads to better search performance. Hence, each vertex is only connected to $K$ nearest neighbors without considering the distribution of neighbors. According to the comparative analysis of \cite{lin2019graph}, if neighbors of a visiting vertex are close to each other, it will guide the search to the same location. That is, it is redundant to compare the query to all neighbors close to each other~\cite{FANNG,HNSW}.

\noindent\underline{\textbf{A9: Diversified Proximity Graph (DPG).}} To overcome the aforementioned issue, DPG~\cite{DPG} practices optimization to control neighbors’ distribution on KGraph. It sets the threshold of the angle between the neighbors of a vertex to make the neighbors evenly distributed in all directions of the vertex. This is only an approximate implementation of RNG from another aspect (see \hyperref[Appendix_C]{Appendix C} for the proof). In addition, to deal with $S$ with a large number of clusters, DPG keeps bi-directed edges on the graph.

\noindent\underline{\textbf{A10: Navigating Spreading-out Graph (NSG).}} Although DPG’s search performance is comparable to HNSW, it suffers from a large index~\cite{NSG}. To settle this problem and further improve search performance, NSG~\cite{NSG} proposes an edge selection strategy based on monotonic RNG (called MRNG), which is actually equivalent to HNSW's (see \hyperref[Appendix_A]{Appendix A} for the proof). Its construction framework is inspired by DPG; that is, to prune edges on KNNG. NSG ensures high construction efficiency by executing ANNS on KGraph to obtain candidate neighbors. NSG has been integrated into Alibaba's Taobao e-commerce platform to combine superior index construction and search performance~\cite{NSG}, and its billion-scale implementation version exceeds the current best FAISS~\cite{FAISS}.

\noindent\underline{\textbf{A11: Navigating Satellite System Graph (NSSG).}} NSSG continues to explore the potential of pruning edges on KNNG, and proposes an edge selection strategy based on SSG~\cite{NSSG}. When obtaining a vertex’s candidate neighbors, instead of conducting the ANNS like NSG, it gets the neighbors and neighbors' neighbors of the vertex on KNNG, which significantly improves construction efficiency. Both SSG and MRNG are approximations to RNG, but SSG is relatively relaxed when cutting redundant neighbors. Therefore, NSSG has a larger out-degree. Although \cite{NSSG} believes that SSG is more beneficial to ANNS than MRNG, we reach the opposite conclusion through a fairer evaluation (see \S \ref{components_evaluation} for details).

\noindent\underline{\textbf{A12: Vamana.}} Microsoft recently proposed Vamana~\cite{DiskANN} to combine with solid-state drives (SSD) for billions of data. It analyzes the construction details of HNSW and NSG to extract and combine the better parts. Its construction framework is motivated by NSG. Instead of using KGraph to initialize like NSG, Vamana initializes randomly. When selecting neighbors, Vamana improves the HNSW's strategy by adding a parameter $\alpha$ to increase the edge selection’s flexibility and executing two passes with different $\alpha$. Experiments show that its result graph has a shorter average path length when searching, which works well with SSD.

\begin{figure*}[!t]
  \centering
  \setlength{\abovecaptionskip}{0.05cm}
  \setlength{\belowcaptionskip}{-0.45cm}
  \includegraphics[width=0.95\linewidth]{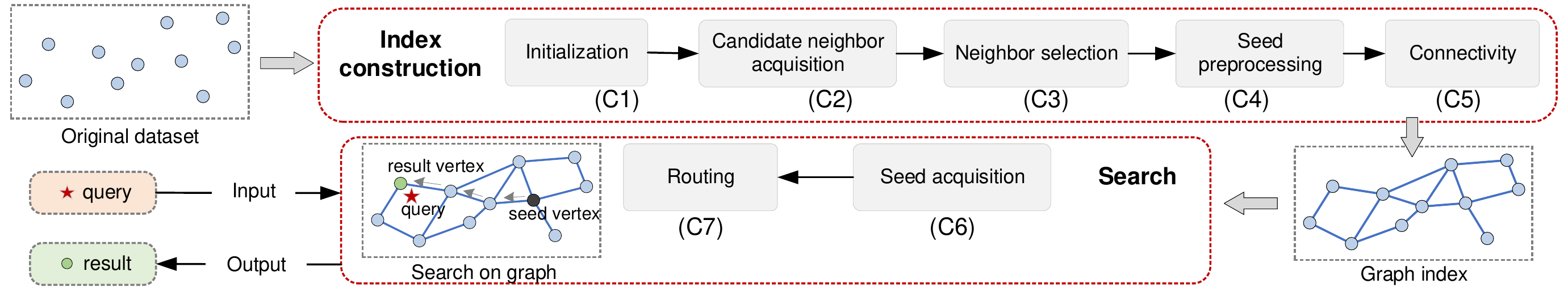}
  \caption{The pipeline of graph-based ANNS algorithms. An algorithm can be divided into two coarse-grained components: index construction, search. We subdivide the index construction into five fine-grained components (C1–C5), the search into two fine-grained components (C6–C7).}
  \label{fig: components}
\end{figure*}

\vspace{0.5em}
\noindent\textbf{MST-based ANNS algorithms (HCNNG).}

\noindent\underline{\textbf{A13: HCNNG.}} Different from the aforementioned techniques, a recent method called Hierarchical Clustering-based Nearest Neighbor Graph (HCNNG)~\cite{HCNNG} uses MST to connect the points on dataset $S$. It uses the same divide-and-conquer framework as SPTAG. The difference is that HCNNG divides $S$ through multiple hierarchical clusters, and all points in each cluster are connected through MST. HCNNG uses multiple global KD-trees to get seeds (like SPTAG and EFANNA). Then to improve search efficiency, rather than using traditional greedy search, it performs an efficient guided search.



\vspace{-0.15cm}
\section {Components’ Analysis} \label{sec4}
Despite the diversity of graph-based ANNS algorithms, they all follow a unified processing pipeline. As \autoref{fig: components} shows, an algorithm can be divided into two coarse-grained components: index construction (top) and search (bottom), which are adopted by most of the current work to analyze algorithms~\cite{NSW,DPG,FANNG,HCNNG}. Recent research has endeavored to take a deeper look at some fine-grained components~\cite{NSG,NSSG}, prompting them to find out which part of an algorithm plays a core role and then propose better algorithms. Motivated by this, we subdivide the index construction and search into seven fine-grained components (\textbf{C1–C7} in \autoref{fig: components}), and compare all 13 graph-based algorithms discussed in this paper by them.

\vspace{-0.2cm}
\subsection{Components for Index Construction} \label{sec4-1}
The purpose of index construction is to organize the dataset $S$ with a graph. Existing algorithms are generally divided into three strategies: \textit{\textbf{Divide-and-conquer}}~\cite{virmajoki2004divide}, \textit{\textbf{Refinement}}~\cite{NNDescent}, and \textit{\textbf{Increment}}~\cite{hajebi2011fast} (see \hyperref[Appendix_E]{Appendix E}). As \autoref{fig: components} (top) show, an algorithm’s index construction can be divided into five detailed components (\textbf{C1–C5}). Among them, \textit{initialization} can be divided into three ways according to different construction strategies.

\noindent\underline{\textbf{C1: Initialization.}}

\noindent\textbf{Overview.} The \textit{initialization} of \textit{\textbf{Divide-and-conquer}} is \textit{dataset division}; it is conducted recursively to generate many subgraphs so that the index is obtained by subgraph merging~\cite{shimomura2019hgraph,chen2009fast}. For \textit{\textbf{Refinement}}, in the \textit{initialization}, it performs \textit{neighbor initialization} to get the initialized graph, then refines the initialized graph to achieve better search performance~\cite{EFANNA,NSG}. While the \textit{\textbf{Increment}} inserts points continuously, the new incoming point is regarded as a query, then it executes ANNS to obtain the query's neighbors on the subgraph constructed by the previously inserted points~\cite{NSW,HNSW}; it therefore implements \textit{seed acquisition} during \textit{initialization}.


\vspace{-0.4em}
\begin{definition}
  \label{def: dataset division}
  \textbf{Dataset Division.} Given dataset $S$, the \textit{dataset division} divides $S$ into $m$ small subsets—i.e., $S_0, S_1, \cdots,S_{m-1}$, and $S_0 \cup S_1 \cdots \cup S_{m-1}=S$.
\end{definition}
\vspace{-0.4em}

\noindent{\textbf{Data division.}} This is a unique \textit{initialization} of the  \textit{\textbf{Divide-and-conquer}} strategy. SPTAG previously adopts a random division scheme, which generates the principal directions over points randomly sampled from $S$, then performs random divisions to make each subset’s diameter small enough~\cite{DBLP:conf/uai/VermaKD09, wang2012scalable}. To achieve better division, SPTAG turns to TP-tree~\cite{wang2013trinary}, in which a partition hyperplane is formed by a linear combination of a few coordinate axes with weights being -1 or 1. HCNNG divides $S$ by iteratively performing hierarchical clustering. Specifically, it randomly takes two points from the set to be divided each time, and performs division by calculating the distance between other points and the two~\cite{HCNNG}.

\vspace{-0.4em}
\begin{definition}
  \label{def: neighbor initialization}
  \textbf{Neighbor Initialization.} Given dataset $S$, for $\forall p \in S$, the \textit{neighbor initialization} gets the subset $C$ from $S \setminus \left \{ p \right \}$, and initializes $N(p)$ with $C$.
\end{definition}
\vspace{-0.4em}

\noindent{\textbf{Neighbor initialization.}} Only the \textit{initialization} of the \textit{\textbf{Refinement}} strategy requires this implementation. Both KGraph and Vamana implement this process by randomly selecting neighbors~\cite{KGraph,DiskANN}. This method offers high efficiency but the initial graph quality is too low. The solution is to initialize neighbors through ANNS based on hash-based~\cite{uno2009efficient} or tree-based~\cite{EFANNA} approaches. EFANNA deploys the latter; it establishes multiple KD-trees on $S$. Then, each point is treated as a query, and get its neighbors through ANNS on multiple KD-trees~\cite{EFANNA}. This approach relies heavily on extra index and increases the cost of index construction. Thus, NSG, DPG, and NSSG deploy the NN-Descent~\cite{NNDescent}; they first randomly select neighbors for each point, and then update each point's neighbors with neighborhood propagation. Finally, they get a high-quality initial graph by a small number of iterations. Specially, FANNG and IEH initialize neighbors via linear scan.

\vspace{-0.5em}
\begin{definition}
  \label{def: seed acquisition}
  \textbf{Seed Acquisition.} Given the index $G(V,E)$, the \textit{seed acquisition} acquires a small subset $\widehat{S}$ from $V$ as the seed set, and ANNS on $G$ starts from $\widehat{S}$.
\end{definition}
\vspace{-0.5em}

\noindent{\textbf{Seed acquisition.}} The \textit{seed acquisition} of the index construction is \textit{\textbf{Increment}} strategy's \textit{initialization}. The other two strategies may also include this process when acquiring candidate neighbors, and this process also is necessary for all graph-based algorithms in the search. For index construction, both NSW and NGT obtain seeds randomly~\cite{NSW,NGT}, while HNSW makes its seed points fixed from the top layer because of its unique hierarchical structure~\cite{HNSW}.

\vspace{-0.5em}
\begin{definition}
  \label{candidate neighbor acquisition}
  \textbf{Candidate Neighbor Acquisition.} Given a finite dataset $S$, point $p \in S$, the \textit{candidate neighbor acquisition} gets a subset $\mathcal{C}$ from $S \setminus \left \{ p \right \}$ as $p$'s candidate neighbors, and $p$ get its neighbors $N(p)$ from $\mathcal{C}$—that is, $N(p) \subset \mathcal{C}$.
\end{definition}
\vspace{-0.5em}

\noindent\underline{\textbf{C2: Candidate neighbor acquisition.}} The graph constructed by the \textit{\textbf{Divide-and-conquer}} generally produce candidate neighbors from a small subset obtained after \textit{dataset division}. For a subset $S_i \subset S$ and a point $p \in S_i$, SPTAG and HCNNG directly take $S_i \setminus \left \{p \right \}$ as candidate neighbors~\cite{wang2012scalable,HCNNG}. Although $\vert S \vert$ may be large, the $\vert S_i \vert$ obtained by the division is generally small. However, \textit{\textbf{Refinement}} and \textit{\textbf{Increment}} do not involve the process of \textit{dataset division}, which leads to low index construction efficiency for IEH and FANNG to adopt the naive method of obtaining candidate neighbors~\cite{IEH,FANNG}. To solve this problem, NGT, NSW, HNSW, NSG, and Vamana all obtain candidate neighbors through ANNS. For a point $p \in S$, the graph $G_{sub}$ (\textit{\textbf{Increment}}) formed by the previously inserted points or the initialized graph $G_{init}$ (\textit{\textbf{Refinement}}), they consider $p$ as a query and execute ANNS on $G_{sub}$ or $G_{init}$, and finally return the query result as candidate neighbors of $p$. This method only needs to access a small subset of $S$. However, according to the analysis of \cite{wang2012scalable}, obtaining candidate neighbors through ANNS is overkill, because the query is in $S$ for index construction, but the ANNS query generally does not belong to $S$. In contrast, KGraph, EFANNA, and NSSG use the neighbors of $p$ and neighbors' neighbors on $G_{init}$ as its candidate neighbors~\cite{NSSG}, which improves index-construction efficiency. DPG directly uses the neighbors of $p$ on $G_{init}$ as candidate neighbors, but to obtain enough candidate neighbors, it generally requires $G_{init}$ with a larger out-degree~\cite{DPG}.

\vspace{-0.5em}
\begin{definition}
  \label{def: neighbor selection}
  \textbf{Neighbor Selection.} Given a point $p$ and its candidate neighbors $\mathcal{C}$, the \textit{neighbor selection} obtains a subset of $\mathcal{C}$ to update $N(p)$.
\end{definition}
\vspace{-0.5em}

\noindent\underline{\textbf{C3: Neighbor selection.}} The current graph-based ANNS algorithms mainly consider two factors for this component: distance and space distribution. Given $p\in S$, the distance factor ensures that the selected neighbors are as close as possible to $p$, while the space distribution factor makes the neighbors distribute as evenly as possible in all directions of $p$. NSW, SPTAG\footnote{\vspace{-0.05cm}This refers to its original version—NGT-panng  for NGT and SPTAG-KDT for SPTAG.\label{original_version}}, NGT\textsuperscript{\ref{original_version}}, KGraph, EFANNA, and IEH only consider the distance factor and aim to build a high-quality graph index~\cite{wang2012scalable,NNDescent}. HNSW\footnote{\vspace{-0.05cm}Although \cite{NSG} distinguishes the \textit{neighbor selection} of HNSW and NSG, we prove the equivalence of the two in \hyperref[Appendix_A]{Appendix A}.\label{HNSW_NSG_neighbors_selection}}, FANNG, SPTAG\footnote{\vspace{-0.1cm}This refers to its optimized version—NGT-onng for NGT and SPTAG-BKT for SPTAG.\label{optimized_version}}, and NSG\textsuperscript{\ref{HNSW_NSG_neighbors_selection}} consider the space distribution factor by evaluating the distance between neighbors, formally, for $x \in \mathcal{C}$, $\forall y \in N(p)$, iff $\delta (x,y)> \delta (y,p)$, $x$ will join $N(p)$~\cite{FANNG,NSG}. To select neighbors more flexibly, Vamana adds the parameter $\alpha$ so that for $x \in \mathcal{C}$, $\forall y \in N(p)$, iff $\alpha \cdot \delta (x,y)> \delta (y,p),(\alpha \geq 1)$, $x$ will be added to $N(p)$~\cite{DiskANN}, so it can control the distribution of neighbors well by adjusting $\alpha$. DPG obtains a subset of $\mathcal{C}$ to minimize the sum of angles between any two points, thereby dispersing the neighbor distribution~\cite{DPG}. NSSG considers the space distribution factor by setting an angle threshold $\theta$, for $x \in \mathcal{C}$, $\forall y \in N(p)$, iff $\arccos(x, y) <\theta$, $x$ will join $N(p)$. NGT\textsuperscript{\ref{optimized_version}} indirectly attains the even distribution of neighbors with path adjustment~\cite{yahoo3}, which updates neighbors by judging whether there is an alternative path between point $p$ and its neighbors on $G_{init}$. HCNNG selects neighbors for $p$ by constructing an MST on $\left \{p \right \} \cup \mathcal{C}$~\cite{HCNNG}. Recently, \cite{zhang2019learning,baranchuk2019towards} perform neighbor selection through learning, but these methods are difficult to apply in practice because of their extremely high training costs.

\noindent\underline{\textbf{C4: Seed preprocessing.}} Different algorithms may exist with different execution sequences between this component and the \textit{connectivity}, such as NSW~\cite{NSW}, NSG~\cite{NSG}. Generally, graph-based ANNS algorithms implement this component in a static or dynamic manner. For the static method, typical representatives are HNSW, NSG, Vamana, and NSSG. HNSW fixes the top vertices as the seeds, NSG and Vamana use the approximate centroid of $S$ as the seed, and the seeds of NSSG are randomly selected vertices. While for the dynamic method, a common practice is to attach other indexes (i.e., for each query, the seeds close to the query are obtained through an additional index). SPTAG, EFANNA, HCNNG, and NGT build additional trees, such as KD-tree~\cite{EFANNA,SPTAG}, balanced k-means tree~\cite{SPTAG}, and VP-tree~\cite{NGT}. IEH prepares for \textit{seed acquisition} through hashing~\cite{IEH}. Then \cite{DBLP:conf/cvpr/DouzeSJ18} compresses the original vector by OPQ ~\cite{DBLP:conf/cvpr/GeHK013} to obtain the seeds by quickly calculating the compressed vector. Random \textit{seed acquisition} is adopted by KGraph, FANNG, NSW, and DPG, and they don't need to implement \textit{seed preprocessing}.

\noindent\underline{\textbf{C5: Connectivity.}} \textit{\textbf{Incremental}} strategy internally ensures connectivity (e.g., NSW). \textit{\textbf{Refinement}} generally attaches depth-first traversal to achieve this~\cite{NSG} (e.g., NSG). \textit{\textbf{Divide-and-conquer}} generally ensures connectivity by multiply performing \textit{dataset division} and subgraph construction (e.g., SPTAG).

\vspace{-0.15cm}
\subsection{Components for Search} \label{sec4-2}
We subdivide the search into two fine-grained components (\textbf{C6–C7}): \textit{seed acquisition} and \textit{routing}.

\noindent\underline{\textbf{C6: Seed acquisition.}} Because the seed has a significant impact on search, this component of the search process is more concerned than the \textit{initialization} of \textit{\textbf{Incremental}} strategy. Some early algorithms obtain the seeds randomly, while state-of-the-art algorithms commonly use \textit{seed preprocessing}. If the fixed seeds are produced in the preprocessing stage, it can be loaded directly at this component. If other index structures are constructed in the preprocessing stage, ANNS returns the seeds with the additional structure.

\begin{figure*}[!t]
  \centering
  \setlength{\abovecaptionskip}{-0.3em}
  \setlength{\belowcaptionskip}{-0.4cm}
  \includegraphics[width=\linewidth]{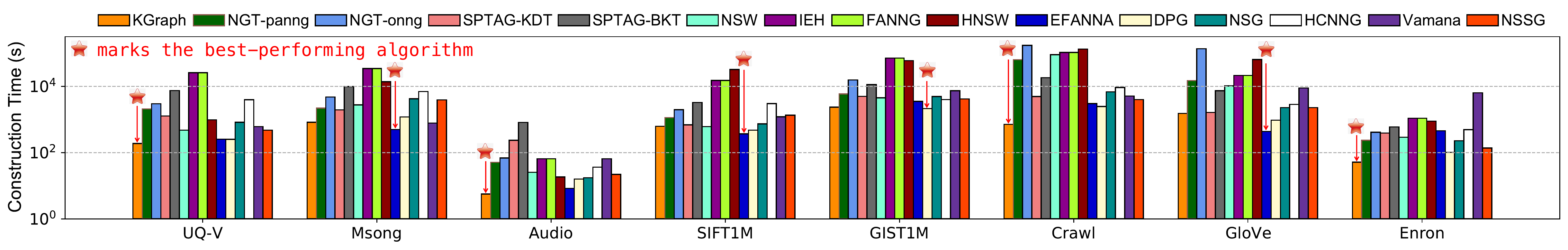}
  \caption{Index construction time of all compared algorithms on real-world datasets (the bar marked with a red star is the best).}
  \label{fig: real_indexing_time}
\end{figure*}

\begin{figure*}[!t]
  \centering
  \setlength{\abovecaptionskip}{-0.3em}
  \setlength{\belowcaptionskip}{-0.45cm}
  \includegraphics[width=\linewidth]{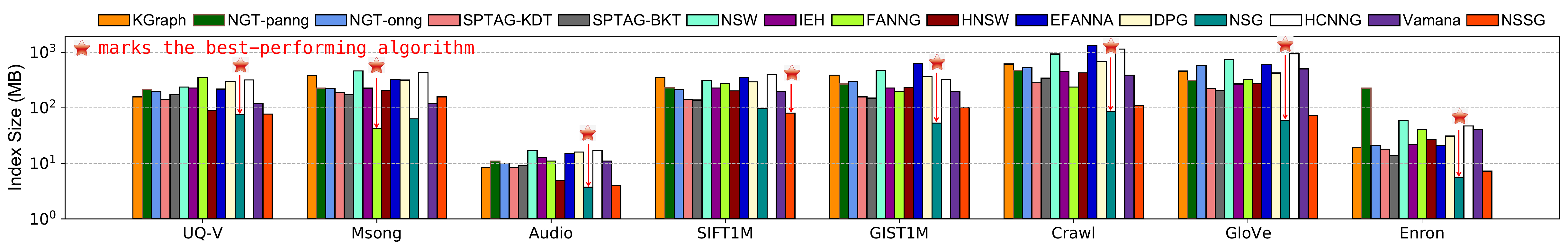}
  \caption{Index size of all compared algorithms on real-world datasets (the bar marked with a red star is the best).}
  \label{fig: real_index_size}
\end{figure*}

\vspace{-0.5em}
\begin{definition}
  \label{routing}
  \textbf{Routing.} Given $G(V,E)$, query $q$, seed set $\widehat{S}$, the \textit{routing} starts from the vertices in $\widehat{S}$, and then converges to $q$ by neighbor propagation along the neighbor $n$ of the visited point with smaller $\delta (n,q)$, until the vertex $r$ so that $\delta (r,q)$ reaches a minimum.
\end{definition}
\vspace{-0.5em}

\vspace{-0.5em}
\begin{definition}
  \label{best_first_search}
  \textbf{Best First Search.} Given $G(V,E)$, query $q$, and vertices to be visited $\mathcal{C}$, its maximum size is $c$ and the result set $\mathcal {R}$. We initialize $\mathcal{C}$ and $\mathcal{R}$ with seed set $\widehat{S}$. For $\hat{x}=\arg\, \min _{x\in \mathcal{C}}{\delta (x,q)}$, best first search access $N(\hat{x})$, then $\mathcal{C} \setminus \left \{\hat{x} \right \}$ and $\mathcal{ C} \cup N(\hat{x})$. To keep $\vert \mathcal{C} \vert=c$, $\hat{y}=\arg\, \max _{y\in \mathcal{C}}{\delta (y,q)}$ will be deleted. $\forall n \in N(\hat{x})$, if $\delta (n,q) <\delta(\hat{z},q), \hat{z}=\arg \, \max _{z \in \mathcal{R}} {\delta (z,q)}$, then $\mathcal{R} \setminus \left \{\hat{z} \right \}$ and $\mathcal {R} \cup \left \{n \right \}$. The aforementioned process is performed iteratively until $\mathcal{R}$ is no longer updated. (see \hyperref[Appendix_F]{Appendix F} for the pseudocode)
\end{definition}
\vspace{-0.5em}

\noindent\underline{\textbf{C7: Routing.}} Almost all graph-based ANNS algorithms are based on a greedy \textit{routing} strategy, including best first search (BFS) and its variants. NSW, HNSW, KGraph, IEH, EFANNA, DPG, NSG, NSSG, and Vamana use the original BFS to perform \textit{routing}. Despite this method being convenient for deployment, it has two shortcomings: susceptibility to local optimum (S1)~\cite{learn_to_route} and low routing efficiency (S2)~\cite{HCNNG}. S1 destroys the search results’ accuracy. For this problem, FANNG adds backtracking to BFS, which slightly improves the search accuracy while significantly increasing the search time~\cite{FANNG}. NGT alleviates S1 by adding a parameter $\epsilon$. On the basis of \hyperref[best_first_search]{Definition 4.7}, it cancels the size restriction on $\mathcal{C}$ and takes $\delta (\hat {y},q)$ as the search radius $r$, for $\forall n \in N(\hat{x})$, if $\delta (n,q) <(1+\epsilon) \cdot r$, then $n$ is added to $\mathcal{C}$. Setting $\epsilon$ to a larger value can alleviate S1, but it will also significantly increase the search time~\cite{yahoo3}. SPTAG solves S1 by iteratively executing BFS. When a certain iteration falls into a local optimum, it will restart the search by selecting new seeds from the KD-tree~\cite{wang2012query}. HCNNG proposes using guided search to alleviate S2 rather than visiting all $N(\hat{x})$ like BFS, so guided search avoids some redundant visits based on the query's location. Recently, some of the literature uses learning methods to perform routing~\cite{learn_to_route,li2020improving,vargas2019genetic}. These methods usually alleviate S1 and S2 simultaneously, but the adverse effect is that this requires extra training, and additional information also increases the memory overhead (see \S \ref{machine_learning_optimization}).

\section{Experimental Evaluation} \label{sec5}
This section presents an abundant experimental study of both individual algorithms (\S \ref{sec3}) and components (\S \ref{sec4}) extracted from the algorithms for graph-based ANNS. Because of space constraints, some of our experimental content is provided in appendix. Our evaluation seeks to answer the following question:

\noindent\textbf{Q1:} How do the algorithms perform in different scenarios? (\S \ref{index_construction_evaluation}–\ref{search_performane})

\noindent\textbf{Q2:} Can an algorithm have the best index construction and search performance at the same time? (\S \ref{index_construction_evaluation}–\ref{search_performane})

\noindent\textbf{Q3:} For an algorithm with the best overall performance, is the performance of each fine-grained component also the best? (\S \ref{components_evaluation})

\noindent\textbf{Q4:} How do machine learning-based optimizations affect the performance of the graph-based algorithms? (\S \ref{machine_learning_optimization})

\noindent\textbf{Q5:} How can we design a better graph-based algorithm based on the experimental observations and verify its performance? (\S \ref{sec6})


\vspace{-0.15cm}
\subsection{Experimental Setting}
\noindent\underline{\textbf{Datasets.}} Our experiment involves eight real-world datasets popularly deployed by existing works, which cover various applications such as video (\textit{\textbf{UQ-V}}~\cite{UQ-V}), audio (\textit{\textbf{Msong}}~\cite{Msong}, \textit{\textbf{Audio}}~\cite{Audio}), text (\textit{\textbf{Crawl}}~\cite{Crawl}, \textit{\textbf{GloVe}}~\cite{GloVe}, \textit{\textbf{Enron}}~\cite{Enron}), and image (\textit{\textbf{SIFT1M}}~\cite{texmex}, \textit{\textbf{GIST1M}}~\cite{texmex}). Their main characteristics are summarized in \autoref{tab: Dataset}. \# Base is the number of elements in the base dataset. LID indicates local intrinsic dimensionality, and a larger LID value implies a ``harder”  dataset~\cite{DPG}. Additionally, 12 synthetic datasets are used to test each algorithm’s scalability to different datasets’ performance (e.g., dimensionality, cardinality, number of clusters, and standard deviation of the distribution in each cluster~\cite{graph_survey}). Out of space considerations, please see the scalability evaluation in \hyperref[Appendix_J]{Appendix J}. All datasets in the experiment are processed into the base dataset, query dataset, and ground-truth dataset.

\setlength{\textfloatsep}{0cm}
\setlength{\floatsep}{0cm}
\begin{table}[!tb]
  \centering
  \setlength{\abovecaptionskip}{0.05cm}
  \setstretch{0.9}
  \fontsize{6.5pt}{3.3mm}\selectfont
  \caption{Statistics of real-world datasets.}
  \label{tab: Dataset}
  \setlength{\tabcolsep}{.025\linewidth}{
  \begin{tabular}{l|l|l|l|l}
    \hline
    \textbf{Dataset} & \textbf{Dimension} & \textbf{\# Base} & \textbf{\# Query} & \textbf{LID}~\cite{DPG,NSSG}\\
    \hline
    \hline
    UQ-V~\cite{UQ-V} & 256 & 1,000,000 & 10,000 & 7.2 \\
    \hline
    Msong~\cite{Msong} & 420 & 992,272 & 200 & 9.5 \\
    \hline
    Audio~\cite{Audio} & 192 & 53,387 & 200 & 5.6\\
    \hline
    SIFT1M~\cite{texmex} & 128 & 1,000,000 & 10,000 & 9.3 \\
    \hline
    GIST1M~\cite{texmex} & 960 & 1,000,000 & 1,000 & 18.9 \\
    \hline
    Crawl~\cite{Crawl} & 300 & 1,989,995 & 10,000 & 15.7 \\
    \hline
    GloVe~\cite{GloVe} & 100 & 1,183,514 & 10,000 & 20.0 \\
    \hline
    Enron~\cite{Enron} & 1,369 & 94,987 & 200 & 11.7 \\
    \hline
  \end{tabular}
  }
\end{table}

\noindent\underline{\textbf{Compared algorithms.}} Our experiment evaluates 13 representative graph-based ANNS algorithms mentioned in \S \ref{sec3}, which are carefully selected from research literature and practical projects. The main attributes and experimental parameters of these algorithms are introduced in \hyperref[Appendix_E]{Appendix E} and \hyperref[Appendix_F]{Appendix F}.

\noindent\underline{\textbf{Evaluation metrics.}} To measure the algorithm’s overall performance, we employ various metrics related to index construction and search. For index construction, we evaluate the index construction efficiency and size. Some index characteristics such as \textit{graph quality}, \textit{average out-degree}, and \textit{the number of connected components} are recorded; they indirectly affect index construction efficiency and size. Given a proximity graph $G^{\prime}(V^{\prime}, E^{\prime})$ (graph index of an algorithm) and the exact graph $G(V, E)$ on the same dataset, we define \textit{graph quality} of an index as $\frac{|E^{\prime} \bigcap E|}{|E|}$~\cite{wang2013fast,chen2009fast,boutet2016being}. For search, we evaluate search efficiency, accuracy, and memory overhead. Search efficiency can be measured by \textit{queries per second (QPS)} and \textit{speedup}. \textit{QPS} is the ratio of the number of queries ($\# q$) to the search time ($t$); i.e., $\frac{\# q}{t}$~\cite{NSG}. \textit{Speedup} is defined as $\frac{\vert S \vert}{NDC}$, where $\vert S \vert$ is the dataset’s size and is also the number of distance calculations of the linear scan for a query, and $NDC$ is the number of distance calculations of an algorithm for a query (equal to $\vert \mathcal{C} \vert$ in \hyperref[def2]{Definition 2.2})~\cite{HCNNG}. We use the \textit{recall} rate to evaluate the search accuracy, which is defined as $\frac{\vert R \cup T \vert}{\vert T \vert}$, where $R$ is an algorithm’s query result set, $T$ is the real result set, and $\vert R \vert = \vert T \vert$. We also measure other indicators that indirectly reflect search performance, such as \textit{the candidate set size} during the search and \textit{the average query path length}.

\noindent\underline{\textbf{Implementation setup.}} We reimplement all algorithms by C++; they were removed by all the SIMD, pre-fetching instructions, and other hardware-specific optimizations. To improve construction efficiency, the parts involving vector calculation are parallelized for index construction of each algorithm~\cite{tellez2021scalable,aumuller2019benchmarking}. All C++ source codes are compiled by g++ 7.3, and MATLAB source codes (only for index construction of a hash table in IEH~\cite{IEH}) are compiled by MATLAB 9.9. All experiments are conducted on a Linux server with a Intel(R) Xeon(R) Gold 5218 CPU at 2.30GHz, and a 125G memory.

\noindent\underline{\textbf{Parameters.}} Because parameters’ adjustment in the entire base dataset may cause overfitting~\cite{NSG}, we randomly sample a certain percentage of data points from the base dataset to form a validation dataset. We search for the optimal value of all the adjustable parameters of each algorithm on each validation dataset, to make the algorithms’ search performance reach the optimal level. Note that high recall areas’ search performance primarily is concerned with the needs of real scenarios.

\vspace{-0.3cm}
\subsection{Index Construction Evaluation}\label{index_construction_evaluation}
We build indexes of all compared algorithms in 32 threads on each real-world dataset. Note that we construct each algorithm with the parameters under optimal search performance.

\noindent\underline{\textbf{Construction efficiency.}} The construction efficiency is mainly affected by the construction strategy, algorithm category, and dataset. In \autoref{fig: real_indexing_time}, the KNNG-based algorithms (e.g., KGraph and EFANNA) constructed by NN-Descent have the smallest construction time among all test algorithms, while the KNNG-based algorithms constructed by divide and conquer (e.g., SPTAG) or brute force (e.g., IEH) have higher construction time. The construction time of RNG-based algorithms vary greatly according to the initial graph. For example, when adding the approximation of RNG on KGraph (e.g., DPG and NSSG), it has a high construction efficiency. However, RNG approximation based on the KNNG built by brute force (e.g., FANNG) has miniscule construction efficiency (close to IEH). Note that Vamana is an exception; its ranking on different datasets has large differences. This is most likely attributable to its neighbor selection parameter $\alpha$ heavily dependent on dataset. The construction time of DG-based algorithms (e.g., NGT and NSW) shows obvious differences with datasets. On some hard datasets (e.g., GloVe), their construction time is even higher than FANNG.

\noindent\underline{\textbf{Index size and average out-degree.}} The index size mainly depends on the average out-degree (AD). Generally, the smaller the AD, the smaller the index size. As \autoref{fig: real_index_size} and \autoref{tab: Index Information} show, RNG-based algorithms (e.g., NSG) have a smaller index size, which is mainly because they cut redundant edges (the lower AD) during RNG approximation. KNNG-, DG-, and MST-based algorithms (e.g., KGraph, NSW, and HCNNG) connect all nearby neighbors without pruning superfluous neighbors, so they always have a larger index size. Additional index structures (e.g., the tree in NGT) will also increase related algorithms’ index size.

\noindent\underline{\textbf{Graph quality.}} The algorithm category and dataset are the main factors that determine graph quality (GQ). In \autoref{tab: Index Information}, the GQ of KNNG-based algorithms (e.g., KGraph) outperform other categories. The approximation to RNG prunes some of the nearest neighbors, thereby destroying RNG-based algorithms’ GQ (e.g., NSG). However, this phenomenon does not happen with DPG, mostly because it undirects all edges. Interestingly, DG- and MST-based algorithms' GQ (e.g., NSW and HCNNG) shows obvious differences with datasets; on simple datasets (e.g., Audio), they have higher GQ, but it degrades on hard datasets (e.g., GIST1M).

\setlength{\textfloatsep}{0cm}
\setlength{\floatsep}{0cm}
\begin{table*}[th!]
\setlength{\abovecaptionskip}{0cm}
\setstretch{0.9}
\fontsize{6.5pt}{3.3mm}\selectfont
    \centering
    \tabcaption{Graph quality (GQ), average out-degree (AD), and \# of connected components (CC) on graph indexes (the bold values are the best).}
    \label{tab: Index Information}
    \setlength{\tabcolsep}{0.00535\linewidth}{
    \begin{tabular}{l|l|l|l|l|l|l|l|l|l|l|l|l|l|l|l|l|l|l|l|l|l|l|l|l}
    \hline
    \multirow{2}*{\textbf{Alg.}} & \multicolumn{3}{c|}{\textbf{UQ-V}} & \multicolumn{3}{c|}{\textbf{Msong}} & \multicolumn{3}{c|}{\textbf{Audio}} & \multicolumn{3}{c|}{\textbf{SIFT1M}} & \multicolumn{3}{c|}{\textbf{GIST1M}} & \multicolumn{3}{c|}{\textbf{Crawl}} & \multicolumn{3}{c|}{\textbf{GloVe}} & \multicolumn{3}{c}{\textbf{Enron}}\\
    \cline{2-25}
    ~ & GQ & AD & CC & GQ & AD & CC & GQ & AD & CC & GQ & AD & CC & GQ & AD & CC & GQ & AD & CC & GQ & AD & CC & GQ & AD & CC \\
    \hline
    \hline
    {KGraph} & 0.974 & 40 & 8,840 & 1.000 & 100 & 3,086 & 0.994 & 40 & 529 & 0.998 & 90 & 331 & 0.995 & 100 & 39,772 & 0.927 & 80 & 290,314 & 0.949 & 100 & 183,837 & 0.992 & 50 & 3,743 \\
    \hline
    {NGT-panng} & 0.770 & 52 & {\color{blue}\textbf{1}} & 0.681 & 56 & {\color{blue}\textbf{1}} & 0.740 & 49 & {\color{blue}\textbf{1}} & 0.762 & 56 & {\color{blue}\textbf{1}} & 0.567 & 67 & {\color{blue}\textbf{1}} & 0.628 & 58 & {\color{blue}\textbf{1}} & 0.589 & 66 & {\color{blue}\textbf{1}} & 0.646 & 55 & {\color{blue}\textbf{1}} \\
    \hline
    {NGT-onng} & 0.431 & 47 & {\color{blue}\textbf{1}} & 0.393 & 55 & {\color{blue}\textbf{1}} & 0.412 & 45 & {\color{blue}\textbf{1}} & 0.424 & 53 & {\color{blue}\textbf{1}} & 0.266 & 75 & {\color{blue}\textbf{1}} & 0.203 & 66 & {\color{blue}\textbf{1}} & 0.220 & 124 & {\color{blue}\textbf{1}} & 0.331 & 53 & {\color{blue}\textbf{1}} \\
    \hline
    {SPTAG-KDT} & 0.957 & 32 & 27,232 & 0.884 & 32 & 110,306 & 0.999 & 32 & 996 & 0.906 & 32 & 23,132 & 0.803 & 32 & 290,953 & 0.821 & 32 & 672,566 & 0.630 & 32 & 594,209 & 0.983 & 32 & 7,500 \\
    \hline
    {SPTAG-BKT} & 0.901 & 32 & 71,719 & 0.907 & 32 & 42,410 & 0.992 & 32 & 61 & 0.763 & 32 & 82,336 & 0.435 & 32 & 45,9529 & 0.381 & 32 & 1,180,072 & 0.330 & 32 & 803,849 & 0.775 & 32 & 20,379 \\
    \hline
    {NSW} & 0.837 & 60 & {\color{blue}\textbf{1}} & 0.767 & 120 & {\color{blue}\textbf{1}} & 0.847 & 80 & {\color{blue}\textbf{1}} & 0.847 & 80 & {\color{blue}\textbf{1}} & 0.601 & 120 & {\color{blue}\textbf{1}} & 0.719 & 120 & {\color{blue}\textbf{1}} & 0.636 & 160 & {\color{blue}\textbf{1}} & 0.796 & 160 & {\color{blue}\textbf{1}} \\
    \hline
    {IEH} & 1.000 & 50 & 24,564 & 1.000 & 50 & 9,133 & 1.000 & 50 & 335 & 1.000 & 50 & 1,211 & 1.000 & 50 & 74,663 & 1.000 & 50 & 289,983 & 1.000 & 50 & 220,192 & 1.000 & 50 & 3,131 \\
    \hline
    {FANNG} & 1.000 & 90 & 3,703 & 0.559 & {\color{blue}\textbf{10}} & 15,375 & 1.000 & 50 & 164 & 0.999 & 70 & 256 & 0.998 & 50 & 47,467 & 0.999 & 30 & 287,098 & 1.000 & 70 & 175,610 & 1.000 & 110 & 1,339 \\
    \hline
    {HNSW} & 0.597 & {\color{blue}\textbf{19}} & 433 & 0.762 & 50 & 36 & 0.571 & 20 & 1 & 0.879 & 49 & 22 & 0.633 & 57 & 122 & 0.726 & 52 & 3,586 & 0.630 & 56 & 624 & 0.833 & 68 & 9 \\
    \hline
    {EFANNA} & 0.975 & 40 & 8,768 & 0.997 & 50 & 10,902 & 0.976 & \textbf{10} & 3,483 & 0.998 & 60 & 832 & 0.981 & 100 & 44,504 & 0.990 & 100 & 227,146 & 0.751 & 100 & 234,745 & 0.999 & 40 & 3,921 \\
    \hline
    {DPG} & 0.973 & 77 & 2 & 1.000 & 82 & {\color{blue}\textbf{1}} & 0.999 & 74 & {\color{blue}\textbf{1}} & 0.998 & 76 & {\color{blue}\textbf{1}} & 0.992 & 94 & {\color{blue}\textbf{1}} & 0.982 & 88 & {\color{blue}\textbf{1}} & 0.872 & 93 & {\color{blue}\textbf{1}} & 0.993 & 84 & {\color{blue}\textbf{1}} \\
    \hline
    {NSG} & 0.562 & {\color{blue}\textbf{19}} & {\color{blue}\textbf{1}} & 0.487 & 16 & {\color{blue}\textbf{1}} & 0.532 & 17 & {\color{blue}\textbf{1}} & 0.551 & 24 & {\color{blue}\textbf{1}} & 0.402 & {\color{blue}\textbf{13}} & {\color{blue}\textbf{1}} & 0.540 & {\color{blue}\textbf{10}} & {\color{blue}\textbf{1}} & 0.526 & {\color{blue}\textbf{12}} & {\color{blue}\textbf{1}} & 0.513 & {\color{blue}\textbf{14}} & {\color{blue}\textbf{1}} \\
    \hline
    {HCNNG} & 0.836 & 41 & {\color{blue}\textbf{1}} & 0.798 & 69 & {\color{blue}\textbf{1}} & 0.847 & 38 & {\color{blue}\textbf{1}} & 0.887 & 61 & {\color{blue}\textbf{1}} & 0.354 & 42 & {\color{blue}\textbf{1}} & 0.503 & 109 & {\color{blue}\textbf{1}} & 0.425 & 167 & {\color{blue}\textbf{1}} & 0.662 & 85 & {\color{blue}\textbf{1}} \\
    \hline
    {Vamana} & 0.034 & 30 & 5,982 & 0.009 & 30 & 2,952 & 0.185 & 50 & {\color{blue}\textbf{1}} & 0.021 & 50 & 82 & 0.016 & 50 & 209 & 0.020 & 50 & 730 & 0.024 & 110 & 3 & 0.234 & 110 & {\color{blue}\textbf{1}} \\
    \hline
    {NSSG} & 0.508 & {\color{blue}\textbf{19}} & {\color{blue}\textbf{1}} & 0.634 & 40 & {\color{blue}\textbf{1}} & 0.494 & 19 & {\color{blue}\textbf{1}} & 0.579 & {\color{blue}\textbf{20}} & {\color{blue}\textbf{1}} & 0.399 & 26 & {\color{blue}\textbf{1}} & 0.580 & 13 & {\color{blue}\textbf{1}} & 0.474 & 15 & {\color{blue}\textbf{1}} & 0.517 & 19 & {\color{blue}\textbf{1}} \\
    \hline
    \end{tabular}
    }\vspace{-0.5cm}

\end{table*}

\noindent\underline{\textbf{Connectivity.}} Connectivity mainly relates to the construction strategy and dataset. \autoref{tab: Index Information} shows that DG- and MST-based algorithms have good connectivity. The former is attributed to the \textbf{\textit{Increment}} construction strategy (e.g., NSW and NGT), and the latter benefits from its approximation to MST. Some RNG-based algorithms perform depth-first search (DFS) to ensure connectivity (e.g., NSG and NSSG). DPG adds reverse edges to make it have good connectivity. Unsurprisingly, KNNG-based algorithms generally have a lot of connected components, especially on hard datasets.

\vspace{-0.2cm}
\subsection{Search Performance}\label{search_performane}
All searches are evaluated on a single thread. The number of nearest neighbors recalled is uniformly set to 10 for each query, and $Recall@10$ represents the corresponding recall rate. Because of space constraints, we only list the representative results in \autoref{fig: real_search_qps} and \hyperref[fig: real_search_speedup]{8}, and the others are displayed in \hyperref[Appendix_O]{Appendix O}. Note that our observations are based on the results on all datasets.

\noindent\underline{\textbf{Accuracy and efficiency.}} As illustrated in \autoref{fig: real_search_qps} and \hyperref[fig: real_search_speedup]{8}, the search performance of different algorithms on the same dataset or the same algorithm on different datasets have large differences. Generally, algorithms capable of obtaining higher speedup also can achieve higher QPS, which demonstrates that the search efficiency of graph-based ANNS algorithms mainly depends on the number of distance evaluations during the search~\cite{SONG}. The search performance of RNG- and MST-based algorithms (e.g., NSG and HCNNG) generally beats other categories by a large margin, especially on hard datasets (e.g., GloVe). KNNG- and DG-based algorithms (e.g., EFANNA and NSW) can only achieve better search performance on simple datasets, their performance drops sharply on hard datasets. Particularly, the search performance of SPTAG decreases dramatically with the increase of LID. This is most likely because it frequently regains entry through the tree during the search~\cite{wang2012query}, we know that the tree has bad \textit{curse of dimensionality}~\cite{DPG}.



\vspace{0.5mm}
\noindent\underline{\textbf{Candidate set size (CS).}} There is a connection between the CS and algorithm category, dataset, and search performance. For most algorithms, we can set CS to obtain the target recall rate, but a few algorithms (e.g., SPTAG) reach the ``ceiling'' before the set recall rate. At this time, the recall rate hardly changes when we increase CS (i.e., a CS value with ``+'' in \autoref{tab: Search requirements}). The elements in a candidate set generally are placed in the cache because of frequent access during the search; so we must constrain the CS to a small value as much as possible because of the capacity’s limitation. Especially in the GPU, the CS will have a greater impact on the search performance~\cite{SONG}. In \autoref{tab: Search requirements}, DG-based and most RNG-based algorithms (e.g., NGT and NSG) require a smaller CS. The CS of KNNG- and MST-based algorithms is related to the dataset, and the harder the dataset, the larger the CS (e.g., SPTAG). In general, algorithms with bad search performance have a larger CS (e.g., FANNG).

\vspace{0.5mm}
\noindent\underline{\textbf{Query path length (PL).}} On large-scale datasets, it generally is necessary to use external storage to store the original data. Normally the PL determines the I/O number, which restricts the corresponding search efficiency~\cite{DiskANN}. From \autoref{fig: real_search_qps} and \autoref{tab: Search requirements}, we see that algorithms with higher search performance generally have smaller PL (e.g., HCNNG), but algorithms with smaller PL do not necessarily have good search performance (e.g., FANNG). In addition, it makes sense that sometimes that an algorithm with a large average out-degree also has a small PL (e.g., NSW).

\begin{figure*}[!t]
  \centering
  \setlength{\abovecaptionskip}{-0.3em}
  \setlength{\belowcaptionskip}{-0.45cm}
  \includegraphics[width=.8\linewidth]{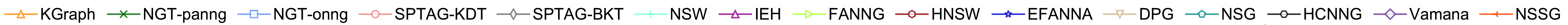}
\end{figure*}

\begin{figure*}
  \vspace{-6mm}
  \centering
  \subfigcapskip=-0.25cm
  \subfigure[Recall@10 (UQ-V)]{ 
    \captionsetup{skip=0pt}
    \vspace{-1.2mm}
    \includegraphics[scale=0.25]{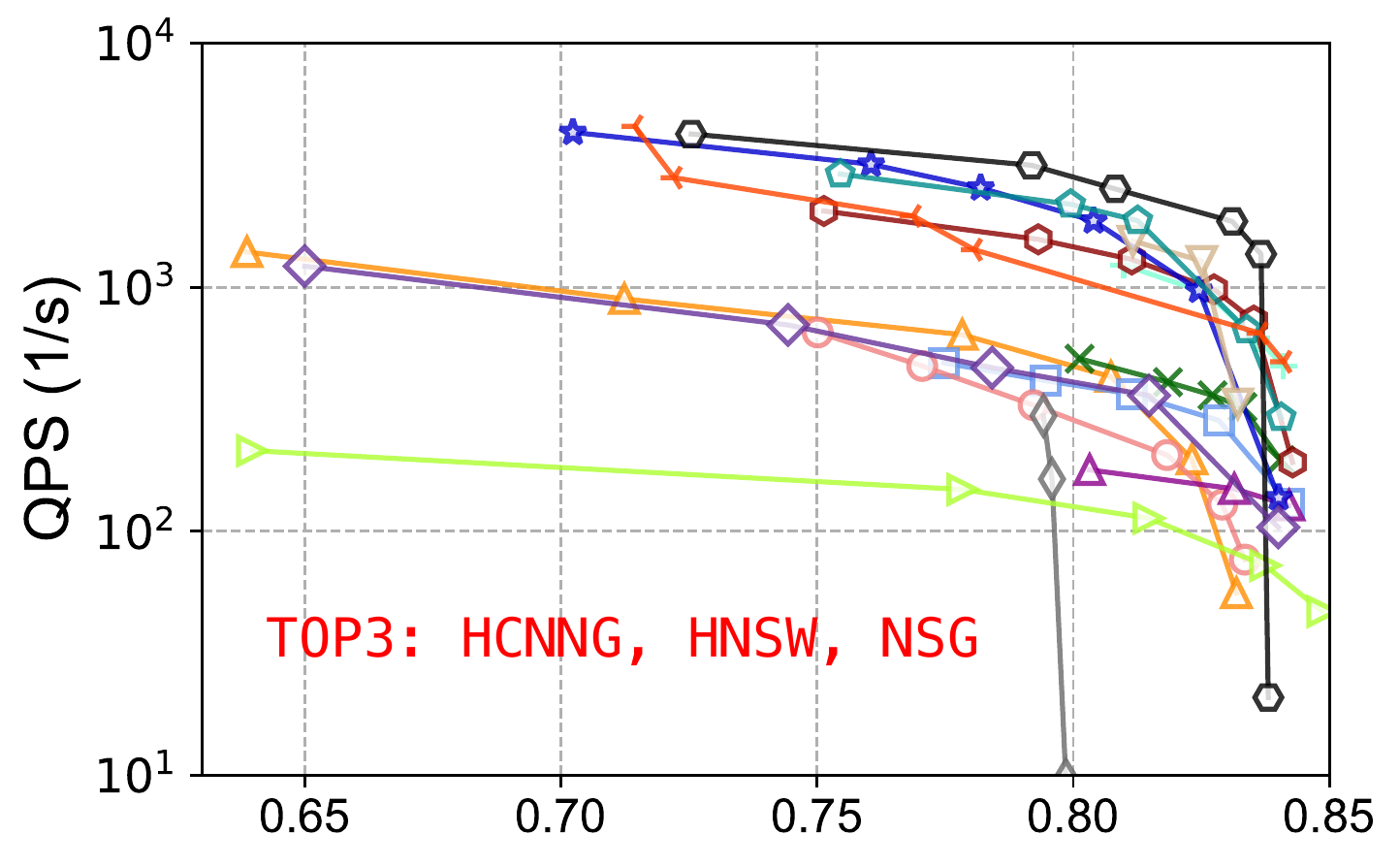}
    \label{fig: qps-vs-recall-uqv}
  }\hspace{-4.15mm}
  \subfigure[Recall@10 (Msong)]{ 
    \captionsetup{skip=0pt}
    \vspace{-1.2mm}
    \includegraphics[scale=0.25]{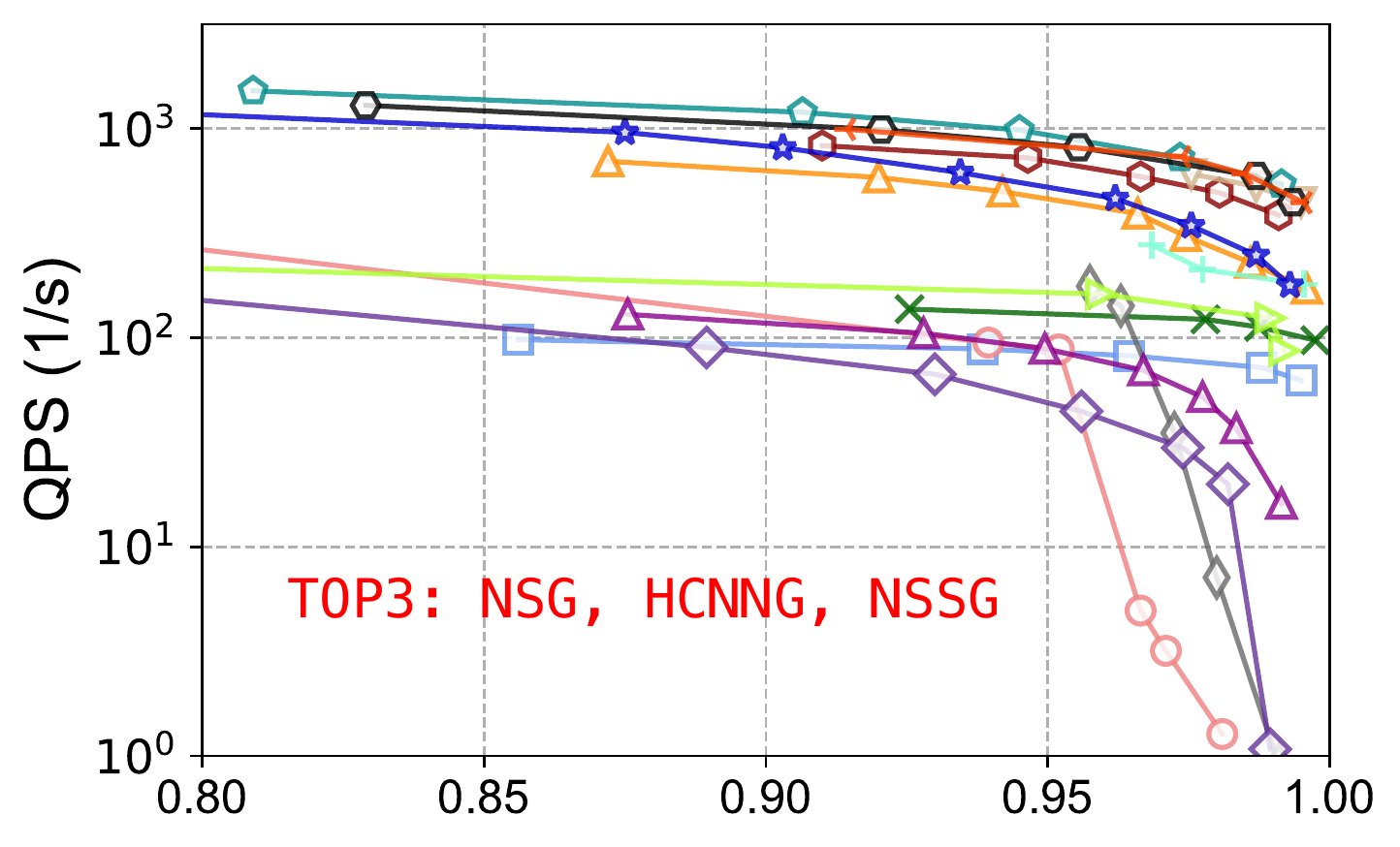}
    \label{fig: qps-vs-recall-msong}
  }\hspace{-4.15mm}
  \subfigure[Recall@10 (SIFT1M)]{ 
    \captionsetup{skip=0pt}
    \vspace{-1mm}
    \includegraphics[scale=0.25]{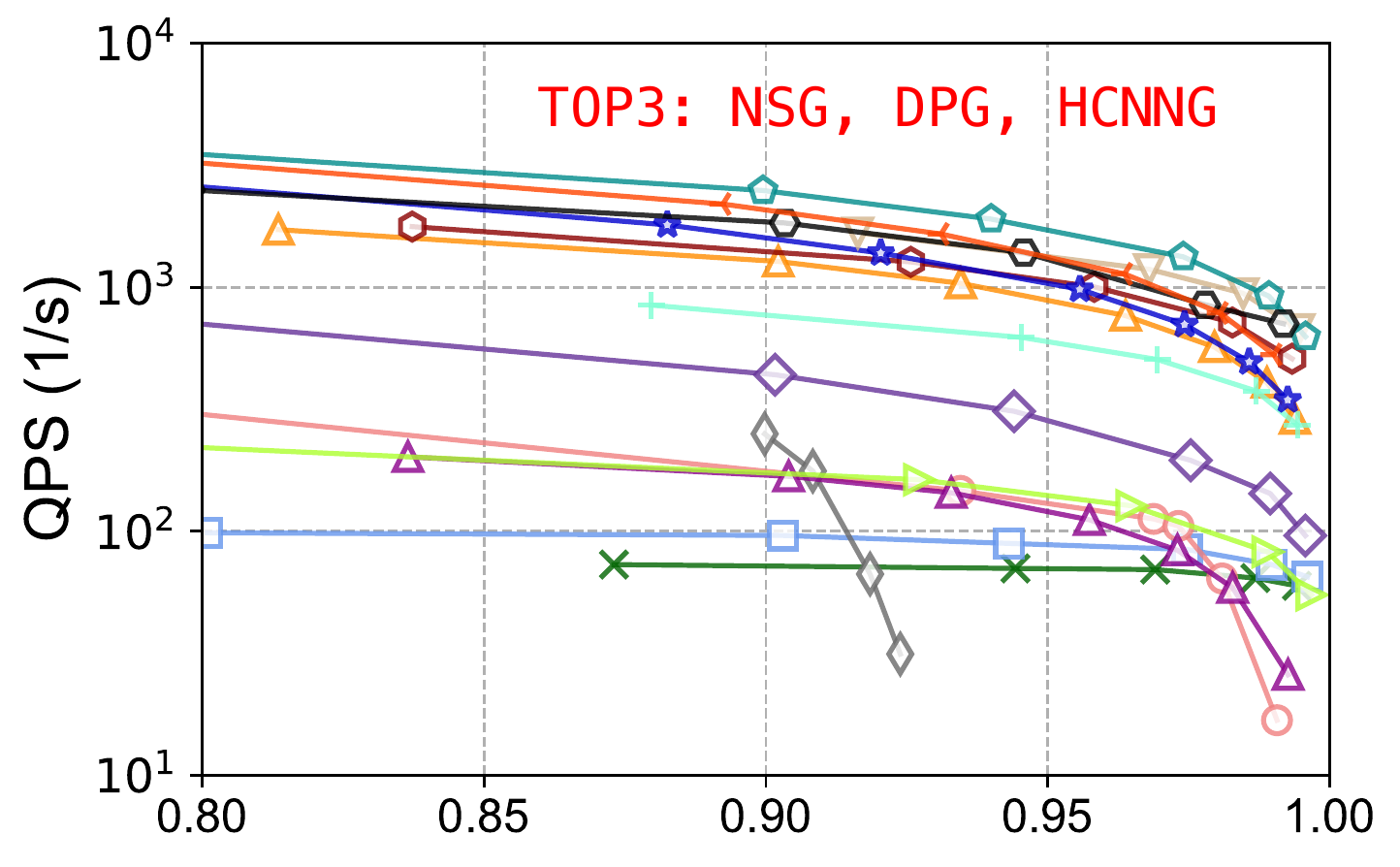}
    \label{fig: qps-vs-recall-sift}
  }\hspace{-4.15mm}
  \subfigure[Recall@10 (GIST1M)]{  
    \captionsetup{skip=0pt}
    \vspace{-1mm}
    \includegraphics[scale=0.25]{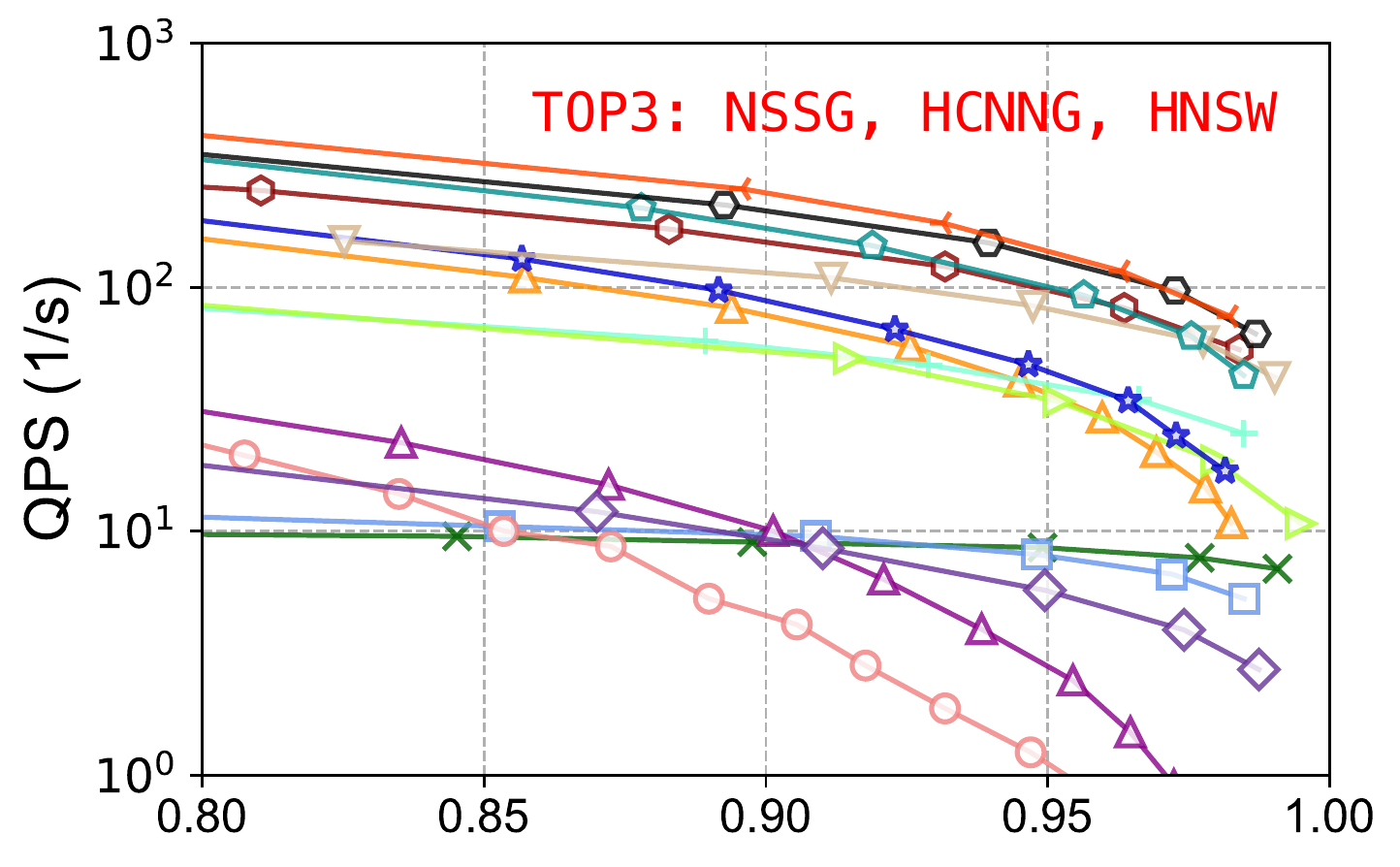}
    \label{fig: qps-vs-recall-gist}
  }\hspace{-4.15mm}
  \subfigure[Recall@10 (GloVe)]{  
    \captionsetup{skip=0pt}
    \vspace{-1mm}
    \includegraphics[scale=0.25]{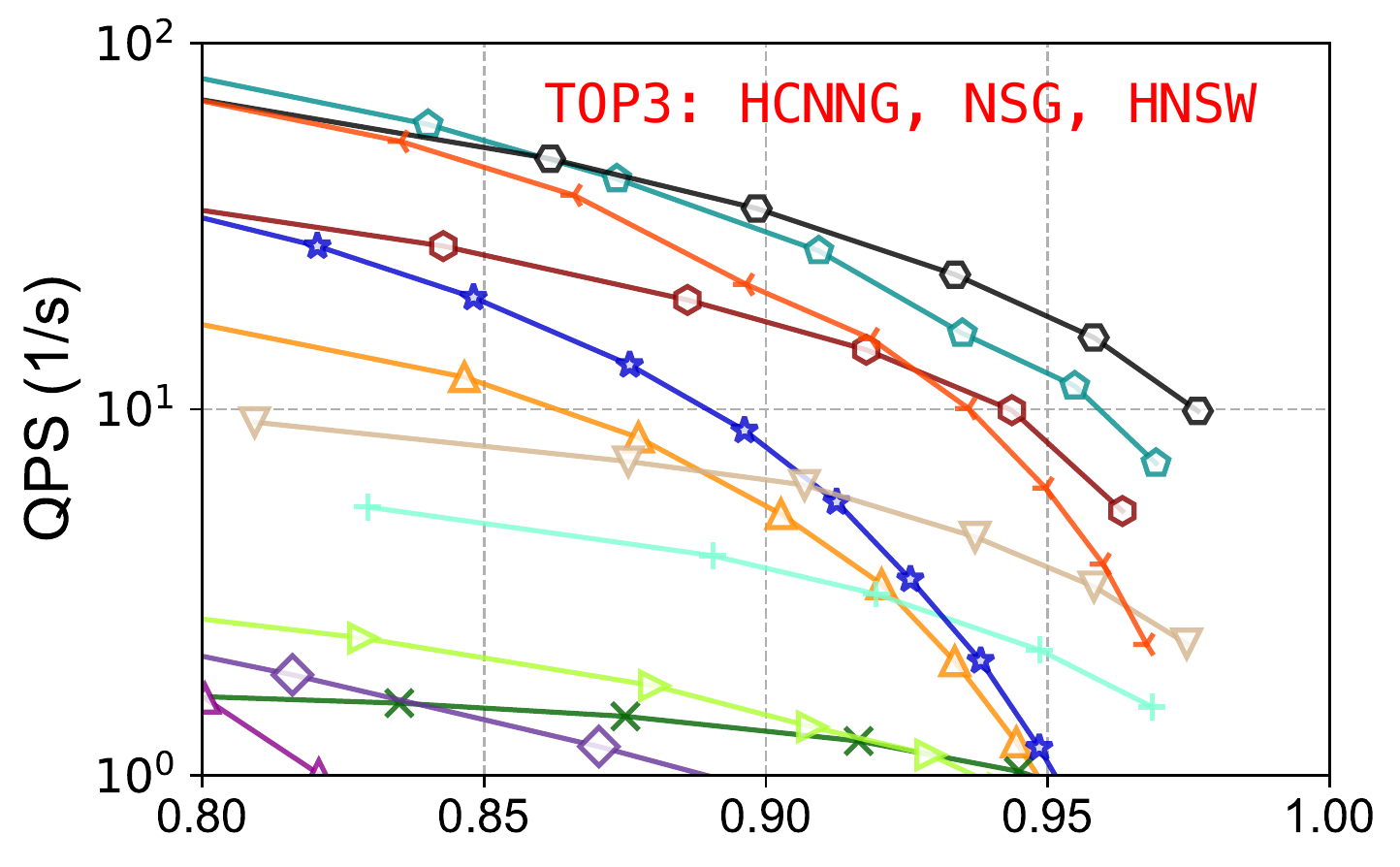}
    \label{fig: qps-vs-recall-glove}
  }

  \vspace{-5mm}
  \caption{The Queries Per Second (QPS) vs Recall@10 of graph-based ANNS algorithms in high-precision region (top right is better).}\vspace{-0.4mm}
  \label{fig: real_search_qps}
\end{figure*}

\begin{figure*}[!t]
  \centering
  \vspace{-0.35cm}
  \setlength{\abovecaptionskip}{-0.3em}
  \setlength{\belowcaptionskip}{-0.45cm}
  \includegraphics[width=.8\linewidth]{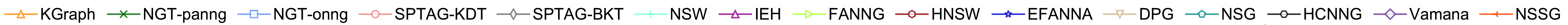}
\end{figure*}

\begin{figure*}
  \vspace{-6mm}
  \centering
  \subfigcapskip=-0.25cm
  \subfigure[Recall@10 (UQ-V)]{ 
    \captionsetup{skip=0pt}
    \vspace{-1.2mm}
    \includegraphics[scale=0.25]{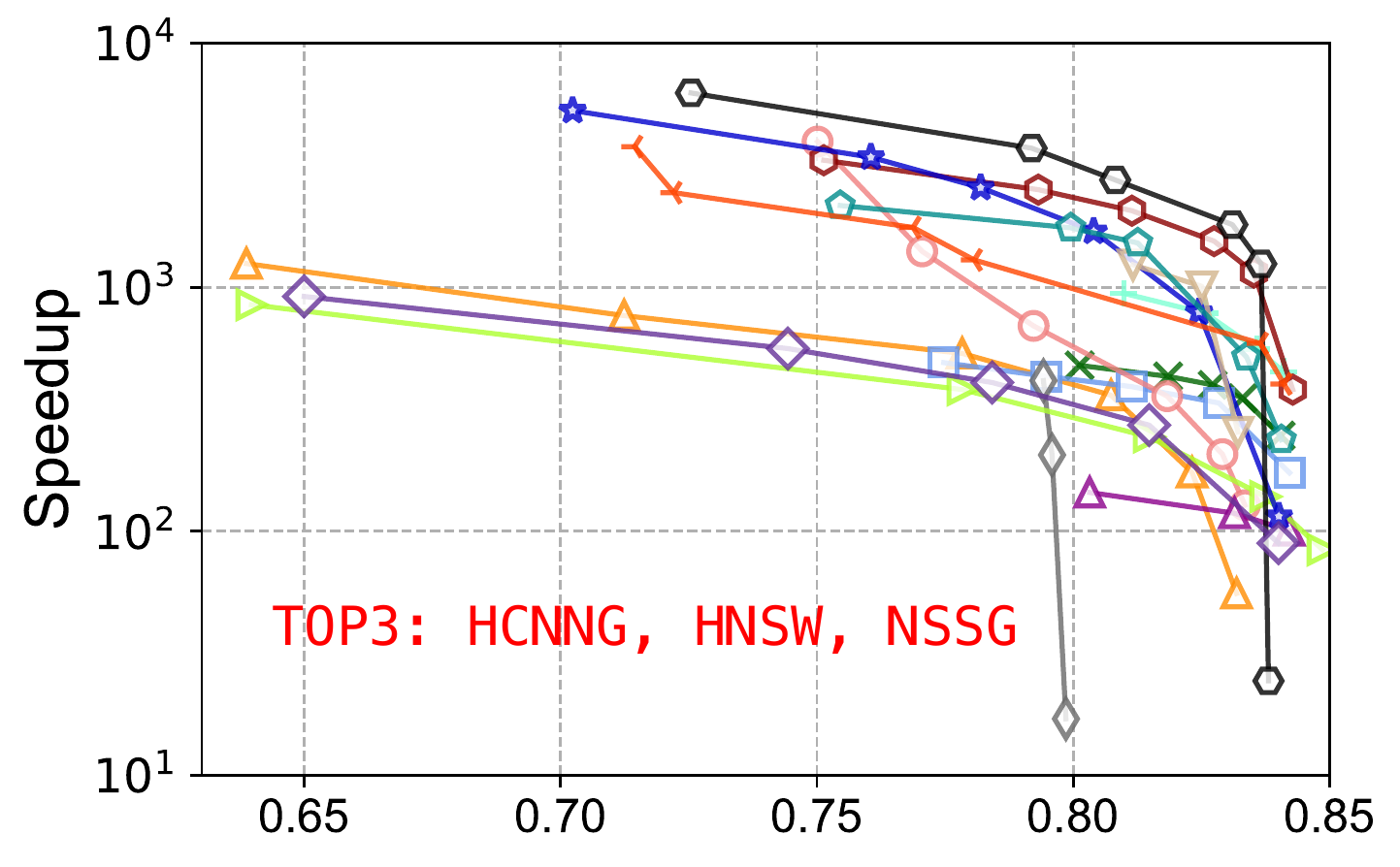}
    \label{fig: speedup-vs-recall-uqv}
  }\hspace{-4.1mm}
  \subfigure[Recall@10 (Msong)]{ 
    \captionsetup{skip=0pt}
    \vspace{-1.2mm}
    \includegraphics[scale=0.25]{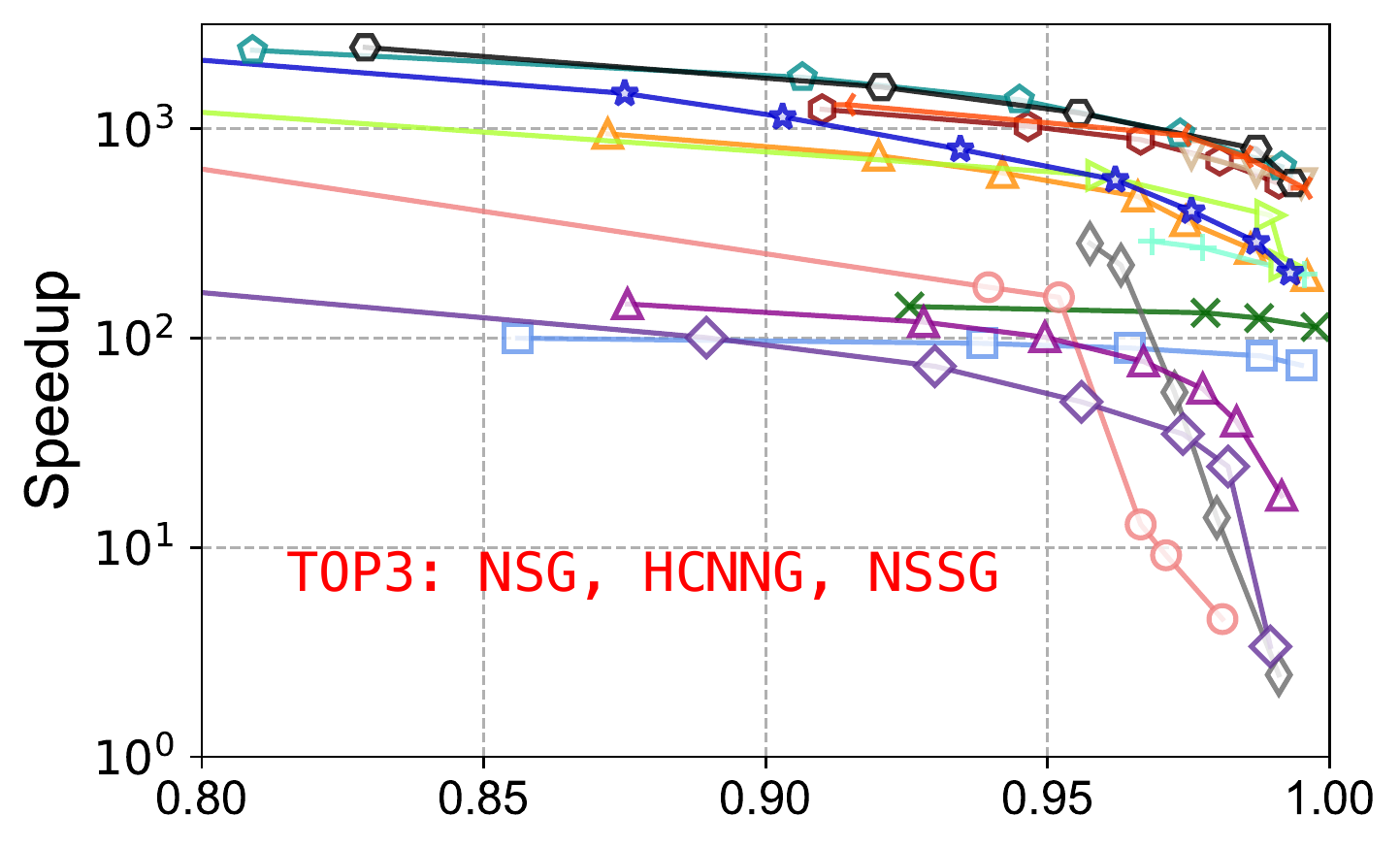}
    \label{fig: speedup-vs-recall-msong}
  }\hspace{-4.1mm}
  \subfigure[Recall@10 (SIFT1M)]{ 
    \captionsetup{skip=0pt}
    \vspace{-1mm}
    \includegraphics[scale=0.25]{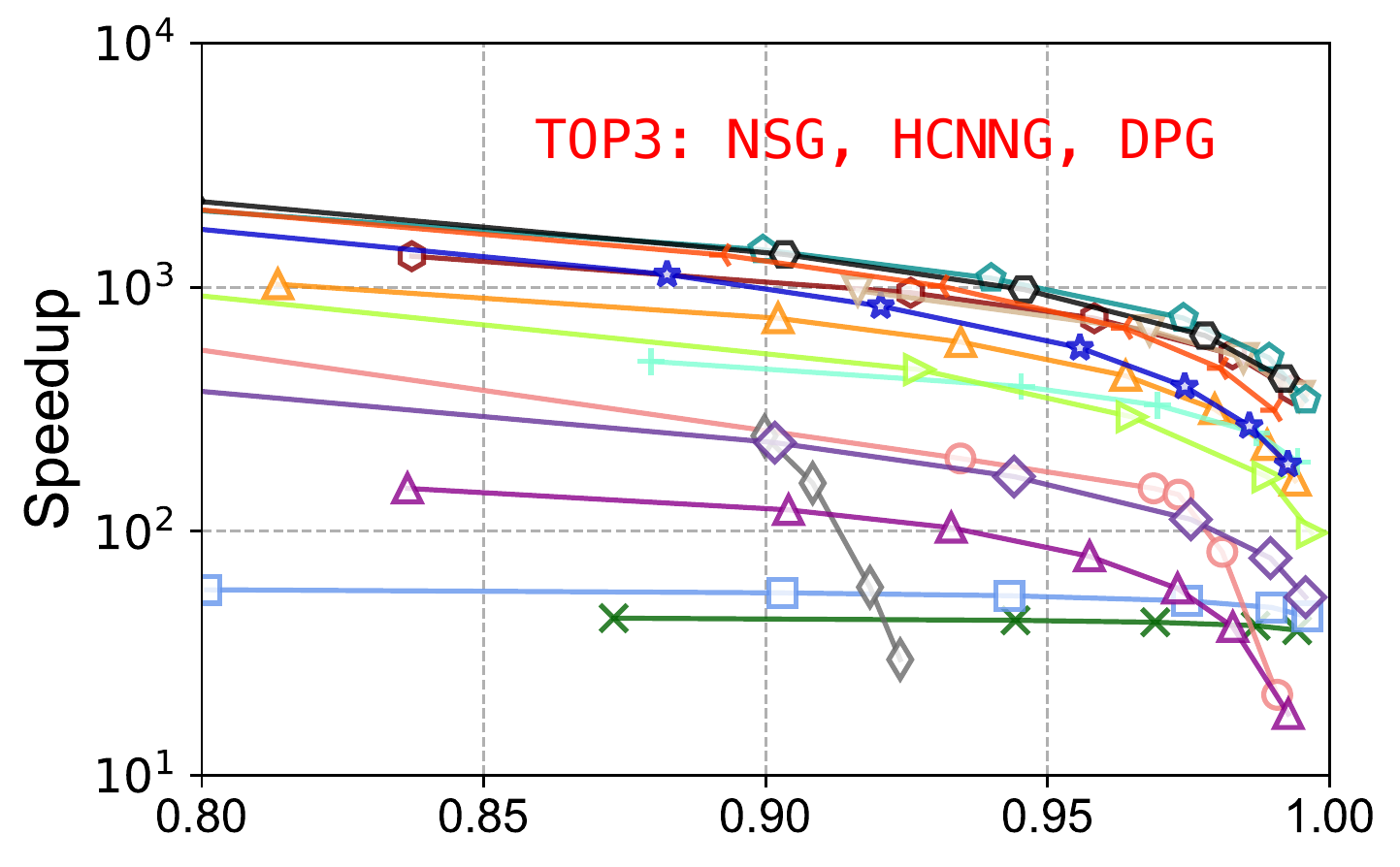}
    \label{fig: speedup-vs-recall-sift}
  }\hspace{-4.1mm}
  \subfigure[Recall@10 (GIST1M)]{  
    \captionsetup{skip=0pt}
    \vspace{-1mm}
    \includegraphics[scale=0.25]{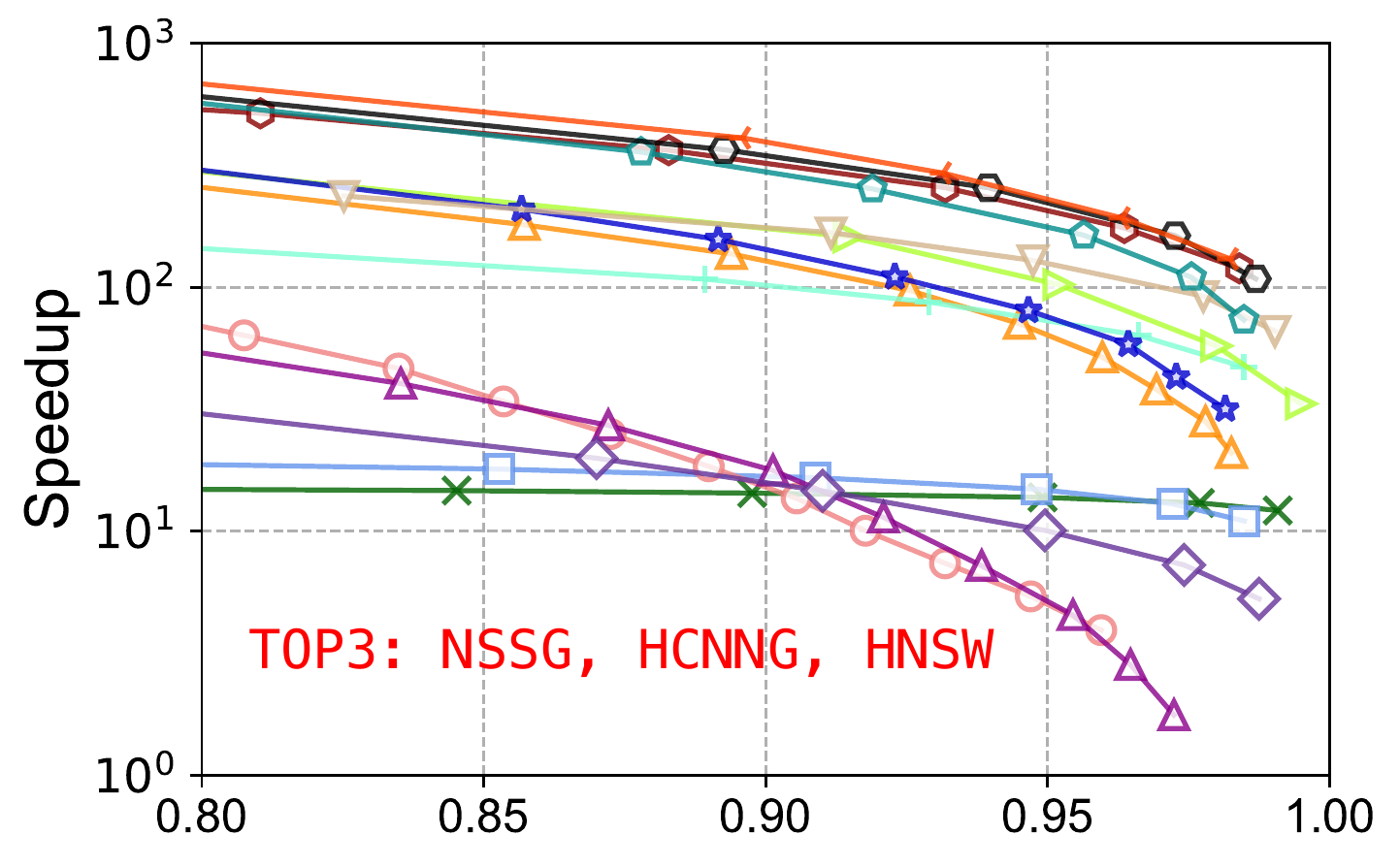}
    \label{fig: speedup-vs-recall-gist}
  }\hspace{-4.1mm}
  \subfigure[Recall@10 (GloVe)]{  
    \captionsetup{skip=0pt}
    \vspace{-1mm}
    \includegraphics[scale=0.25]{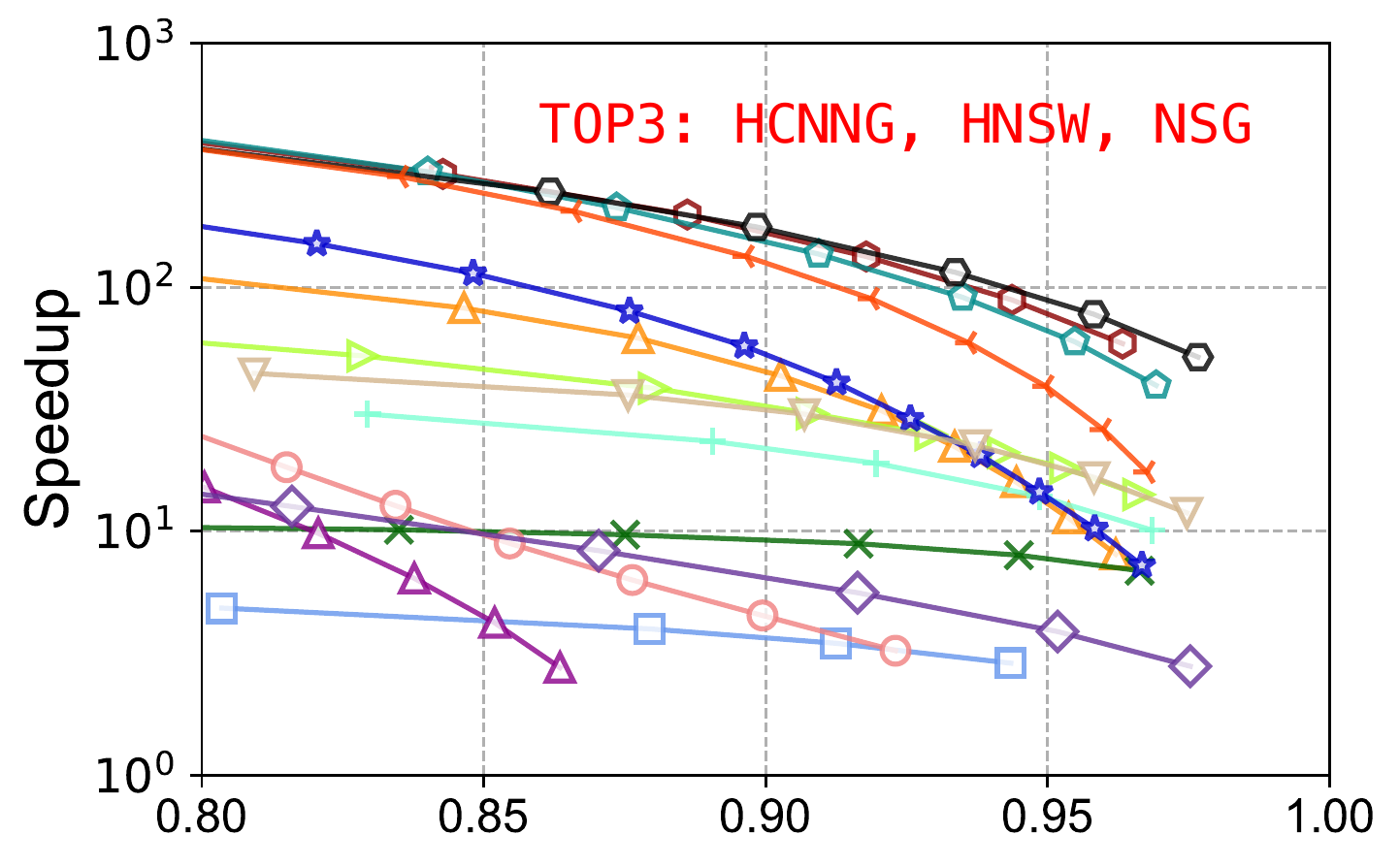}
    \label{fig: speedup-vs-recall-glove}
  }

  \vspace{-5mm}
  \caption{The Speedup vs Recall@10 of graph-based ANNS algorithms in high-precision region (top right is better).}
  \label{fig: real_search_speedup}
\end{figure*}

\vspace{0.5mm}
\noindent\underline{\textbf{Memory overhead (MO).}} As \autoref{tab: Search requirements} show, RNG-based algorithms generally have the smallest memory overhead (e.g., NSG and NSSG). Some algorithms with additional index structures have high memory overhead (e.g., SPTAG and IEH). Larger AD and CS values also will increase the algorithms’ memory overhead (e.g., NSW and SPTAG-BKT). Overall, the smaller the algorithm’s index size, the smaller the memory overhead during search.

\vspace{-0.3cm}
\subsection{Components’ Evaluation} \label{components_evaluation}
In this subsection, we evaluate representative components of graph-based algorithms on two real-world datasets with different difficulty. According to the aforementioned experiments, algorithms based on the \textbf{\textit{Refinement}} construction strategy generally have better comprehensive performance. Therefore, we design a unified evaluation framework based on this strategy and the pipline in \autoref{fig: components}. Each component in the evaluation framework is set for a certain implementation to form a benchmark algorithm (see \hyperref[Appendix_K]{Appendix K} for detailed settings). We use the \textit{C\# + algorithm name} to indicate the corresponding component’s specific implementation. For example, \textit{C1\_NSG} indicates that we use the initialization (C1) of NSG, i.e., the initial graph is constructed through NN-Descent.

Note that many algorithms have the same implementation for the same component (e.g., \textit{C3\_NSG}, \textit{C3\_HNSW}, and \textit{C3\_FANNG}). We randomly select an algorithm to represent this implementation (e.g., \textit{C3\_HNSW}). The impact of different components on search performance and construction time are depicted in \autoref{fig: component_search} and \hyperref[Appendix_M]{Appendix M}, respectively.

\setlength{\textfloatsep}{0cm}
\setlength{\floatsep}{0cm}
\begin{table*}[th!]
\setlength{\abovecaptionskip}{0cm}
\setlength{\belowcaptionskip}{-0.3cm}
\setstretch{0.9}
\fontsize{6.5pt}{3.3mm}\selectfont
    \centering
    \tabcaption{The candidate set size (CS), query path length (PL), and peak memory overhead (MO) during the search (the bold values are the best).}
    \label{tab: Search requirements}
    \setlength{\tabcolsep}{0.00315\linewidth}{
    \begin{tabular}{l|l|l|l|l|l|l|l|l|l|l|l|l|l|l|l|l|l|l|l|l|l|l|l|l}
    \hline
    \multirow{2}*{\textbf{Alg.}} & \multicolumn{3}{c|}{\textbf{UQ-V}} & \multicolumn{3}{c|}{\textbf{Msong}} & \multicolumn{3}{c|}{\textbf{Audio}} & \multicolumn{3}{c|}{\textbf{SIFT1M}} & \multicolumn{3}{c|}{\textbf{GIST1M}} & \multicolumn{3}{c|}{\textbf{Crawl}} & \multicolumn{3}{c|}{\textbf{GloVe}} & \multicolumn{3}{c}{\textbf{Enron}}\\
    \cline{2-25}
    ~ & CS & PL & MO & CS & PL & MO & CS & PL &MO & CS & PL &MO & CS & PL &MO & CS & PL &MO & CS & PL &MO & CS & PL &MO \\
    \hline
    \hline
    {KGraph} & 15,000+ & 1,375 & 1,211 & 50,442 & 1,943 & 2,036 & 701 & 105 & 55 & 139 & 52 & 900 & 2,838 & 411 & 4,115 & 50,000+ & 3,741 & 3,031 & 24,318 & 1,333 & 991 & 9,870 & 607 & 525 \\
    \hline
    {NGT-panng} & 65 & 79 & 1,423 & {\color{blue}\textbf{10}} & 144 & 1,927 & {\color{blue}\textbf{10}} & 33 & 63 & {\color{blue}\textbf{20}} & 438 & 933 & {\color{blue}\textbf{10}} & 1,172 & 4,111 & {\color{blue}\textbf{10}} & 5,132 & 3,111 & {\color{blue}\textbf{10}} & 2,281 & 928 & {\color{blue}\textbf{10}} & 83 & 535 \\
    \hline
    {NGT-onng} & 1,002 & 431 & 1,411 & 20 & 227 & 2,007 & 15 & 45 & 63 & 33 & 392 & 859 & 33 & 1,110 & 4,088 & 157 & 244 & 3,147 & 74 & 388 & 1,331 & 25 & 131 & 533 \\
    \hline
    {SPTAG-KDT} & 37,097 & 2,259 & 2,631 & 50,000+ & 11,441 & 2,091 & 61 & 107 & 91 & 7,690 & 1,227 & 1,048 & 50,000+ & 15,162 & 5,643 & 50,000+ & 12,293 & 8,872 & 50,000+ & 10,916 & 11,131 & 10,291 & 592 & 569 \\
    \hline
    {SPTAG-BKT} & 15,000+ & 10,719 & 5,114 & 97,089 & 11,119 & 1933 & 10 & 61 & 91 & 50,000+ & 8,882 & 4,587 & 50,000+ & 7,685 & 4,299 & 50,000+ & 10,851 & 6,643 & 50,000+ & 7,941 & 4,625 & 93,294 & 5,126 & 629 \\
    \hline
    {NSW} & 85 & {\color{blue}\textbf{38}} & 1,857 & 20 & 35 & 3,122 & 18 & 17 & 101 & 58 & 54 & 1,574 & 69 & 161 & 5,180 & 36 & 435 & 5,217 & 65 & 634 & 2,782 & 21 & 29 & 690 \\
    \hline
    {IEH} & {\color{blue}\textbf{29}} & 196 & 5,166 & 301 & 1,007 & 6,326 & 53 & 269 & 253 & 238 & 816 & 4,170 & 9,458 & 24,339 & 10,508 & 15,000+ & 5,928 & 10,913 & 15,000+ & 3,620 & 4,681 & 274 & 893 & 1,302 \\
    \hline
    {FANNG} & 1,072 & 86 & 1,395 & 594 & 245 & {\color{blue}\textbf{1,687}} & 1,462 & 195 & 58 & 1,377 & 92 & 825 & 3,007 & 269 & 3,917 & 8,214 & 2,152 & 2,639 & 9,000 & 1,062 & 850 & 2,084 & 152 & 548 \\
    \hline
    {HNSW} & 927 & 296 & 1,424 & 43 & 35 & 2,370 & 51 & 37 & 67 & 66 & 47 & 1,206 & 181 & 130 & 4,372 & 61 & 108 & 3,950 & 505 & 334 & 1,294 & 59 & 32 & 595 \\
    \hline
    {EFANNA} & 1,446 & 217 & 1,297 & 312 & 85 & 2,030 & 800 & 283 & 67 & 204 & 76 & 967 & 1,652 & 292 & 4,473 & 1311 & 180 & 3,848 & 22,349 & 1241 & 1,188 & 2,180 & 254 & 531 \\
    \hline
    {DPG} & 15,000+ & 1,007 & 1,352 & 16 & {\color{blue}\textbf{20}} & 1,965 & {\color{blue}\textbf{10}} & {\color{blue}\textbf{10}} & 62 & 37 & {\color{blue}\textbf{30}} & 851 & 55 & {\color{blue}\textbf{124}} & 4,091 & 67 & 761 & 3,089 & 84 & 792 & 956 & 89 & 60 & 538 \\
    \hline
    {NSG} & 354 & 156 & {\color{blue}\textbf{1,127}} & 106 & 90 & 1,714 & 63 & 47 & {\color{blue}\textbf{51}} & 101 & 85 & 653 & 867 & 826 & {\color{blue}\textbf{3,781}} & 345 & 723 & {\color{blue}\textbf{2,499}} & 814 & 1,875 & {\color{blue}\textbf{594}} & 118 & 138 & {\color{blue}\textbf{513}} \\
    \hline
    {HCNNG} & 15,000+ & 1,398 & 1,472 & 62 & 21 & 2,200 & 35 & 12 & 69 & 97 & 37 & 1,056 & 371 & 179 & 4,159 & 173 & {\color{blue}\textbf{61}} & 3,753 & 217 & {\color{blue}\textbf{95}} & 1,590 & 62 & {\color{blue}\textbf{20}} & 564 \\
    \hline
    {Vamana} & 1,049 & 346 & 1,164 & 40,596 & 7,155 & 1,763 & 68 & 30 & 57 & 493 & 263 & 748 & 8,360 & 3,127 & 3,916 & 53,206 & 7,465 & 2,786 & 22,446 & 2,157 & 1026 & 526 & 103 & 547 \\
    \hline
    {NSSG} & 310 & 122 & 1,129 & 39 & 40 & 1,807 & 65 & 42 & {\color{blue}\textbf{51}} & 255 & 157 & {\color{blue}\textbf{640}} & 280 & 270 & 3,829 & 13,810 & 12,892 & 2,524 & 3,846 & 3,047 & 605 & 458 & 236 & 514 \\
    \hline
    \end{tabular}
    }\vspace{-0.4cm}
\end{table*}

\noindent\underline{\textbf{C1: Initialization.}} \autoref{fig: component_search}(a) reports the impact of different graph index initialization methods on search performance. The search performance of \textit{C1\_NSG} is much better than \textit{C1\_EFANNA} and \textit{C1\_KGraph}; and although \textit{C1\_NSG} needs more construction time, it is worthwhile for such a large performance improvement. Moreover, a larger gap exists between \textit{C1\_NSG} and the other two on GIST1M (harder), which shows that it has better scalability.

\noindent\underline{\textbf{C2: Candidate neighbor acquisition.}} As shown in \autoref{fig: component_search}(b), different candidate neighbor acquisition methods vary slightly. \textit{C2\_NSW} has the best search performance, especially on GIST1M, with the price being more construction time. \textit{C2\_NSSG} obtains better search performance than \textit{C2\_DPG} under a similar construction time. It is worth noting that although DPG’s search performance on SIFT1M is better than HNSW’s in \autoref{fig: real_search_qps}, the search performance of \textit{C2\_HNSW} (i.e., \textit{C2\_NSW}) exceeds that of \textit{C2\_DPG}.

\noindent\underline{\textbf{C3: Neighbor selection.}} \autoref{fig: component_search}(c) depicts the impact of different neighbor selection schemes on search performance. Obviously, it shows better search performance for algorithms that consider the distribution of neighbors (e.g., \textit{C3\_HNSW}, \textit{C3\_NSSG}, \textit{C3\_DPG}, \textit{C3\_Vamana}) than those that do not consider this (e.g., \textit{C3\_KGraph}). Note that \textit{C3\_Vamana}’s performance is no better than \textit{C3\_HNSW}’s, as claimed in the paper~\cite{DiskANN}. NSSG\cite{NSSG} appears to have better search performance than NSG in their experiment, so the researchers believe that \textit{C3\_NSSG} is better than \textit{C3\_NSG} (i.e., \textit{C3\_HNSW}). However, the researchers do not control the consistency of other components during the evaluation, which is unfair.

\setlength{\textfloatsep}{0cm}
\setlength{\floatsep}{0cm}
\begin{figure}
\setlength{\abovecaptionskip}{0cm}
\setstretch{0.9}
\fontsize{6.5pt}{3.3mm}\selectfont
\begin{minipage}{0.26\textwidth}
    \centering
    \fontsize{6.5pt}{3.3mm}\selectfont
    \tabcaption{Index processing time (IPT) and memory consumption (MC) of ML1-based optimization.}
    \label{tab: index_build_ml}
    \setlength{\tabcolsep}{.017\linewidth}{
    \begin{tabular}{l|l|l|l}
      \hline
      \multicolumn{2}{c|}{\textbf{Method}} & {\textbf{SIFT100K}} & {\textbf{GIST100K}}\\
      \hline
      \hline
      \multirow{2}*{\textbf{IPT(s)}}& NSG & {\color{blue}\textbf{55}} & {\color{blue}\textbf{142}}\\
      \cline{2-4}
      ~ & NSG+ML1 & 55+67,260 & 142+45,600 \\
      \hline
      \hline
      \multirow{2}*{\textbf{MC(GB)}}& NSG & {\color{blue}\textbf{0.37}} & {\color{blue}\textbf{0.68}} \\
      \cline{2-4}
      ~ & NSG+ML1 & 23.8 & 58.7 \\
      \hline
    \end{tabular}
    }
\end{minipage}
\begin{minipage}{0.21\textwidth}
  \centering
  \vspace{-1mm}
  \includegraphics[width=0.9\linewidth]{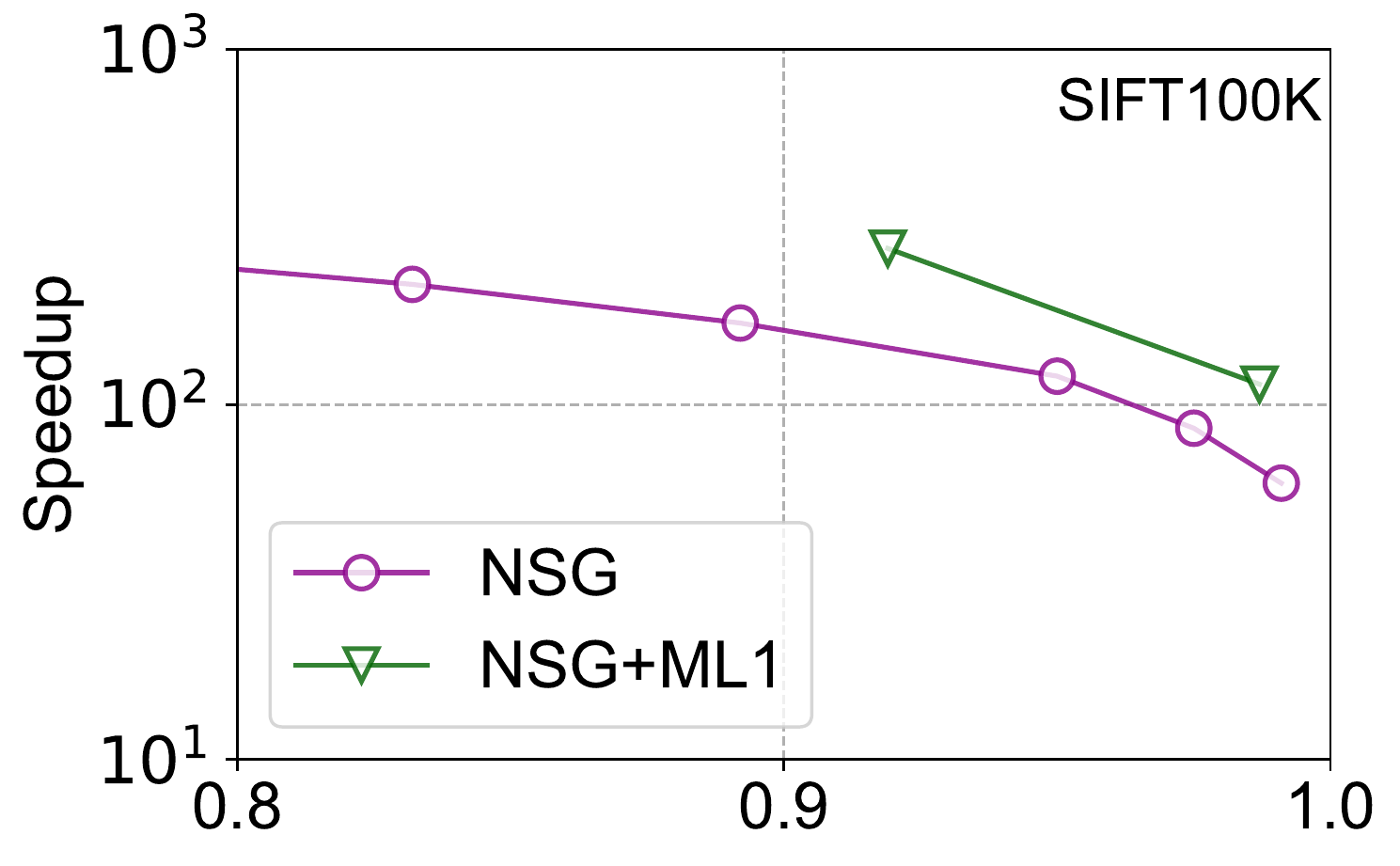}
  \figcaption{Speedup vs Recall@1 of NSG under ML1-based optimizations.}
  \label{fig: t_search_ml}
\end{minipage}
\vspace{-0.1cm}

\end{figure}



\begin{figure*}
  \centering
  \subfigcapskip=-0.25cm
  \subfigure[Recall@10 on SIFT1M (left) and GIST1M (right)]{  
    \captionsetup{skip=0pt}
    \vspace{-1mm}
    \includegraphics[scale=0.29]{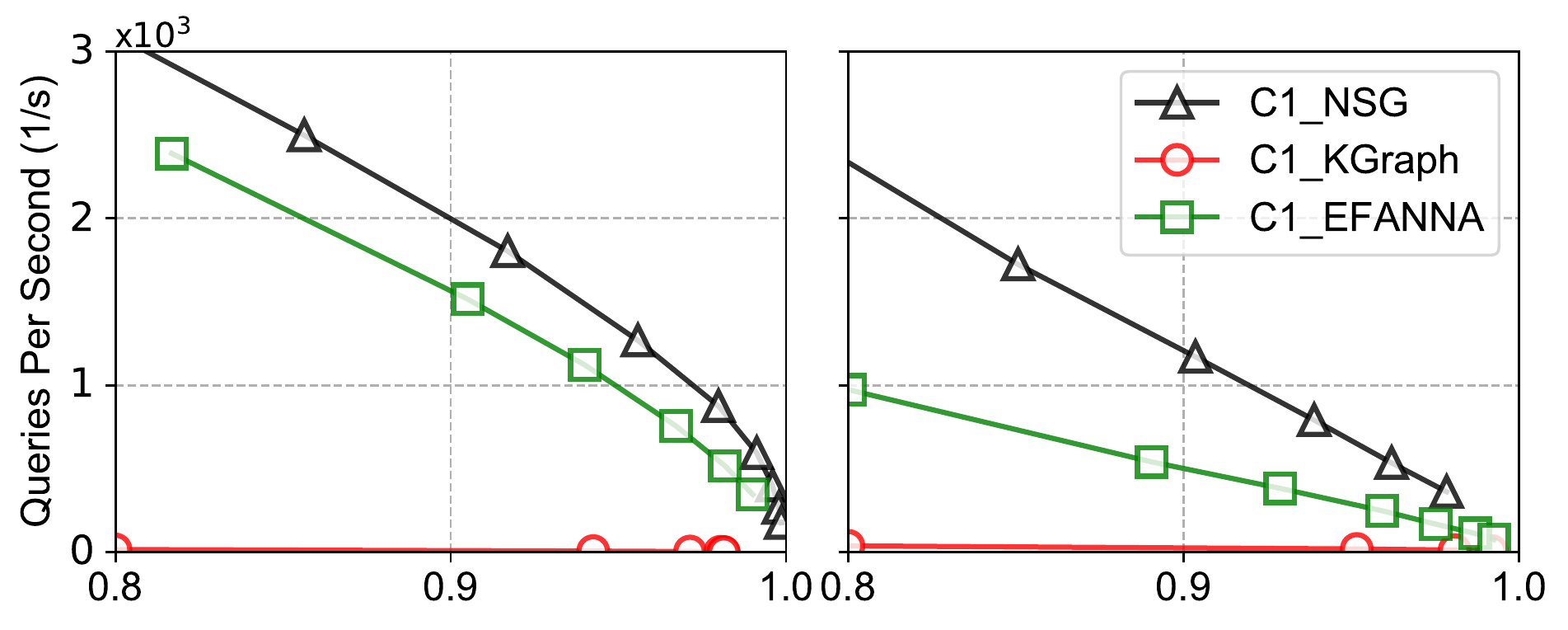}
    \label{fig: c1}
  }
  \subfigure[Recall@10 on SIFT1M (left) and GIST1M (right)]{  
  \captionsetup{skip=0pt}
  \vspace{-1mm}
  \includegraphics[scale=0.29]{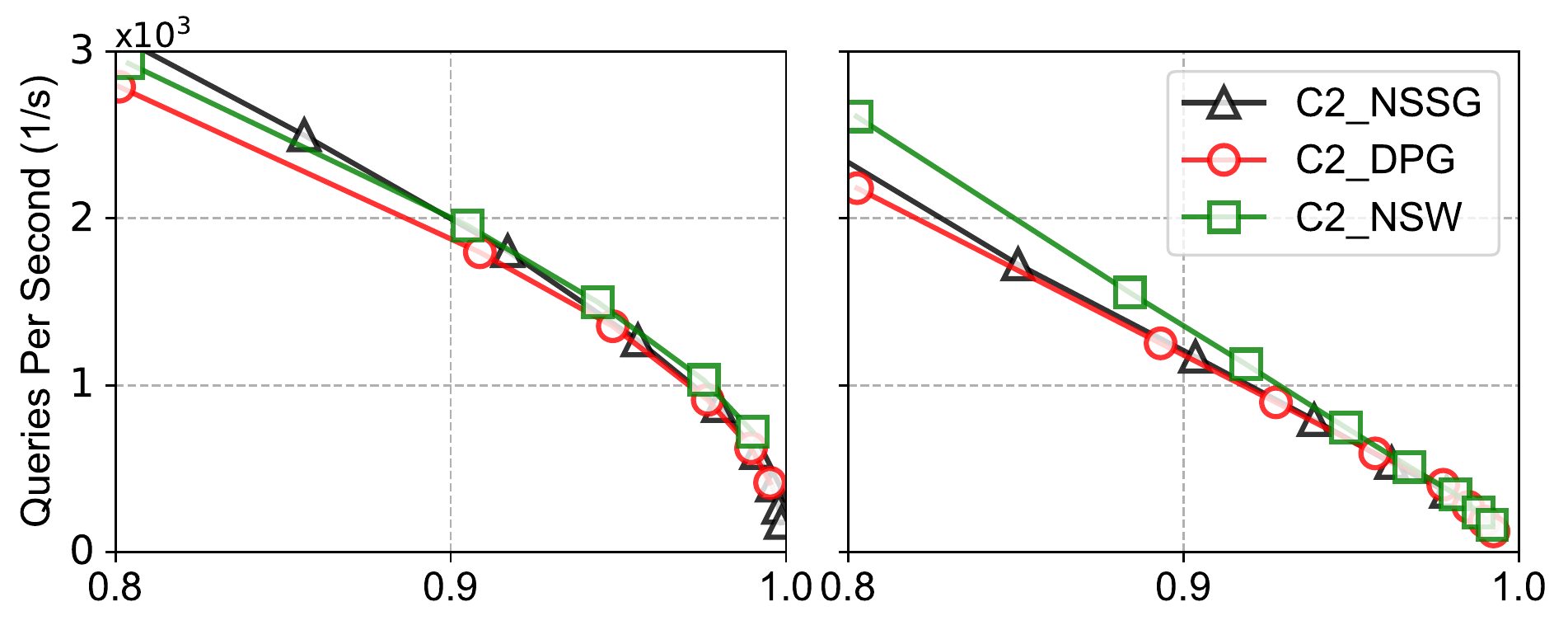}
  \label{fig: c2}
  }
  \subfigure[Recall@10 on SIFT1M (left) and GIST1M (right)]{  
  \captionsetup{skip=0pt}
  \vspace{-1mm}
  \includegraphics[scale=0.29]{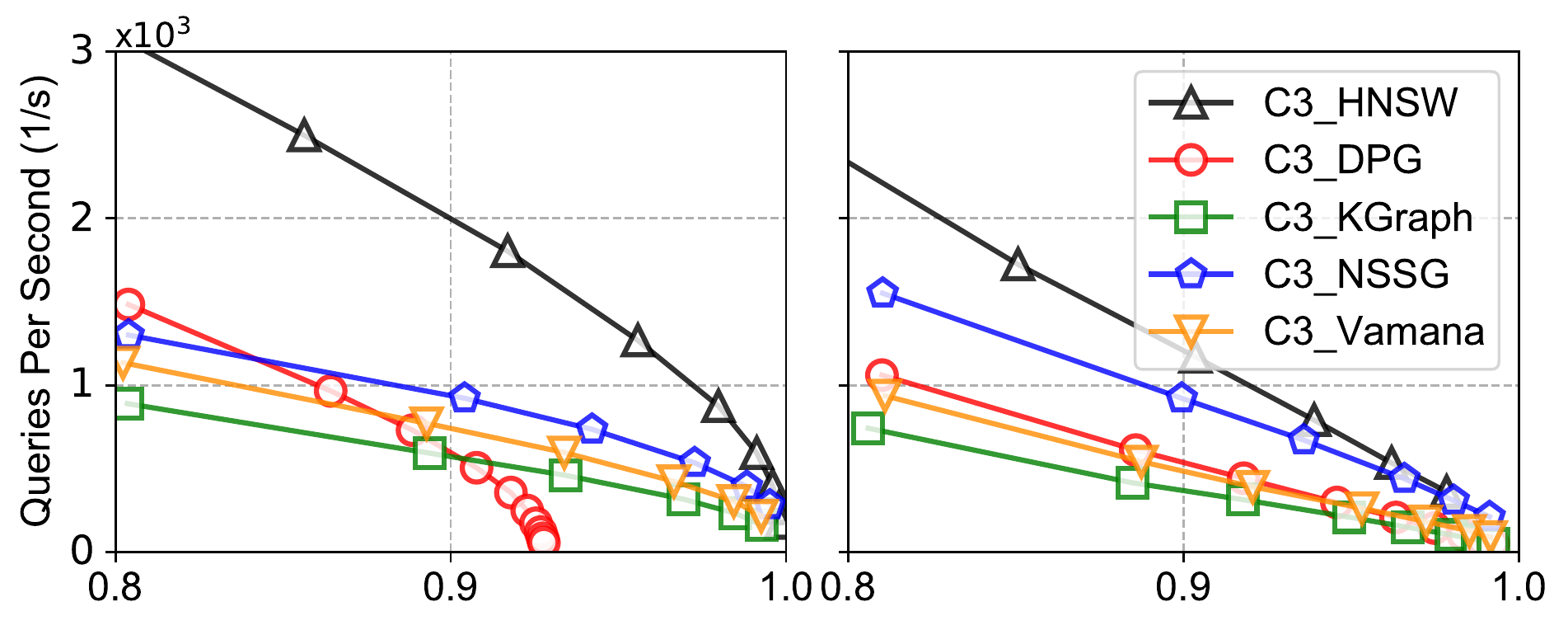}
  \label{fig: c3}
  }\vspace{-0.4cm}
  \subfigure[Recall@10 on SIFT1M (left) and GIST1M (right)]{ 
  \captionsetup{skip=0pt}
  \vspace{-1mm}
  \includegraphics[scale=0.29]{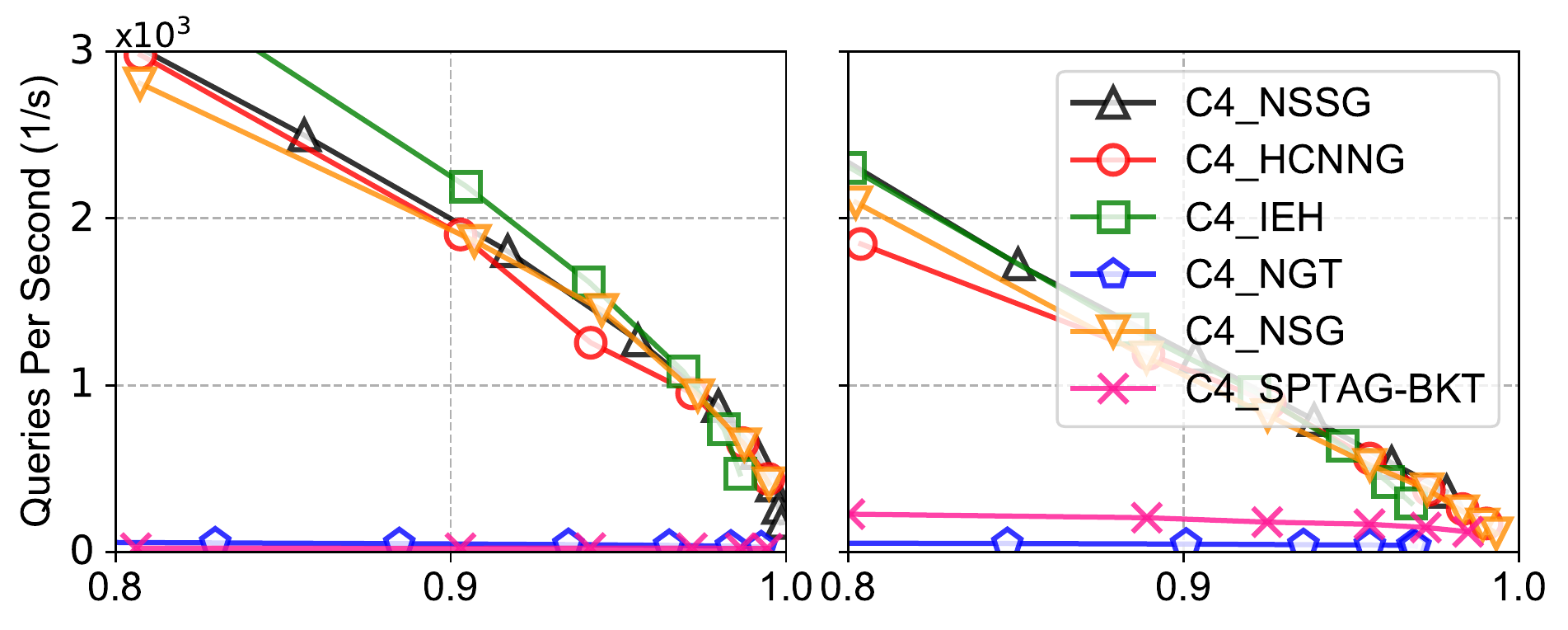}
  \label{fig: c4}
  }
  \subfigure[Recall@10 on SIFT1M (left) and GIST1M (right)]{ 
  \captionsetup{skip=0pt}
  \vspace{-1mm}
  \includegraphics[scale=0.29]{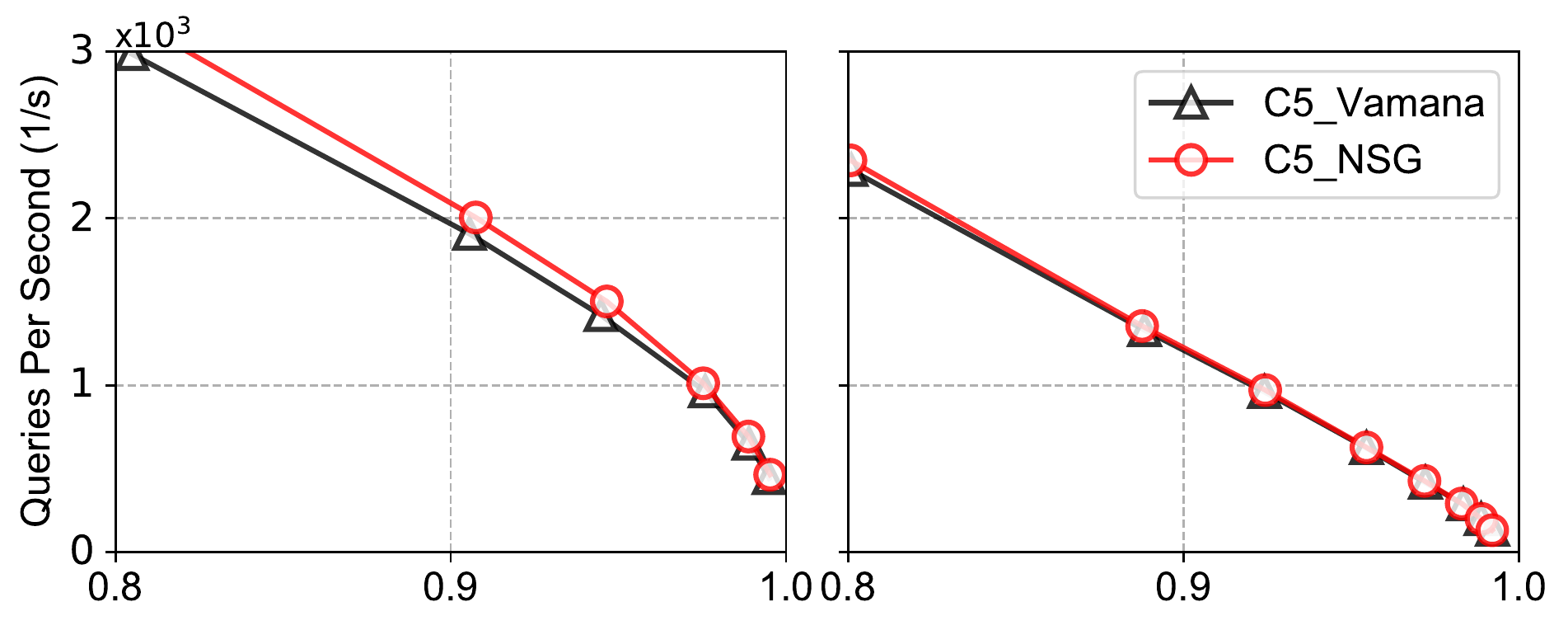}
  \label{fig: c5}
  }
  \subfigure[Recall@10 on SIFT1M (left) and GIST1M (right)]{ 
  \captionsetup{skip=0pt}
  \vspace{-1mm}
  \includegraphics[scale=0.29]{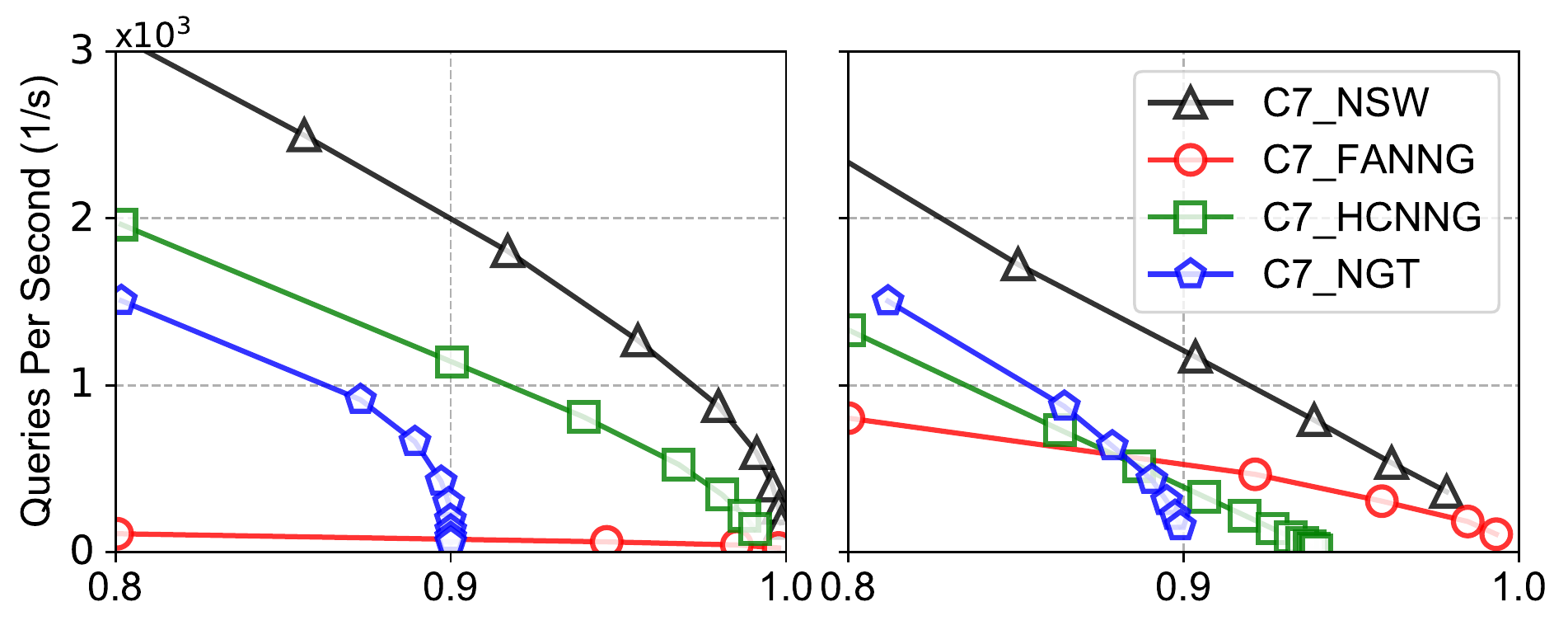}
  \label{fig: c7}
  }\vspace{-0.35cm}
  \vspace{-0.2cm}
  \caption{Components’ search performance under a unified framework on SIFT1M (a simple dataset) and GIST1M (a hard dataset).}\vspace{-0.5cm}
  \label{fig: component_search}
\end{figure*}

\noindent\underline{\textbf{C4: Seed preprocessing and C6: Seed acquisition.}} The \textbf{C4} and \textbf{C6} components are interrelated in all compared algorithms; that is, after specifying \textbf{C4}, \textbf{C6} is also determined. Briefly, we use \textit{C4\_NSSG} to indicate \textit{C6\_NSSG}. As \autoref{fig: component_search}(d) shows, the extra index structure to get the entry significantly impacts search performance. \textit{C4\_NGT} and \textit{C4\_SPTAG-BKT} have the worst search performance; they both obtain entry by performing distance calculations on an additional tree (we know that the tree index has a serious \textit{curse of dimensionality}). Although \textit{C4\_HCNNG} also obtains entry through a tree, it only needs value comparison and no distance calculation on the KD-Tree, so it shows better search performance than \textit{C4\_NGT} and \textit{C4\_SPTAG-BKT}. \textit{C4\_IEH} adds the hash table to obtain entry, yielding the best search performance. This may be because the hash can obtain entry close to the query more quickly than the tree. Meanwhile, \textit{C4\_NSSG} and \textit{C4\_NSG} still achieve high search performance without additional index. Note that there is no significant difference in index construction time for these methods.

\noindent\underline{\textbf{C5: Connectivity.}} \autoref{fig: component_search}(e) shows the algorithm with guaranteed connectivity has better search performance (e.g., \textit{C5\_NSG}) than that without connectivity assurance (e.g., \textit{C5\_Vamana}).

\noindent\underline{\textbf{C7: Routing.}} \autoref{fig: component_search}(f) shows different routing strategies’ impact on search performance. \textit{C7\_NSW}’s search performance is the best, and it is used by most algorithms (e.g., HNSW and NSG). \textit{C7\_NGT} has a precision ``ceiling'' because of the $\epsilon$ parameter’s limitation, which can be alleviated by increasing $\epsilon$, but search efficiency will decrease. \textit{C7\_FANNG} can achieve high accuracy through backtracking, but backtracking also limits search efficiency. \textit{C7\_HCNNG} avoids some redundant calculations based on the query position, however, this negatively affects search accuracy.

\setlength{\textfloatsep}{0cm}
\setlength{\floatsep}{0cm}
\begin{table}[!tb]
  \centering
  \setlength{\abovecaptionskip}{0.05cm}
  \setstretch{0.9}
  \fontsize{6.5pt}{3.3mm}\selectfont
  \caption{Recommendation of the algorithms in different scenarios.}
  \label{tab: recommendation_algorithms}
  \setlength{\tabcolsep}{.017\linewidth}{
  \begin{tabular}{l|l}
    \hline
    \textbf{Scenario} & \textbf{Algorithm}\\
    \hline
    \hline
    \textbf{S1}: A large amount of data updated frequently & NSG, NSSG\\
    \hline
    \textbf{S2}: Rapid construction of KNNG & KGraph, EFANNA, DPG\\
    \hline
    \textbf{S3}: Data is stored in external memory & DPG, HCNNG\\
    \hline
    \textbf{S4}: Search on hard datasets & HNSW, NSG, HCNNG\\
    \hline
    \textbf{S5}: Search on simple datasets & DPG, NSG, HCNNG, NSSG\\
    \hline
    \textbf{S6}: GPU acceleration & NGT\\
    \hline
    \textbf{S7}: Limited memory resources & NSG, NSSG\\
    \hline
  \end{tabular}
  }
\end{table}

\vspace{-0.3cm}
\subsection{Machine Learning-Based Optimizations} \label{machine_learning_optimization}
Recently, machine learning (ML)-based methods are proposed to improve the speedup vs recall trade-off of the algorithms~\cite{learn_to_route,li2020improving,prokhorenkova2020graph}. In general, they can be viewed as some optimizations on graph-based algorithms discussed above (such as NSG and NSW). We evaluate three ML-based optimizations on NSG and HNSW, i.e., ML1~\cite{learn_to_route}, ML2~\cite{li2020improving}, and ML3~\cite{prokhorenkova2020graph}. Because of space limitations, we only show the test results on ML1 in \hyperref[tab: index_build_ml]{Table 6} and \autoref{fig: t_search_ml}, others share similar feature (see \hyperref[Appendix_R]{Appendix R} for more details).

\noindent\underline{\textbf{Analysis.}} ML-based optimizations generally obtain better speedup vs recall tradeoff at the expense of more time and memory. For example, the original NSG takes 55s and maximum memory consumption of 0.37 GB for index construction on SIFT100K; however, NSG optimized by ML1 takes 67,315s to process the index (even if we use the GPU for speedup), and the memory consumption is up to 23.8 GB. In summary, current ML-based optimizations have high hardware requirements and time cost, so their wide application is limited. Considering that most of the original graph-based algorithms can return query results in < 5ms, some high-tech companies (such as Alibaba, Facebook, and ZILLIZ) only deploy NSG without ML-based optimizations in real business scenarios~\cite{NSG,FAISS,milvus}.

\vspace{-0.2cm}
\section{Discussion} \label{sec6}
According to the behaviors of algorithms and components on real-world and synthetic datasets, we discuss our findings as follows.

\noindent\underline{\textbf{Recommendations.}} In \autoref{tab: recommendation_algorithms}, our evaluation selects algorithms based on best performance under different scenarios. NSG and NSSG have the smallest construction time and index size, so they are suitable for \textbf{S1}. KGraph, EFANNA, and DPG achieve the highest graph quality with lower construction time, so they are recommended for \textbf{S2}. For \textbf{S3} (such as SSD~\cite{DiskANN}), DPG and HCNNG are the best choices because their smaller average path length can reduce I/O times. On hard datasets (\textbf{S4}, large LID), HNSW, NSG, and HCNNG show competitive search performance, while on simple datasets (\textbf{S5}), DPG, NSG, HCNNG, and NSSG offer better search performance. For \textbf{S6}, we need a smaller candidate set size because of the cache’s limitation~\cite{SONG}; for now, NGT appears more advantageous. NSG and NSSG offer the smallest out-degree and memory overhead, so they are the best option for \textbf{S7}.

\noindent\underline{\textbf{Guidelines.}} Intuitively, a practical graph-based ANNS algorithm should have: \textbf{(H1) high construction efficiency}; \textbf{(H2) high routing efficiency}; \textbf{(H3) high search accuracy}; and \textbf{(L4) low memory overhead}. For \textbf{H1}, we should not spend too much time improving graph quality, because the best graph quality is not necessary to achieve the best search performance. For \textbf{H2}, we should control the appropriate out-degree, diversify neighbors’ distribution (such as \textit{C3\_HNSW}), and reduce the cost of obtaining entries (like \textit{C4\_IEH}), to navigate quickly to the query’s nearest neighbors with a small number of distance calculations. In addition, we should avoid redundant distance calculations by optimizing the routing strategy (such as \textit{C7\_HCNNG}) in the routing process. In terms of \textbf{H3}, to improve the search’s immunity from falling into the local optimum~\cite{learn_to_route}, we should reasonably design the distribution of neighbors during index construction, ensure connectivity (such as \textit{C5\_NSG}), and optimize the routing strategy~\cite{vargas2019genetic}. For \textbf{L4}, we can start by reducing the out-degree and candidate set size, and this can be achieved by improving the neighbor selection strategy (such as \textit{C3\_NSG}) and routing strategy (like \textit{C7\_HCNNG}).

\begin{figure}[!t]
  \centering
  \vspace{1mm}
  \setlength{\abovecaptionskip}{-0.3em}
  \includegraphics[width=.8\linewidth]{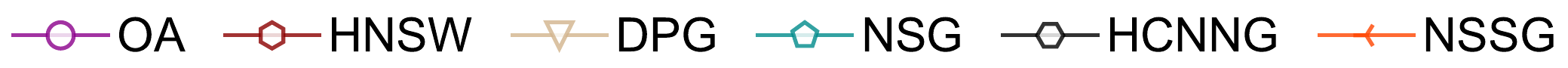}
\end{figure}

\begin{figure}
  \vspace{-2mm}
  \centering
  \subfigcapskip=-0.25cm
  \subfigure[Recall@10 (Msong)]{ 
    \captionsetup{skip=0pt}
    \vspace{-1.2mm}
    \includegraphics[scale=0.28]{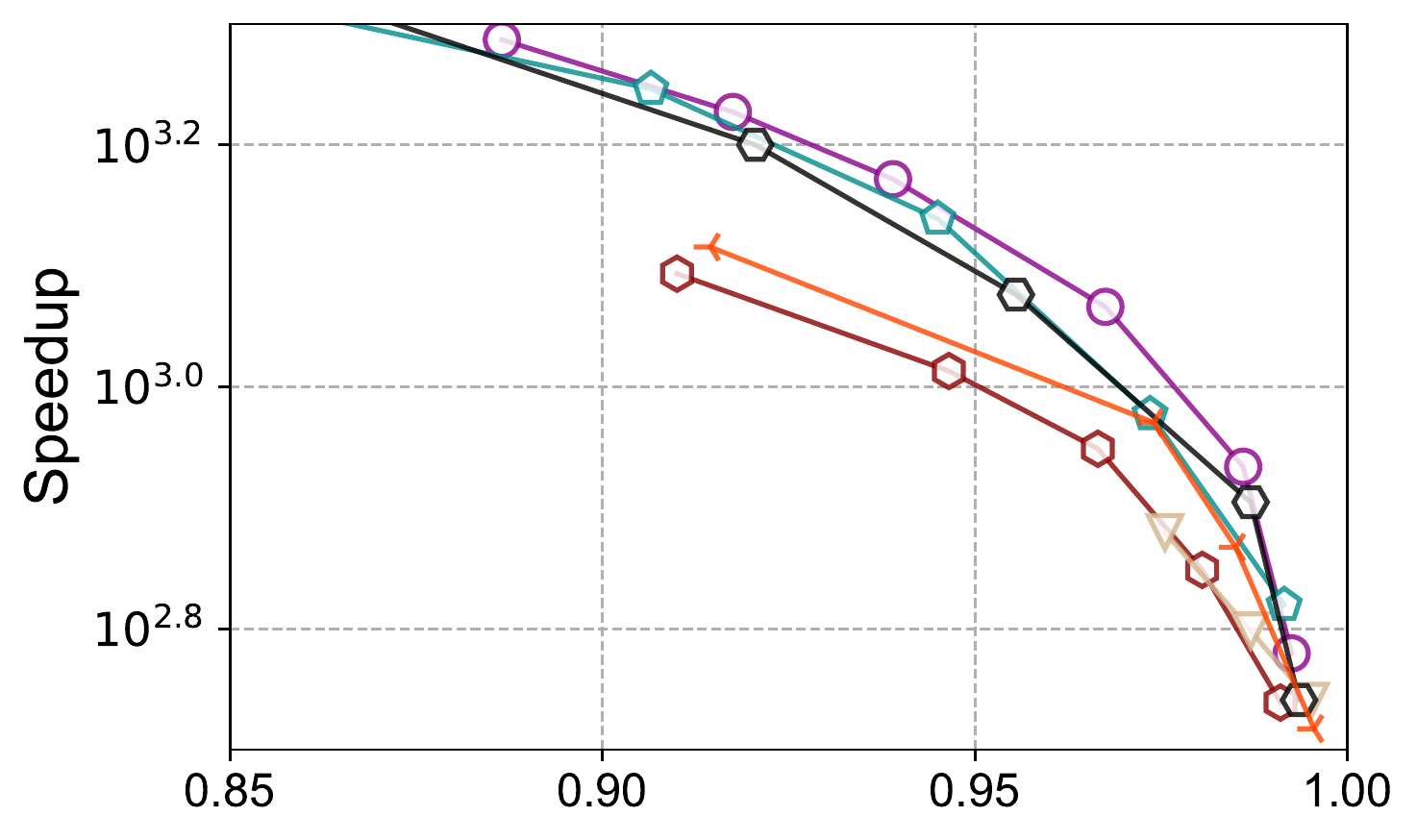}
    \label{fig: test_sift1M}
  }\hspace{-4mm}
  \subfigure[Recall@10 (SIFT1M)]{ 
    \captionsetup{skip=0pt}
    \vspace{-1.2mm}
    \includegraphics[scale=0.28]{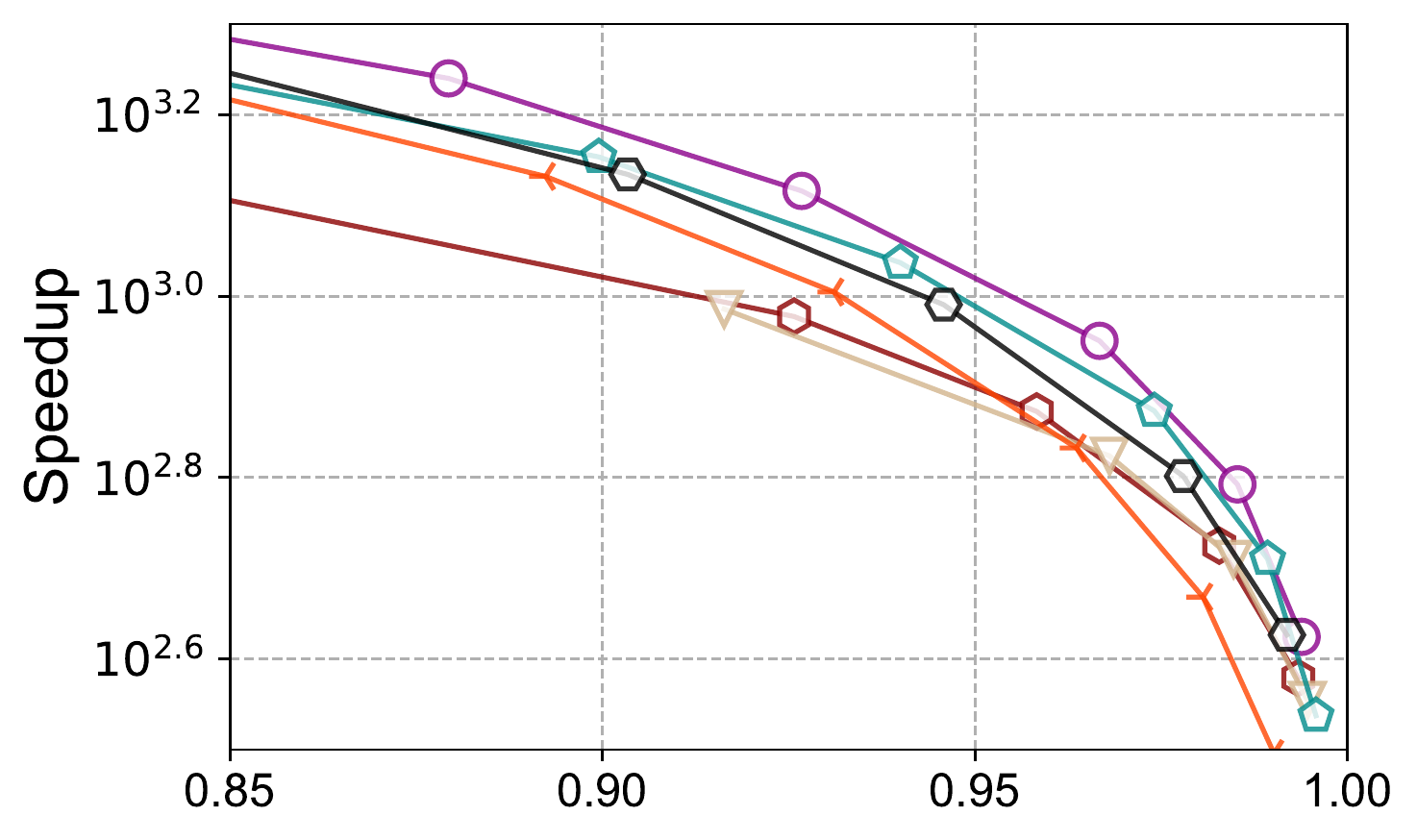}
    \label{fig: test_gist}
  }

  \vspace{-0.5cm}
  \caption{Speedup vs Recall@10 of the optimized algorithm (OA).}\vspace{1mm}
  \label{fig: t_test_oa}
\end{figure}

\noindent\underline{\textbf{Improvement.}} Based on our observations and \textbf{Guidelines}, we design an optimized algorithm that addresses \textbf{H1}, \textbf{H2}, \textbf{H3}, and \textbf{L4} simultaneously. In the index construction phase, it initializes a graph with appropriate quality by NN-Descent (C1), quickly obtains candidate neighbors with \textit{C2\_NSSG} (C2), uses \textit{C3\_NSG} to trim redundant neighbors (C3), randomly selects a certain number of entries (C4), and ensures connectivity through depth-first traversal (C5); in the search phase, it starts from the random entries (C6), and performs two-stage routing through \textit{C7\_HCNNG} and \textit{C7\_NSW} in turn. As shown in \autoref{fig: t_test_oa}, the optimized algorithm surpasses the state-of-the-art algorithms in terms of efficiency vs accuracy trade-off, while ensuring high construction efficiency and low memory overhead (see \hyperref[Appendix_P]{Appendix P} for more details).

\noindent\underline{\textbf{Tendencies.}} Over the last decade, graph-based ANNS algorithms have ranged from simple approximation of four classic base graphs (e.g., KGraph and NSW) to ANNS’s optimization (e.g., HNSW and NSG). Along the way, their performance—especially their search performance—has improved qualitatively. It is worth noting that almost all state-of-the-art algorithms are based on RNG (e.g., HNSW and NSG), and thus many approaches add an approximation of RNG on the basis of KNNG- or DG-based algorithms (see \autoref{fig: roadmap}). The RNG-based category is still a promising research direction for graph-based ANNS. The MST-based algorithm recently was applied to graph-based ANNS, and it also achieves excellent results in our evaluation, especially on hard datasets. On the basis of the core algorithm discussed in this paper, researchers are refining and improving graph-based ANNS algorithms’ performance via hardware~\cite{DiskANN,SONG,jieren2020hm}. Other literatures add quantitative or distributed schemes to cope with data increases~\cite{DBLP:conf/cvpr/DouzeSJ18,deng2019pyramid}. To meet hybrid query requirements, the latest research adds structured attribute constraints to the search process of graph-based algorithms~\cite{DBLP:journals/pvldb/WeiWWLZ0C20,multiattribute2020}.

\noindent\underline{\textbf{Challenges.}} At present, almost all graph-based algorithms are oriented to raw data, which is the main reason why these algorithms have high memory usage. Determining how to organically combine data encoding or other methods with graph-based ANNS algorithms is a problem worth exploring. Compared with tree, hashing, and quantization, the graph-based algorithms have the highest index construction time~\cite{DPG}, which adds difficulty with updating the graph index in real time. Also, figuring how to combine GPU acceleration or other methods with the graph-based ANNS algorithm to realize the real-time update of the graph index is worthy of an in-depth study. For data with different characteristics, the graph-based algorithms have different adaptability, and thus exhibit different performance levels. Finally, a major outstanding challenge is discerning how to adaptively select the optimal graph-based algorithm according to the dataset’s characteristics by learning.

\vspace{-0.15cm}
\section{Conclusions} \label{sec7}
In this paper, we consider 13 representative graph-based ANNS algorithms from a new taxonomy. We then divide all the algorithms into seven components for in-depth analysis. Next, we comprehensively evaluate and discuss all the algorithms’ performance on eight real-world datasets and 12 synthetic datasets. We also fairly evaluate each algorithm’s important components through a unified framework. In some ways, this work validates many previous empirical conclusions while leading to novel discoveries that will aid future researchers and practitioners. We also provide some rule-of-thumb recommendations about promising research directions and insightful principles to optimize algorithms.

Finally, we want to note that because of various constraints, our study only investigates core algorithms based on the main memory. Going forward, we will consider hardware (e.g., SSD and GPU) and machine learning optimizations, deploy distributed implementations, and add structured attribute constraints to ANNS.

\vspace{-0.3cm}
\begin{acks}
The National Key Research \& Development Program (Number 2017YFC0820503) and National Natural Science Foundation of China (Number 62072149) supported this work. Additional funding was provided by the Primary Research \& Development Plan of Zhejiang Province (Number 2021C03156 and 2021C02004), and the Public Welfare Research Program of Zhejiang (Number LGG19F020017).
\end{acks}


\bibliographystyle{ACM-Reference-Format}
\bibliography{mybibli.bib}

\section*{Appendix}
\subsection*{Appendix A. Proof for the equivalence of the neighbor selection strategies of HNSW and NSG}
\label{Appendix_A}

\noindent\underline{\textbf{Notations.}} Given any point $p$ on dataset $S$, the candidate neighbor set of $p$ obtained before neighbor selection is marked as $\mathcal{C}$ (see \hyperref[candidate neighbor acquisition]{Definition 4.4} for the definition of \textit{candidate neighbor acquisition}), and the \textit{neighbor selection} (\hyperref[def: neighbor selection]{Definition 4.5}) gets the final neighbor set $N(p)$ from $\mathcal{C}$ for $p$. $\mathcal{B}(p,r)$ denotes an open sphere such that $\mathcal{B}(p,r)=\left \{ x | \delta(p,x) <r, x \in S \right \}$, where $r$ is a constant. $lune_{pm}$ denotes a region such that $lune_{pm}=\mathcal{B}(p,\delta(p,m)) \cap \mathcal{B}(m,\delta(m,p) )$.

\vspace{0.5em}
\noindent\underline{\textbf{The neighbor selection strategy of HNSW.}} In the original paper of HNSW~\cite{HNSW}, the \textit{neighbor selection} strategy is called heuristic neighbor selection. When selecting neighbors for the inserted point $p$, HNSW regards $p$ as a query to perform ANNS on the constructed partial graph index to obtain a certain amount of its nearest neighbors as candidate neighbors $\mathcal{ C}$. Next, the heuristic \textit{neighbor selection} iteratively gets the unvisited point $m$ that has the smallest $\delta(m,p)$ from $\mathcal{C}$, if $\forall n \in N(p)$, $\delta (m,n)> \delta (m,p)$ (\textit{\textbf{Condition 1}}), then $N(p) \cup \left \{ m \right \}$, otherwise, $m$ will be discarded. For more details, please see the Algorithm 4 in the original publication of HNSW~\cite{HNSW}.

\vspace{0.5em}
\noindent\underline{\textbf{The neighbor selection strategy of NSG.}} In the original paper of NSG~\cite{NSG}, the \textit{neighbor selection} strategy is called edge selection strategy of Monotonic Relative Neighborhood Graph (MRNG). When selecting neighbors for $p$, MRNG gets the unvisited point $m$ with the smallest $\delta(m,p)$ from $\mathcal{C}$. Iff $lune_{pm} \cap \mathcal{C}=\emptyset$ or $\forall u \in (lune_{pm} \cap \mathcal{C})$, $ u \not\in N(p)$ (\textit{\textbf{Condition 2}}), then $N (p) \cup \left \{ m \right \}$. For more details, please refer to the Algorithm 2 in the original publication of NSG~\cite{NSG}.

Below we prove the equivalence of the \textit{neighbor selection} of the two.
\begin{proof}
First, we prove that the \textit{neighbor selection} of NSG can be derived from the \textit{neighbor selection} of HNSW. For any point $m \in \mathcal{C}$ that can be added to $N(p)$, we only need to prove that if \textit{\textbf{Condition 1}} is satisfied, then \textit{\textbf{Condition 2}} must be satisfied. For \textit{\textbf{Condition 1}}: $\forall n \in N(p)$, $\delta (m,n)> \delta (m,p)$, we can infer that $\forall n \in N(p)$ must satisfy $n \not\in\mathcal{B}(m,\delta(m,p))$, otherwise, there will $\exists n \in N(p)$ makes $\delta (m,n) <\delta (m,p)$. Thus, we have
\begin{align}
  \nonumber lune_{pm} \cap N(p) &= \mathcal{B}(p,\delta(p,m)) \cap \mathcal{B}(m,\delta(m,p)) \cap  N(p)\\
  \nonumber &=\mathcal{B}(p,\delta(p,m)) \cap \emptyset \\
  \nonumber &=\emptyset
\end{align}
Since $N(p)$ are all selected from $\mathcal{C}$, that is, $N(p) \subset \mathcal{C}$, below, we will discuss whether $lune_{pm} \cap (\mathcal {C} \setminus N(p))$ is $\emptyset$.

\noindent(1) Suppose $lune_{pm} \cap (\mathcal{C} \setminus N(p))=\emptyset$. Then, we have
\begin{align}
  \nonumber lune_{pm} \cap \mathcal{C} &=lune_{pm} \cap ((\mathcal{C} \setminus N(p)) \cup N(p)) \\
  \nonumber &=(lune_{pm} \cap (\mathcal{C} \setminus N(p))) \cup (lune_{pm} \cap N(p))\\
  \nonumber &=\emptyset \cup \emptyset \\
  \nonumber &= \emptyset
\end{align}
\textit{\textbf{Condition 1}} is satisfied.

\noindent(2) Suppose $lune_{pm} \cap (\mathcal{C} \setminus N(p))\not = \emptyset$. Obviously, $lune_{pm} \cap \mathcal{C}\not =\emptyset$, because $lune_{pm} \cap N(p)= \emptyset$, so $\forall u \in (lune_{pm} \cap \mathcal{C})$ must have
\begin{align}
  \nonumber u \in lune_{pm} \cap (\mathcal{C} \setminus N(p))=(lune_{pm} \cap \mathcal{C}) \setminus N(p)
\end{align}
That is, $u \not\in N(p)$. \textit{\textbf{Condition 1}} is satisfied.

Therefore,  if \textit{\textbf{Condition 1}} is established, then \textit{\textbf{Condition 2}} must be established.

Next, we prove that the \textit{neighbor selection} of HNSW can be derived from the \textit{neighbor selection} of NSG. For any point $m \in \mathcal{C}$ that can be added to $N(p)$, we only need to prove that if \textit{\textbf{Condition 2}} is satisfied, then \textit{\textbf{Condition 1}} must be satisfied. For \textit{\textbf{Condition 2}}: $lune_{pm} \cap \mathcal{C}=\emptyset$ or $\forall u \in (lune_{pm} \cap \mathcal{C})$, $ u \not\in N(p)$, we discuss the two cases of $lune_{pm} \cap \mathcal{C}=\emptyset$ and $\forall u \in (lune_{pm} \cap \mathcal{C})$, $ u \not\in N(p)$ separately. 

\noindent(1) When $lune_{pm} \cap \mathcal{C}=\emptyset$ is established, because $N(p) \subset \mathcal{C}$, therefore, $lune_{pm} \cap N(p)=\emptyset$. Since the unvisited point $m$ with the smallest $\delta(m,p)$ is taken from $\mathcal{C}$ every time, that is, $\forall n \in N(p)$, $\delta(n ,p) <\delta(m,p)$, thus, $\forall n \in N(p)$, $n \in \mathcal{B}(p,\delta(p,m))$, and we have
\begin{align}
  \nonumber n \not\in &\mathcal{B}(m,\delta(m,p))\setminus \mathcal{B}(p,\delta(p,m)) \\
  \nonumber &=\mathcal{B}(m,\delta(m,p)) \setminus (\mathcal{B}(m,\delta(m,p)) \cap \mathcal{B}(p,\delta(p,m))) \\
  \nonumber &=\mathcal{B}(m,\delta(m,p)) \setminus lune_{pm}
\end{align}
Since $lune_{pm} \cap N(p)=\emptyset$, then $\forall n \in N(p)$, $n \not\in lune_{pm}$, so
\begin{align}
  \nonumber n \not\in(\mathcal{B}(m,\delta(m,p)) \setminus lune_{pm}) \cup lune_{pm}=\mathcal{B}(m,\delta(m,p))
\end{align}
Thus, $\forall n \in N(p)$, $\delta (m,n)> \delta (m,p)$, \textit{\textbf{Condition 1}} is satisfied.

\noindent(2) When $\forall u \in (lune_{pm} \cap \mathcal{C})$, $ u \not\in N(p)$, it is easy to know that $\forall n \in N(p) $, $n \not\in (lune_{pm} \cap \mathcal{C})$. Thus, $n \not\in lune_{pm}$, otherwise, if $n \in lune_{pm}$, $n \in \mathcal{C}$ is known, then $\exists n \in N(p )$ makes $n \in (lune_{pm} \cap \mathcal{C})$, which contradicts the known. Since $n \not\in \mathcal{B}(m,\delta(m,p)) \setminus lune_{pm}$, we have
\begin{align}
  \nonumber n \not\in lune_{pm} \cup (\mathcal{B}(m,\delta(m,p)) \setminus lune_{pm})=\mathcal{B}(m,\delta(m,p))
\end{align}
Thus, $\forall n \in N(p)$, $\delta (m,n)> \delta (m,p)$, \textit{\textbf{Condition 1}} is satisfied.

Therefore, if \textit{\textbf{Condition 2}} is established, then \textit{\textbf{Condition 1}} must be established.

In summary, the \textit{neighbor selection} strategies of HNSW and NSG are equivalent.
\end{proof}

\subsection*{Appendix B. Proof for the path adjustment of NGT is an approximation to the neighbor selection of RNG}
\label{Appendix_B}
\begin{figure}[!t]
  \centering	
  \includegraphics[width=\linewidth]{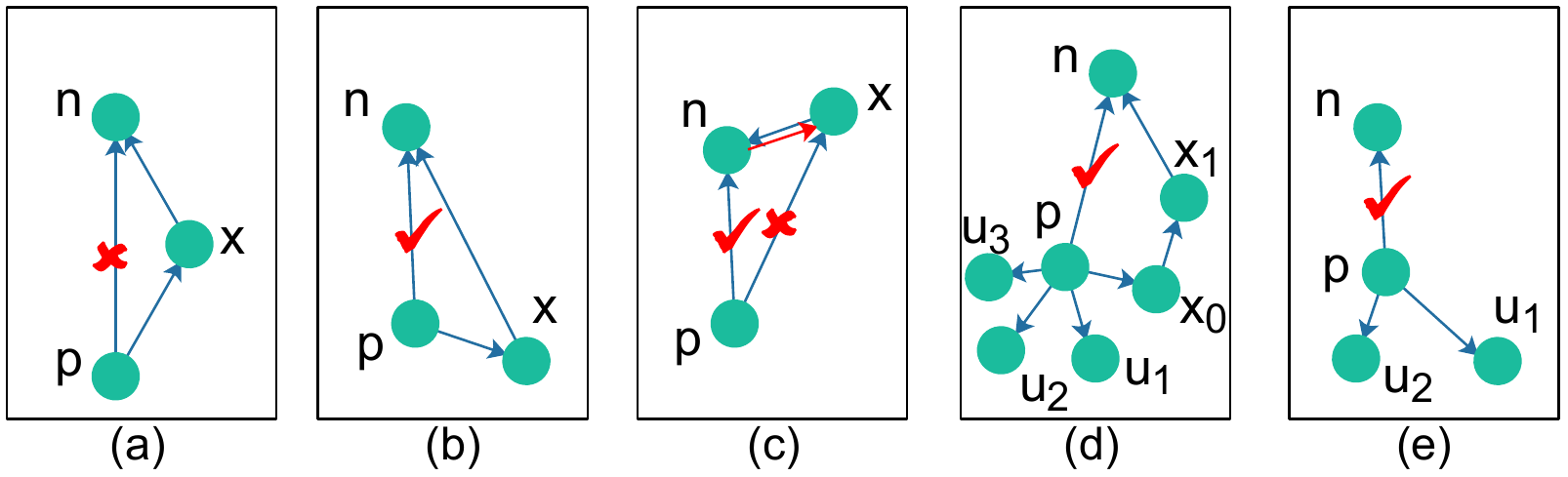}
  \caption{The path adjustment of NGT~\cite{yahoo3}.}
  \label{fig: NGT_path_adjustment}
\end{figure}

\noindent\underline{\textbf{Path adjustment.}} We can formally describe the path adjustment as follows. Given graph $G(V,E)$, $N(p)$ is the current neighbor set of $p \in V$. As shown in \autoref{fig: NGT_path_adjustment}, if there is an alternative path $p \rightarrow x \rightarrow n$ with length $l=2$ (the number of edges) between $p$ and $n$, where $n \in N(p)$, then do the following: If $\max \left \{ \delta(p,x), \delta(x,n) \right \} < \delta (p,n)$, then delete $n$ from $N(p)$, otherwise keep $n$. If there is no alternative path between $p$ and $n$ or the length $l \not = 2$ of the alternative path, then $n$ is also reserved. For more details about path adjustment, please refer to the original paper~\cite{yahoo3}.

\vspace{0.5em}
\noindent\underline{\textbf{The \textit{neighbor selection} of RNG.}} Given any point $p$ on dataset $S$, the candidate neighbor set of $p$ is $\mathcal{C}$, gets the unvisited point $m$ that has the smallest $\delta(m,p)$ from $\mathcal{C}$, if $\forall n \in N(p)$, $\delta (m,n)> \delta (m,p)$, then $N(p) \cup \left \{ m \right \}$, otherwise, $m$ will be discarded~\cite{FANNG}.

Next, we prove that the above path adjustment operation is an approximate implementation of \textit{neighbor selection} of RNG. Given vertex $p$ on $G(V,E)$, the neighbors $N(p)$ of $p$ is sorted in ascending order according to the distance between each neighbor and $p$, and the neighbors are visited in this order when conducting path adjustment. Therefore, the selection way of visited vertex for path adjustment is consistent with \textit{neighbor selection} of RNG. We only need to prove that the judgment criteria of the visited vertex cut or retained is an approximation to that of RNG.
\begin{proof}
  We conduct the discussion based on the alternative path length $l$ between $p$ and $n$.

  \noindent(1) As shown in \autoref{fig: NGT_path_adjustment} (a), (b), and (c), when the length $l=2$ of alternative path $p \rightarrow x \rightarrow n$ between $p$ and $n$, there are two situations:
  \begin{itemize}
  \item If $\max \left \{ \delta(p,x), \delta(x,n) \right \} < \delta (p,n)$ (\autoref{fig: NGT_path_adjustment} (a)), then $n$ will be deleted. It is easy to know that $\delta (p,n)> \delta (x,n)$, which is consistent with the \textit{neighbor selection} of RNG.
  \item If $\max \left \{ \delta(p,x), \delta(x,n) \right \} > \delta (p,n)$ (\autoref{fig: NGT_path_adjustment} (b) and (c)), then $n$ will be reserved. At this time, when $\delta (p,n) < \delta (x,n)$ (\autoref{fig: NGT_path_adjustment}(b)), the \textit{neighbor selection} of RNG is met; When $\delta (p,n)> \delta (x,n)$ (\autoref{fig: NGT_path_adjustment}(c)), then $\delta (p,n) <\delta (p,x)$. Since $\delta (x,n)$ is small, $n \in N(x)$ has a high probability of occurrence. We know that $x \in N(p)$, when judging whether there is an alternative path between $p$ and $x$, since the alternative path $(p \rightarrow n \rightarrow x)$ exists, and satisfies $\delta (p,n) <\delta (p,x)$ and $\delta (n,x) <\delta (p,x)$, that is, $\max \left \{ \delta(p ,n), \delta(n,x) \right \} <\delta (p,x)$, thus, $x$ needs to be deleted. At this time, keeping $n$ and deleting $x$ is consistent with the result of the \textit{neighbor selection} of RNG.
  \end{itemize}

  \noindent(2) As shown in \autoref{fig: NGT_path_adjustment}(d), when there is an alternative path with length $l > 2$ between $p$ and $n$, it means that the distance between $n$ and $p$ is likely to be farther, so that most of the neighbors of $n$ and $p$ are far away. Therefore, $\delta (n,p) <\delta (n,u)$ is easy to hold for most $u$, and $u$ is one of the neighbors of $p$ except $n$. In this case, the results of keeping $n$ is consistent with the \textit{neighbor selection} of RNG.

  \noindent(3) As shown in  \autoref{fig: NGT_path_adjustment}(e), if there is no alternative path between $p$ and $n$, it means that there is a high probability that $p$ is closer to $n$, so $\delta (p,n) <\delta (u,n) $ for most $u$, $u$ is one of the neighbors of $p$ except $n$. In this case, the results of keeping $n$ is consistent with the \textit{neighbor selection} of RNG.

  In summary, the path adjustment of NGT is an approximation to the neighbor selection of RNG.
\end{proof}

\subsection*{Appendix C. Proof for the neighbor selection of DPG is an approximation to that of RNG}
\label{Appendix_C}

\noindent\underline{\textbf{Overview.}} According to \hyperref[Appendix_A]{Appendix A} and \hyperref[Appendix_B]{Appendix B}, the \textit{neighbor selection} of RNG can be approximately described by the \textit{neighbor selection} of HNSW. Due to the equivalence of the neighbor selection strategies of HNSW and NSG, we can represent the \textit{neighbor selection} of RNG by the \textit{neighbor selection} of NSG (\hyperref[Appendix_A]{Appendix A}). Therefore, we only need to prove that the \textit{neighbor selection} of DPG is an approximation to that of NSG.

\vspace{0.5em}
\noindent\underline{\textbf{The \textit{neighbor selection} of DPG.}} The construction of DPG is a diversification of the KGraph~\cite{KGraph}, followed by adding reverse neighbors. Given any a vertex $p$ on $G(V,E)$, $\mathcal{C}$ is the candidate neighbor set of $p$, $\theta (x,y)$ denotes $\measuredangle xpy$, where $x,y \in \mathcal{C}$. The \textit{neighbor selection} of DPG aims to choose a subset $N(p)$ of $\kappa$ vertices from $\mathcal{C}$ so that $N(p) = \arg \max _{N(p) \subseteq \mathcal{C}} \sum _{x,y \in \mathcal{C}} \theta (x,y)$. For full details, please see the original publication of DPG~\cite{DPG}.

\begin{figure}[!t]
  \centering	
  \includegraphics[width=\linewidth]{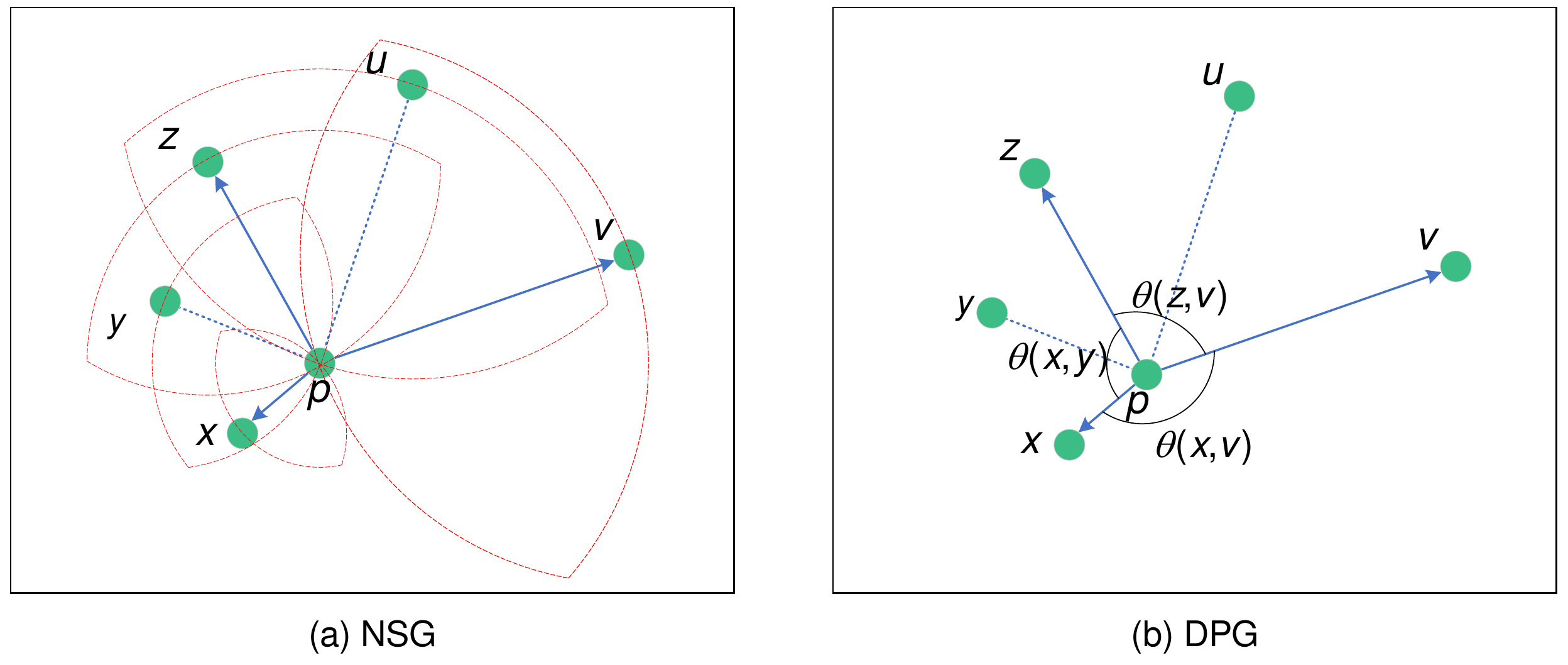}
  \caption{The \textit{neighbor selection} of NSG and DPG.}
  \label{fig: NSG_DPG_neighbor_selection}
\end{figure}

\begin{lemma}
  \label{lemma: NSG_angle}
  Given the $G(V,E)$ constructed by the \textit{neighbor selection} of NSG, any a vertex $p \in V$ and the neighbor set $N(p)$ of $p$ on $G(V,E)$. $\forall x,y \in N(p)$, $\measuredangle xpy \geq 60^{\circ}$.
\end{lemma}

We prove \hyperref[lemma: NSG_angle]{Lemma 7.1} as follows.
\begin{proof}
  As shown in \autoref{fig: NSG_DPG_neighbor_selection}(a), for the $N(p)$ obtained from $\mathcal{C}$ by the \textit{neighbor selection} of NSG (\hyperref[Appendix_A]{Appendix A}), if $\exists x, y \in N(p)$, and $\measuredangle xpy < 60^{\circ}$, then we can know that $\measuredangle pyx + \measuredangle pxy > 120^{\circ}$ in $\triangle xpy$, thus, there must be $\measuredangle pyx > 60^{\circ}$ or $\measuredangle pxy > 60^{\circ}$.

  Suppose $\delta (p,y)> \delta (p,x)$, then $\measuredangle pxy> 60^{\circ}$ (i.e., the situation shown in \autoref{fig: NSG_DPG_neighbor_selection}(a)), we know that $\measuredangle pxy > \measuredangle xpy$, so $\delta (p,y)>\delta (x,y)$, we have $x \in \mathcal{B}(y,\delta (p,y))$. Since $\delta (p,y)> \delta (p,x)$, so $x \in \mathcal{B}(p,\delta (p,y))$, it is easy to know that $x \in \mathcal{B }(y,\delta (p,y)) \cap \mathcal{B}(p,\delta (p,y)) = lune_{py}$. However, $x \in N(p)$, according to the \textit{neighbor selection} strategy of NSG (\hyperref[Appendix_A]{Appendix A}), $y$ cannot be added to $N(p)$, which contradicts the known ($y \in N(p)$).

  When $\delta (p,y) <\delta (p,x)$, we can just swap the positions of $x$ and $y$ above, and then get the same conclusion.

  Therefore, \hyperref[lemma: NSG_angle]{Lemma 7.1} is proved.
\end{proof}

Now we prove that the \textit{neighbor selection} of DPG is an approximation to that of NSG.

\begin{proof}
  As shown in \autoref{fig: NSG_DPG_neighbor_selection}(a), when selecting neighbors for $p$, it needs to determine whether a point in $\mathcal{C} = \left \{ x, y, z,u,v \right \}$ can be added to $N(p)$ according to the \textit{neighbor selection} of NSG, and finally $N(p)= \left \{ x,z,v \right \}$. It can be seen that $\forall x,y \in N(p)$, $\measuredangle xpy \geq 60^{\circ}$ from \hyperref[lemma: NSG_angle]{Lemma 7.1}.

  \autoref{fig: NSG_DPG_neighbor_selection}(b) is based on the \textit{neighbor selection} strategy of DPG, which selects $\kappa = 3$ neighbors to $N(p)$ from $\mathcal{C} = \left \{ x, y, z,u,v \right \}$ to maximize the sum of angles between neighbors. From this, it can be inferred that $\exists \hat {\theta} \in [0^{\circ},180^{\circ}]$, for $\forall v_1,v_2 \in N(p)$, $\theta (v_1,v_2) \geq \hat{\theta}$. When $\hat{\theta} = 60^{\circ}$, that is, $\hat{\theta} \geq 60^{\circ}$, which ensures the result of \textit{neighbor selection} of NSG.
  
  Therefore, we can say that the \textit{neighbor selection} of DPG is an approximation to that of NSG, and thus an approximation to that of RNG.
\end{proof}

\subsection*{Appendix D. Complexity analysis}
\label{Appendix_D}
Since the construction or search complexity of some graph-based ANNS algorithms were not informed by the authors, we deduce the approximate complexity based on the description of the algorithm and our experimental evaluation. In order to make the evaluation results more general, we carefully select the characteristics of the dataset, which are shown in \autoref{tab: characteristics_dataset_complexity_evaluation}. The standard deviation is the standard deviation of the distribution in each cluster.

For index construction complexity, we record the index construction time under different scales and calculate the functional relationship between the two. As for search complexity, we count the number of distance evaluations under a given recall rate, considering that distance evaluation occupies the main search time~\cite{SONG}. Note that the relevant parameters used in the evaluation process are the optimal parameters obtained by grid search (see our online publication\footnote{https://github.com/Lsyhprum/WEAVESS \label{github_code}} for more details). All experimental evaluation codes remove parallel commands and use single-threaded program execution on a Linux server with Intel(R) Core(TM) i9-10900X CPU at 3.70GHz, and 125G memory.

\setlength{\textfloatsep}{0cm}
\setlength{\floatsep}{0cm}
\begin{table}
\setlength{\abovecaptionskip}{0.05cm}
\setstretch{0.9}
\fontsize{8pt}{4mm}\selectfont
    \centering
    \tabcaption{Characteristics of the dataset for complexity evaluation}
    \label{tab: characteristics_dataset_complexity_evaluation}
    \setlength{\tabcolsep}{.015\linewidth}{
    \begin{tabular}{l|l|l|l|l}
    \hline
    \textbf{\# Base} & \textbf{\# Query} & \textbf{Dimension} & \textbf{\# Cluster} & \textbf{Standard deviation} \\
    \hline
    \hline
    $10^{5} \sim 10^{6}$ & $10^{3}$ & 32 & 10 & 5 \\
    \hline
    \end{tabular}
    }

\end{table}


\begin{figure*}
  \vspace{-4mm}
  \centering
  \subfigcapskip=-0.25cm
  \subfigure[Search complexity of KGraph]{  
    \captionsetup{skip=0pt}
    \vspace{-1mm}
    \includegraphics[scale=0.3]{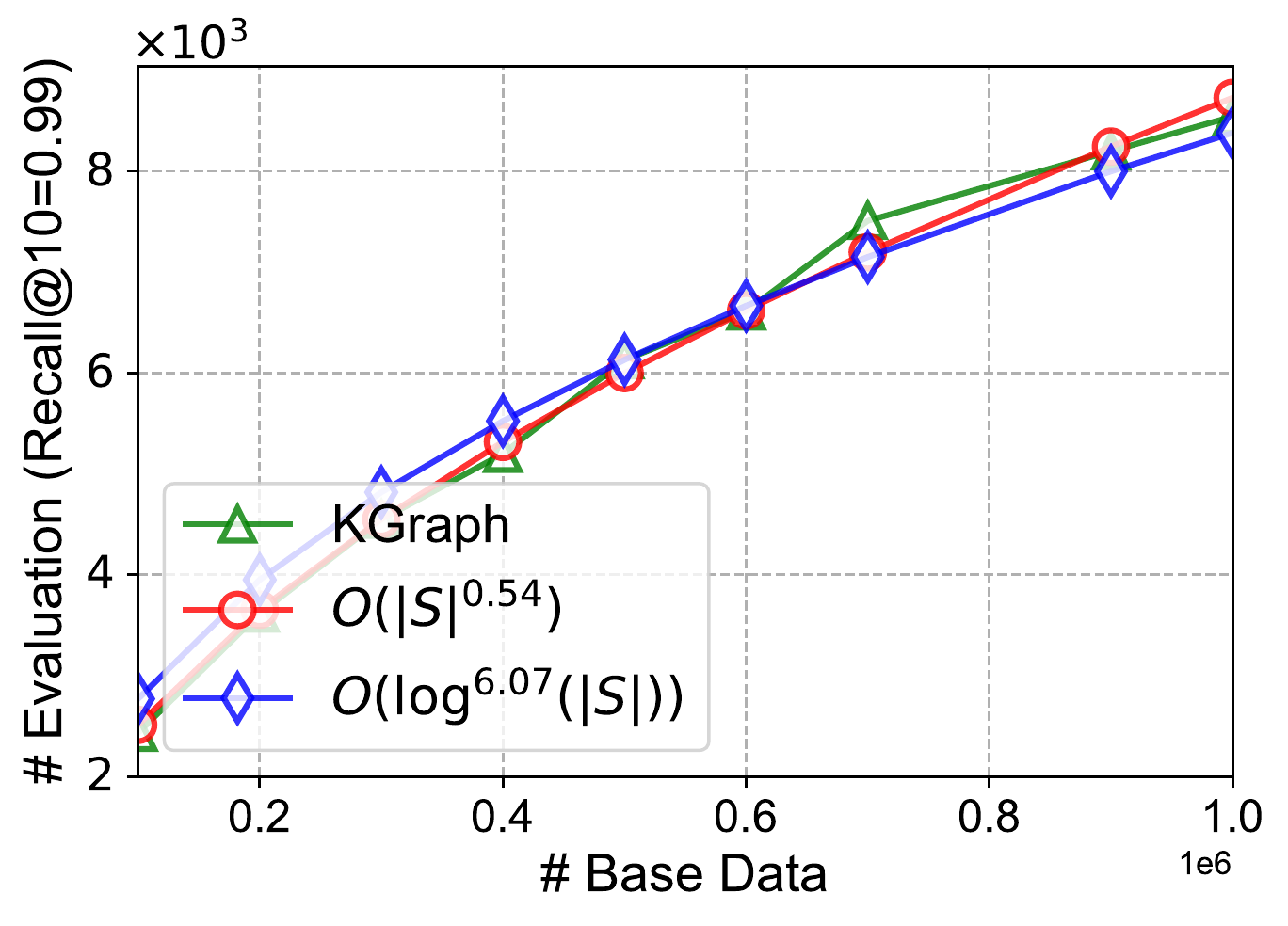}
    \label{fig: kgraph_search_complexity}
  }
  \subfigure[Construction complexity of NGT]{ 
    \captionsetup{skip=0pt}
    \vspace{-1mm}
    \includegraphics[scale=0.3]{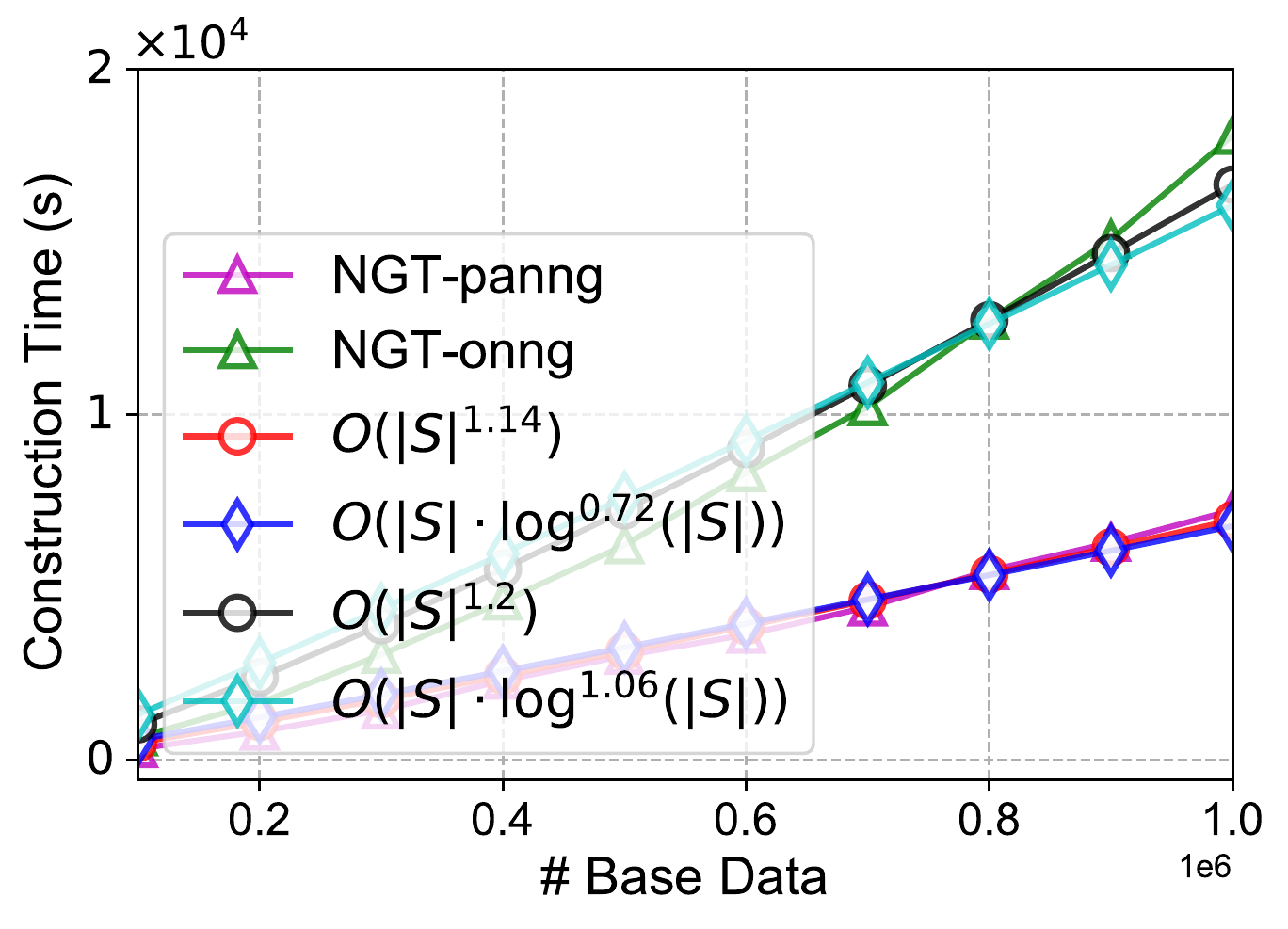}
    \label{fig: ngt_construction_complexity}
  }
  \subfigure[Search complexity of NGT]{ 
    \captionsetup{skip=0pt}
    \vspace{-1mm}
    \includegraphics[scale=0.3]{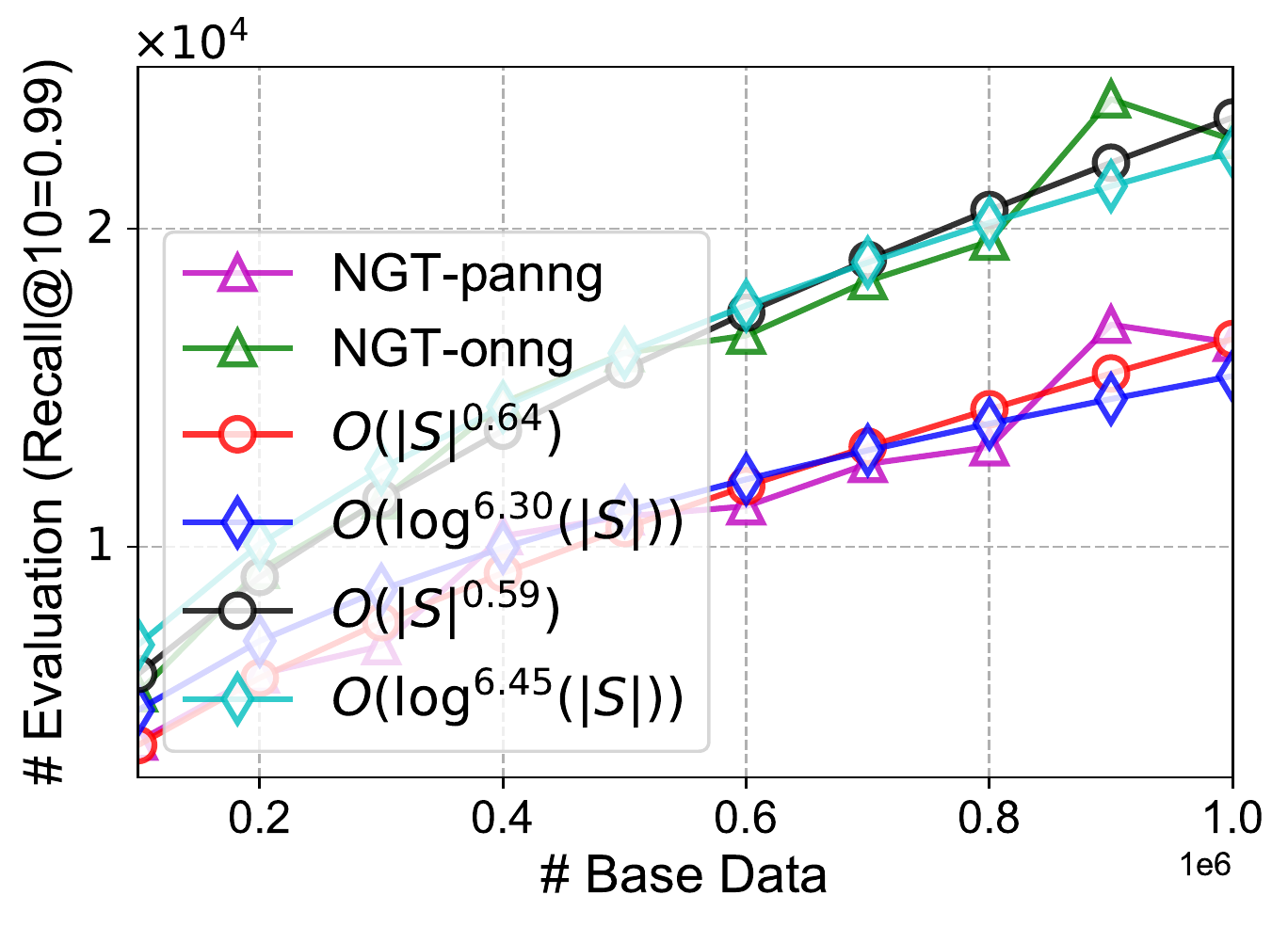}
    \label{fig: ngt_search_complexity}
  }
  \subfigure[Search complexity of SPTAG]{ 
    \captionsetup{skip=0pt}
    \vspace{-1mm}
    \includegraphics[scale=0.3]{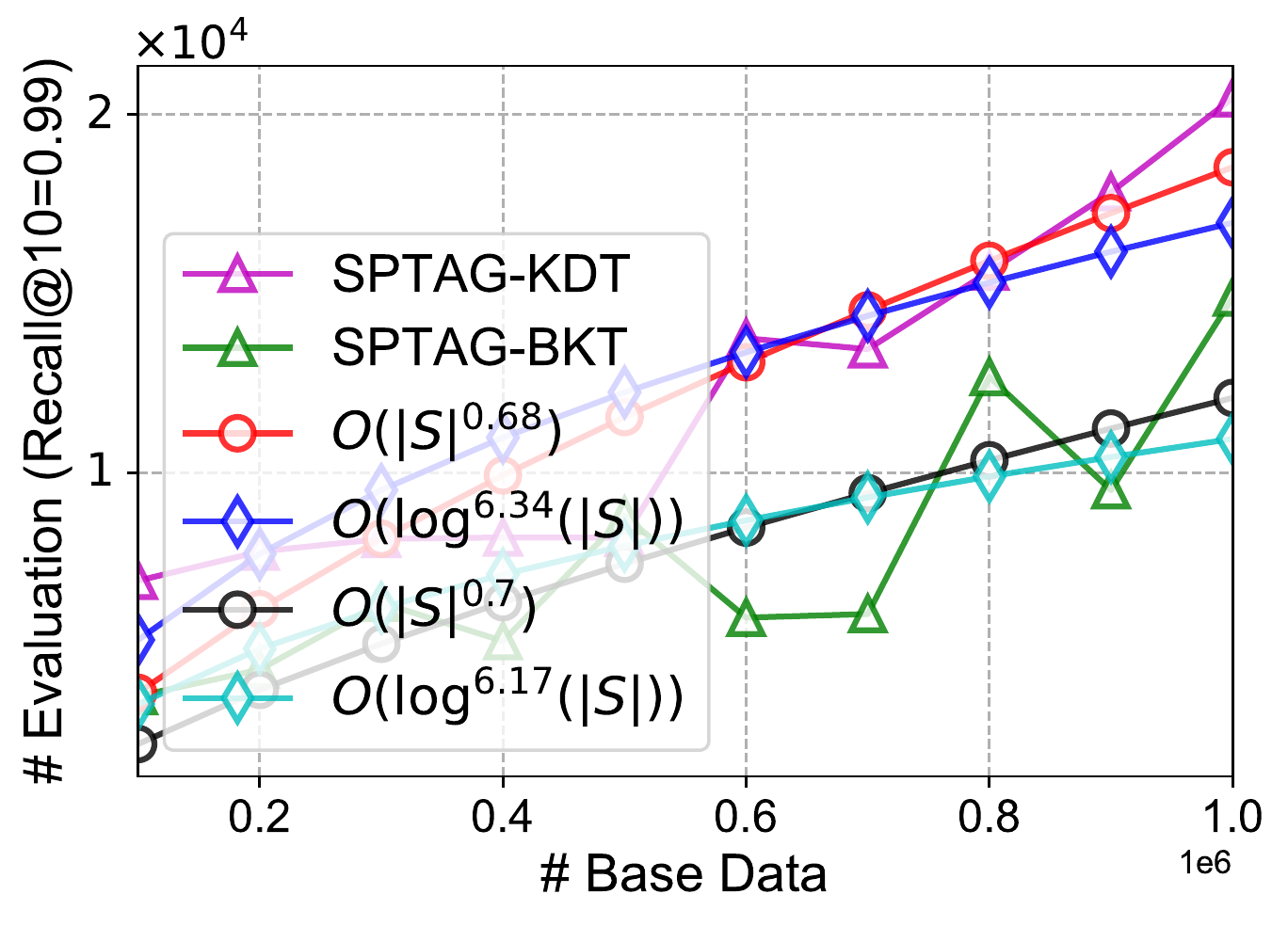}
    \label{fig: sptag_search_complexity}
  }\vspace{-0.3cm}
  \subfigure[Search complexity of IEH]{ 
    \captionsetup{skip=0pt}
    \vspace{-1mm}
    \includegraphics[scale=0.3]{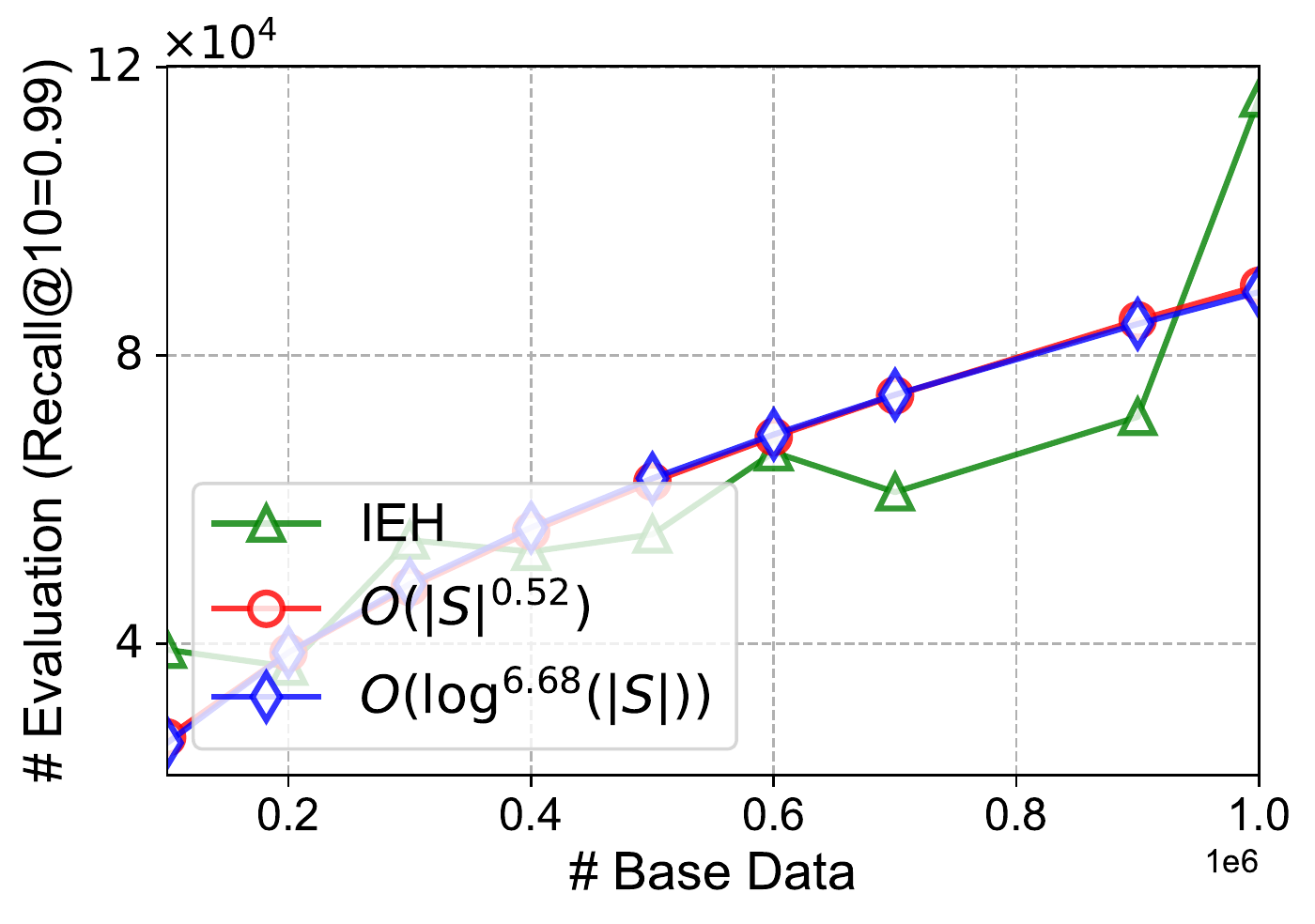}
    \label{fig: ieh_search_complexity}
  }
  \subfigure[Construction complexity of EFANNA]{  
    \captionsetup{skip=0pt}
    \vspace{-1mm}
    \includegraphics[scale=0.3]{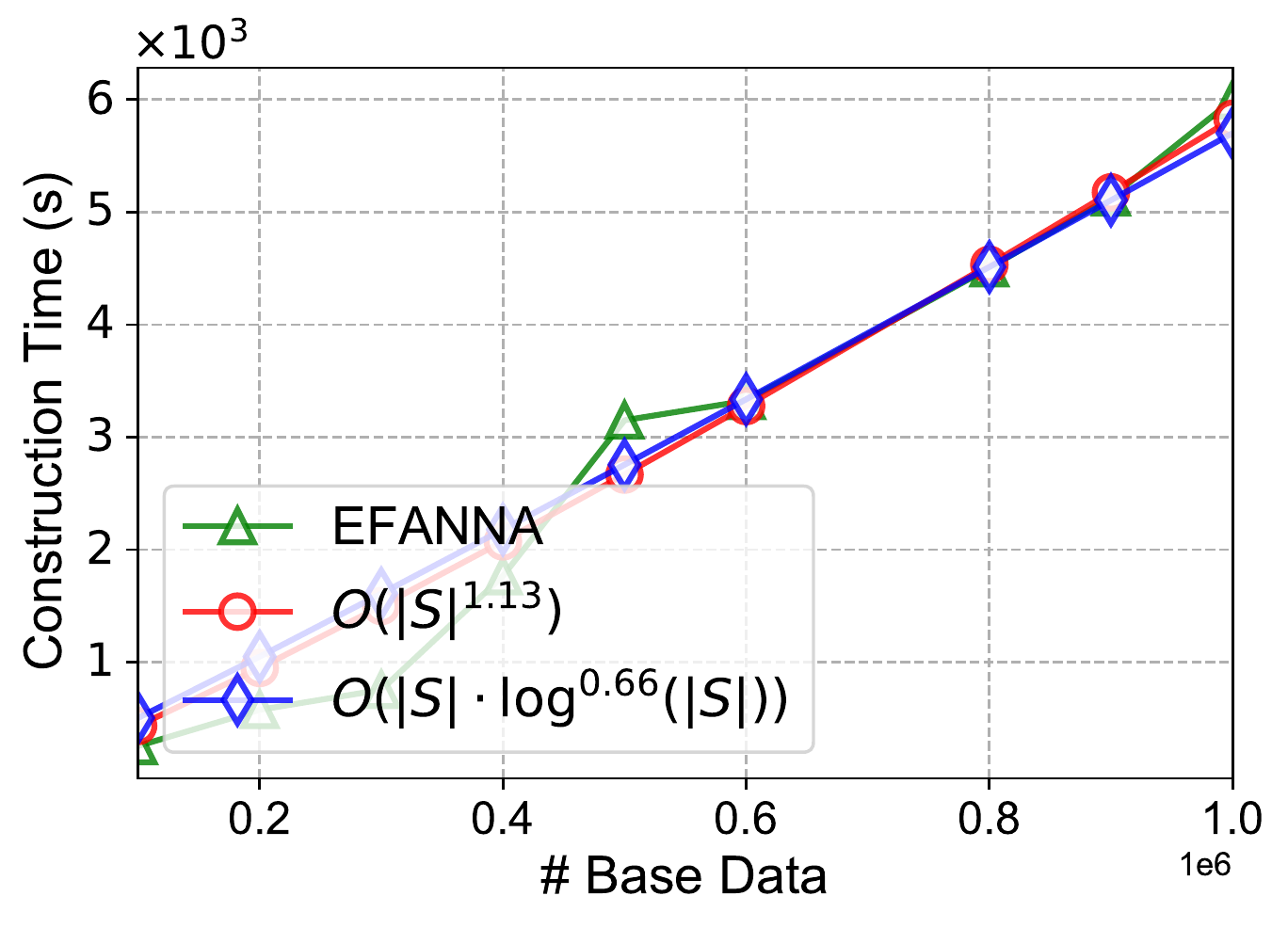}
    \label{fig: efanna_construction_complexity}
  }
  \subfigure[Search complexity of EFANNA]{  
    \captionsetup{skip=0pt}
    \vspace{-1mm}
    \includegraphics[scale=0.3]{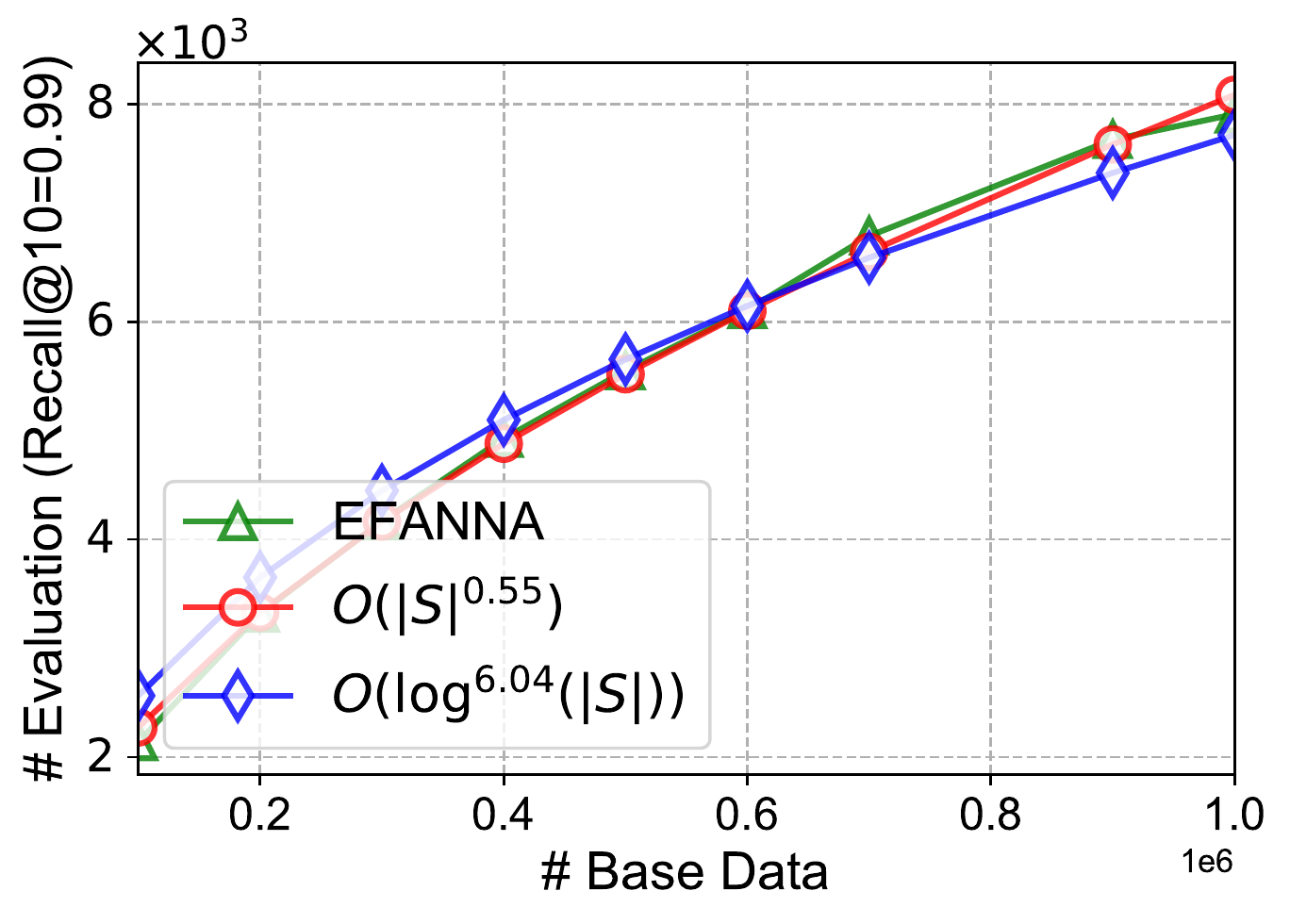}
    \label{fig: efanna_search_complexity}
  }
  \subfigure[Search complexity of DPG]{  
    \captionsetup{skip=0pt}
    \vspace{-1mm}
    \includegraphics[scale=0.3]{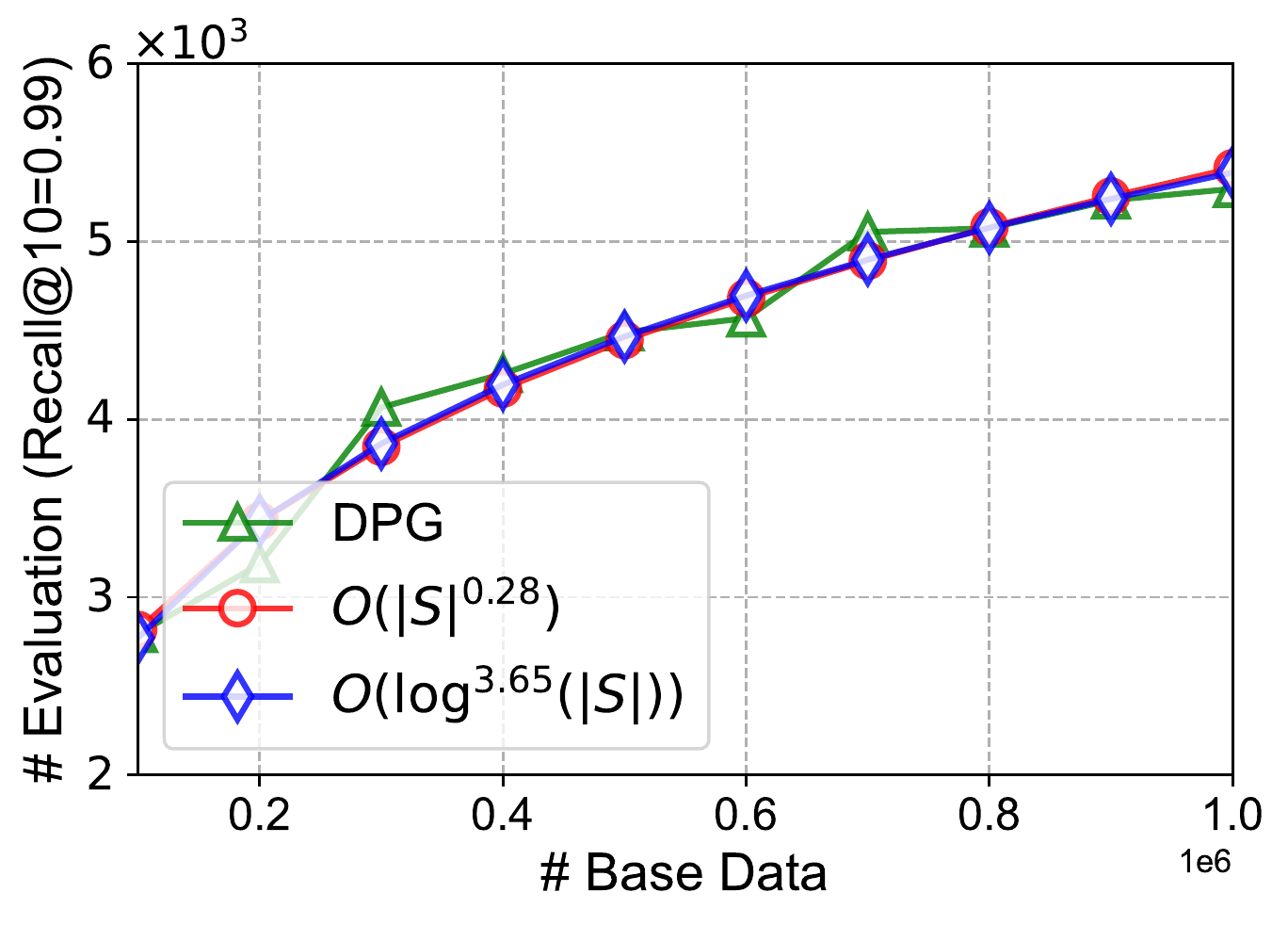}
    \label{fig: dpg_search_complexity}
  }\vspace{-0.3cm}
  \subfigure[Search complexity of HCNNG]{ 
    \captionsetup{skip=0pt}
    \vspace{-1mm}
    \includegraphics[scale=0.3]{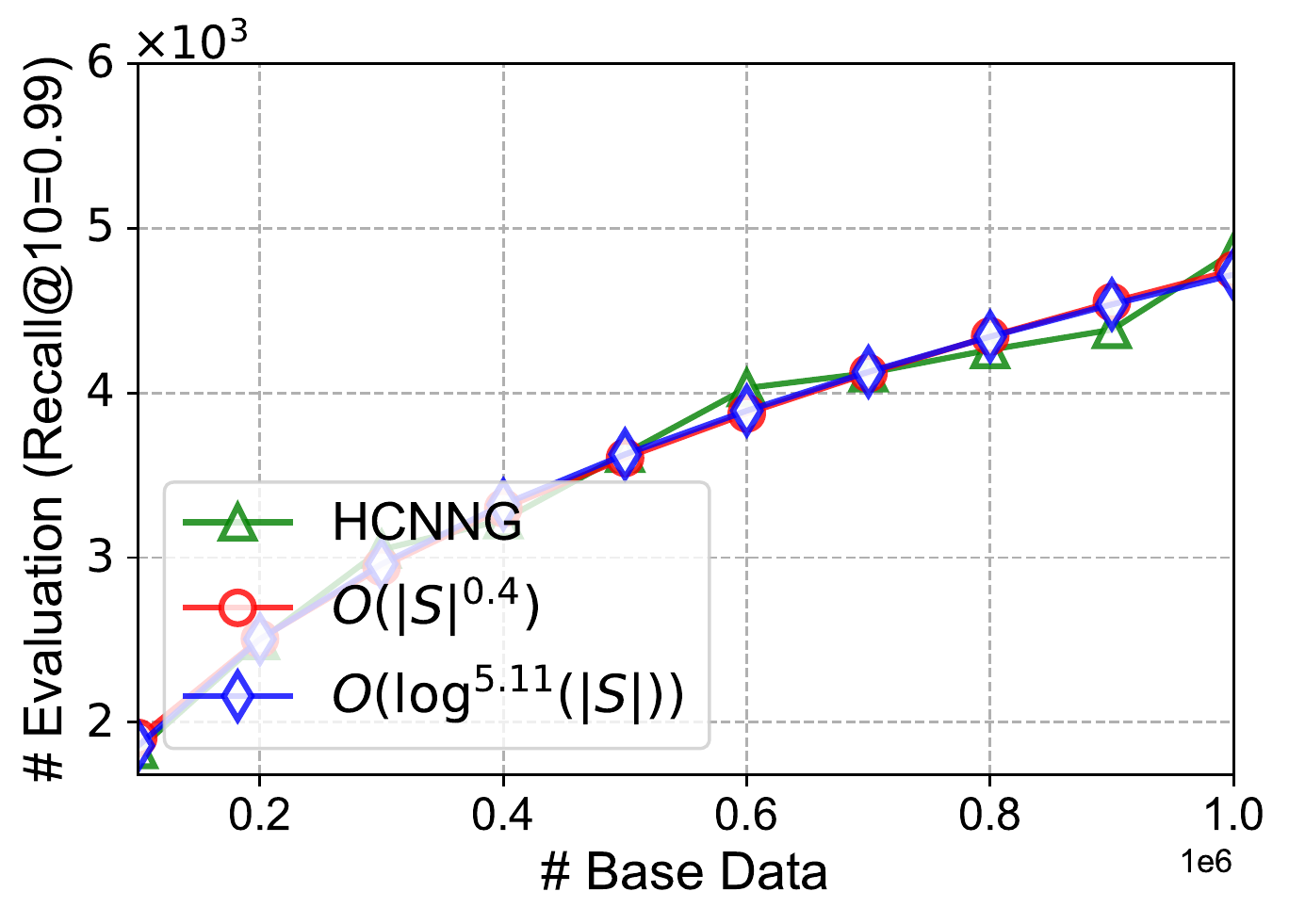}
    \label{fig: hcnng_search_complexity}
  }
  \subfigure[Construction complexity of Vamana]{  
    \captionsetup{skip=0pt}
    \vspace{-1mm}
    \includegraphics[scale=0.3]{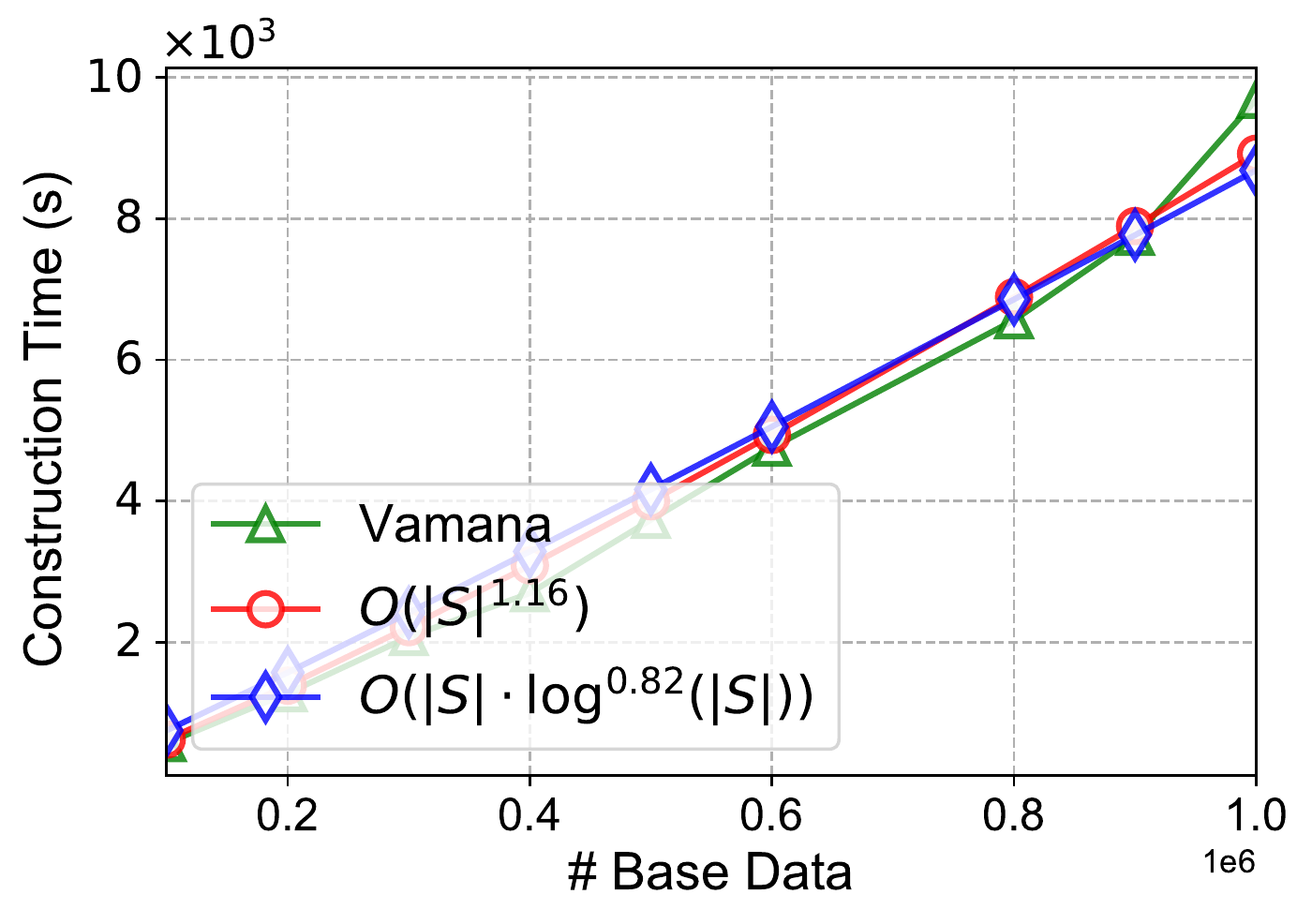}
    \label{fig: vamana_construction_complexity}
  }
  \subfigure[Search complexity of Vamana]{  
    \captionsetup{skip=0pt}
    \vspace{-1mm}
    \includegraphics[scale=0.3]{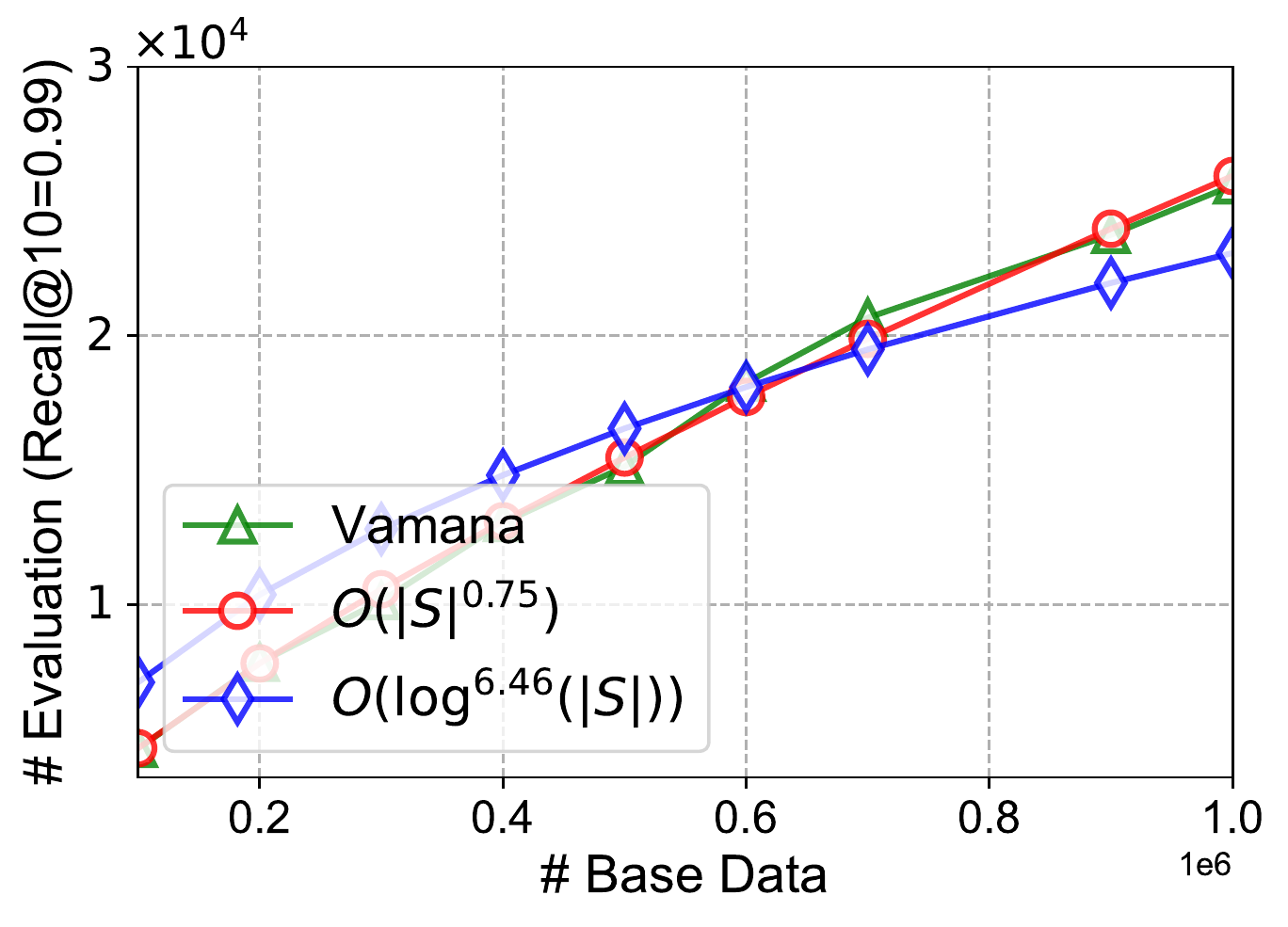}
    \label{fig: vamana_search_complexity}
  }
  \subfigure[Search complexity of k-DR]{  
  \captionsetup{skip=0pt}
  \vspace{-1mm}
  \includegraphics[scale=0.3]{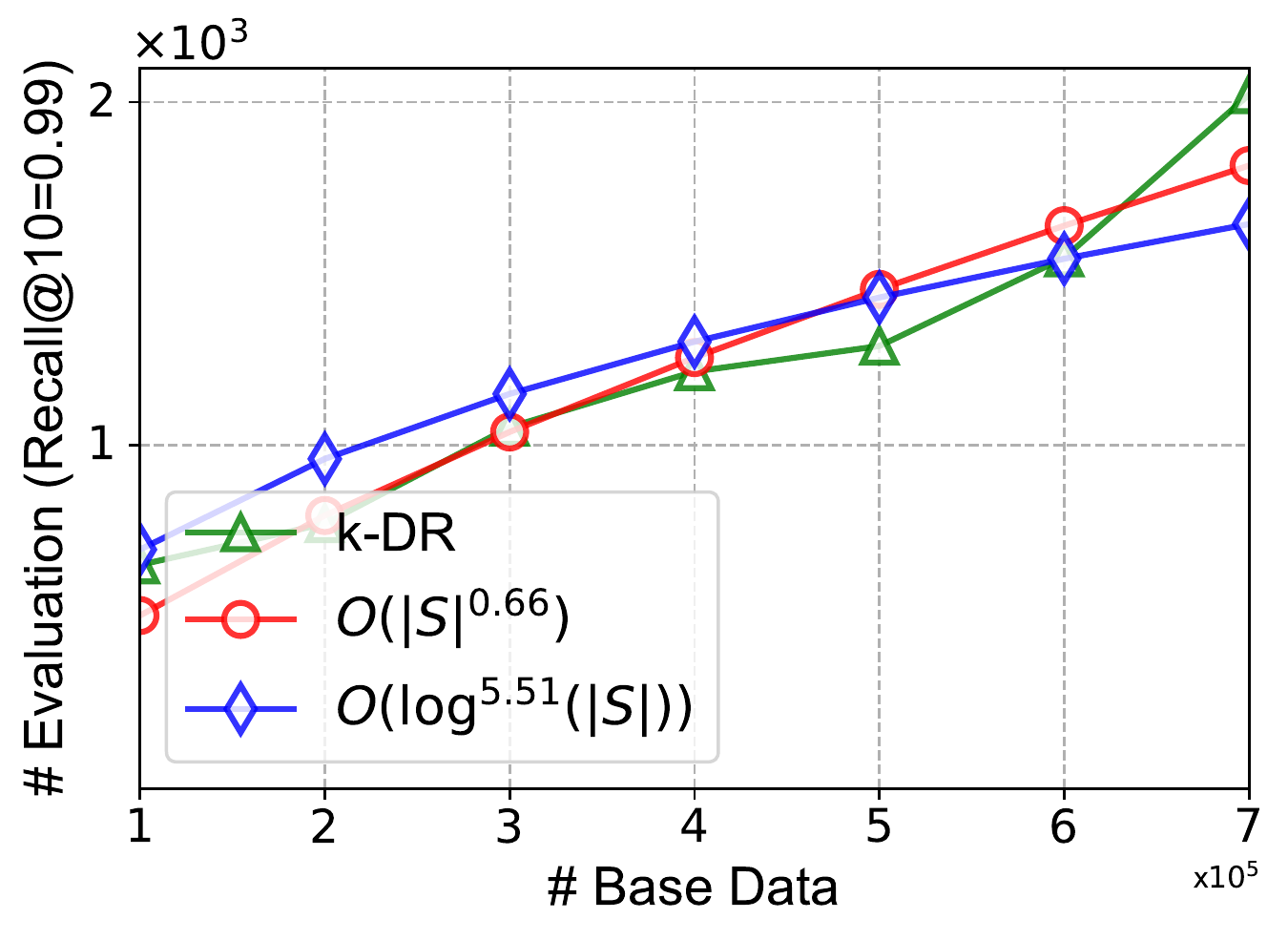}
  \label{fig: kdrg_search_complexity}
}
  \vspace{-0.5cm}
  \caption{Complexity evaluation. \# Base is the size of base dataset, and \# Evaluation is the number of distance evaluations.}\vspace{-0.4mm}
  \label{fig: complexity}
\end{figure*}

\noindent\underline{\textbf{KGraph.}} We set $Recall@10=0.99$ and evaluate how the number of distance evaluations with the data size. As shown in \autoref{fig: complexity}(a), the complexity of search on KGraph is about $O(\vert S \vert ^{0.54})$. When the size of the dataset is small ($< 0.6 \times 10^{6}$), the search complexity of KGraph is lower than that of $O(\log^{6.07}(\vert S \vert))$. However, as the size of the dataset increases, the search complexity of KGraph is slightly higher than $O(\log^{6.07}(\vert S \vert))$.

\noindent\underline{\textbf{NGT.}} There are two main versions of NGT, NGT-panng and NGT-onng respectively. The main difference between the two is that NGT-onng has one extra step that out-degree and in-degree adjustment than NGT-panng~\cite{NGT}. According to the evaluation results in \autoref{fig: complexity}(b) and (c), NGT-onng requires a larger construction time than NGT-panng due to additional out-degree and in-degree adjustment. However, NGT-onng has not received better search performance than NGT-panng, the search complexity of the two is very close, and NGT-panng has higher efficiency in the evaluation dataset, which shows that the effectiveness of out-degree and in-degree adjustment needs to be further verified.

\noindent\underline{\textbf{SPTAG.}} We implement two versions of SPTAG, SPTAG-KDT and SPTAG-BKT respectively. SPTAG-KDT is the original version, which corresponds to the description of the paper~\cite{wang2012query,wang2012scalable,wang2013trinary}. SPTAG-BKT is an optimized version based on SPTAG-KDT, the specific description can refer to the online project ~\cite{SPTAG}. As shown in \autoref{fig: complexity}(d), SPTAG-KDT ($O((\vert S \vert ^{0.68})$) and SPTAG-KMT ($O((\vert S \vert ^{0.7})$) have approximately the same search time complexity. However, under the same dataset size, SPTAG-BKT requires fewer distance evaluation times, that is, SPTAG-BKT mainly optimizes the constant part of search complexity.

\noindent\underline{\textbf{NSW.}} In \cite{NSW}, the authors give that the search complexity of NSW is $O(\log^{2}(\vert S \vert))$ by experiment. For index construction, NSW is based on the idea that graph structure is assembled by inserting elements one by one and connects them with a certain amount of nearest neighbors at each step. These nearest neighbors connected to each vertex is obtained by greedy search on the already constructed graph. As the time complexity of inserting each vertex is $O(\log^{2}(\vert S \vert))$, the complexity of constructing NSW on the dataset $S$ is $O(\vert S \vert \cdot \log^{2}(\vert S \vert))$.

\noindent\underline{\textbf{IEH.}} In \cite{IEH}, the authors report that the construction complexity of KNN table is $O(\vert S \vert ^2 \cdot (d + \log{(\vert S \vert)})$, where $d$ is the dimensionality and $d \ll \vert S \vert$. As we all know, the complexity of building a hash bucket is much less than $O(\vert S \vert ^2 \cdot (\log{(\vert S \vert)})$. Therefore, the construction complexity of IEH is $O(\vert S \vert ^2 \cdot \log(\vert S \vert) + \vert S \vert ^{2})$. In \autoref{fig: complexity}(e), the search complexity of IEH is about $O(\vert S \vert ^{0.52})$. However, under the same dataset size, IEH requires more distance evaluations than most other algorithms.

\noindent\underline{\textbf{EFANNA.}} The construction process of EFANNA is very similar to KGraph except for the \textit{initialization}. KGraph initializes the neighbors of each vertex randomly, while EFANNA uses KD-trees to initialize the neighbors more accurately. From \autoref{fig: complexity}(f), we can see that the construction complexity of EFANNA is about $O(\vert S \vert ^{1.13})$, which is very close to KGraph. It shows that EFANNA's optimization to KGraph only changes the constant factor of the construction complexity. Since both are constructing a KNNG with high quality, the search complexity of KGraph ($O(\vert S \vert ^{0.54})$) and EFANNA ($O(\vert S \vert ^{0.55})$) are also very similar in \autoref{fig: complexity}(a) and (g).

\noindent\underline{\textbf{DPG.}} The construction process of DPG includes two steps: (1) a KNNG construction and (2) the diversification of the KNNG. For the first step, the time complexity of constructing KNNG through NN-Descent on the dataset $S$ is $O(\vert S \vert ^{1.14})$. The second step is the process of pruning edges on the KNNG constructed in the previous step to maximize the angle between neighbors, followed by adding reverse edges (\hyperref[Appendix_C]{Appendix C}). For the vertex $p$ on KNNG, $p$'s neighbor set is $\mathcal{C}=\left \{ v_{0},v_{1},\cdots,v_{\kappa-1} \right \}$, in which elements are sorted in ascending order of distance from $p$, and $\vert \mathcal{C} \vert =\kappa$, the result neighbor set is $N(p)$ and initialized to $\emptyset$ for DPG. At the beginning of selecting neighbors for $p$ (the first iteration), we add $v_{0}$ to $N(p)$ and $\mathcal{C} \setminus \left \{ v_{0} \right \}$ ($\vert \mathcal{C} \vert = \kappa -1$, $\vert N(p) \vert = 1$); In the second iteration, we select $v_i \in \mathcal{C}$ to maximize $\measuredangle v_{i}pv_{0}$, then let $N(p)\cup \left \{v_{i} \right \}$ and $\mathcal{C} \setminus \left \{ v_{i} \right \}$ ($\vert \mathcal{C} \vert = \kappa -2$, $\vert N(p) \vert = 2$), it requires $\kappa -1$ calculations for selecting such a $v_{i}$; In the third iteration, we select $v_{j} \in \mathcal{C}$ so that $\measuredangle v_{j}pv_{0} +\measuredangle v_{j}pv_{i}$ is maximized, then let $ N(p)\cup \left \{v_{j}\right \}$ and $\mathcal{C} \setminus \left \{ v_{j} \right \}$ ($\vert \mathcal{C} \vert = \kappa -3$, $\vert N(p) \vert = 3$), we need $2\cdot(\kappa-2)$ calculations for obtaining $v_{j}$; ......

Therefore, if we select $c$ points from $\mathcal{C}$ to $N(p)$, the total number of calculations is
\begin{align}
  \nonumber \sum_{m=1}^{c-1}m \cdot(\kappa-m) &= \sum_{m=1}^{c-1}m \cdot \kappa - \sum_{m=1}^{c-1}m^{2} \\
  \nonumber &=\frac{c(c-1)}{2} \kappa - \frac{c(c-1)(2c-1)}{6}
\end{align}
Thereby,
\begin{align}
  \nonumber O(\frac{c(c-1)}{2} \kappa - \frac{c(c-1)(2c-1)}{6}) &=O(c^{2}\cdot \kappa)-O(c^{3})\\
  \nonumber &=O(c^{2}\cdot \kappa)
\end{align}

The time complexity of executing the above process for all $\vert S \vert$ points is $O(c^{2}\cdot \kappa \cdot \vert S \vert)=O(\vert S \vert)$ ($c^{2}\cdot \kappa \ll \vert S \vert$). Therefore, the construction complexity of DPG is $O(\vert S \vert ^{1.14} + \vert S \vert)$.

We set $Recall@10=0.99$ and evaluate how the number of distance evaluations with the data size. As shown in \autoref{fig: complexity}(h), the search complexity of DPG is about $O(\vert S \vert ^{0.28})$, which is obviously lower than KGraph. This confirms the effectiveness of diversification on DPG.

\noindent\underline{\textbf{HCNNG.}} As shown in \autoref{fig: complexity}(i), the search complexity of HCNNG is about $O(\vert S \vert ^{0.4})$. Note that HCNNG's routing strategy uses guided search, which improves routing efficiency through directional access to neighbors.

\noindent\underline{\textbf{Vamana.}} As shown in \autoref{fig: complexity}(j), the construction complexity of Vamana is about $O(\vert S \vert ^{1.16})$, which is close to KGraph and EFANNA. Among the algorithms that approximate RNG, Vamana achieves the lowest construction complexity. The search complexity of Vamana is about $O(\vert S \vert ^{0.75})$ from \autoref{fig: complexity}(k), which is even lower than some algorithms that only approximate KNNG (like KGraph). We do not receive the results achieved in the original paper~\cite{DiskANN}.

\noindent\underline{\textbf{k-DG.}} The time complexity of k-DR to construct an exact KNNG through linear scanning is $O(|S|^2)$, and on this basis, the time complexity of deleting neighbors reachable through alternative paths is $O(k\cdot |S|)$, where $k$ is the number of neighbors of each vertex on KNNG, and $k \ll |S|$. Therefore, the construction complexity of k-DR is $O(|S|^2 +k \cdot |S|)$. As shown in \autoref{fig: complexity}(l), the search complexity of k-DR is about $O(\log^{5.51}(|S|))$, which is lower than NGT (the constant factor of k-DG's complexity is an order of magnitude lower than NGT). In the experimental evaluation, we can also see that the search performance of k-DG is better than NGT on all real world datasets.

\setlength{\textfloatsep}{0cm}
\setlength{\floatsep}{0cm}
\begin{table*}
\setlength{\abovecaptionskip}{0.05cm}
\setstretch{0.9}
\fontsize{8pt}{4mm}\selectfont
    \centering
    \tabcaption{Characteristics of components within graph-based ANNS algorithms}
    \label{tab: Characteristics_algorithms}
    \begin{tabular}{p{35pt}|l|l|l|l|p{43pt}|p{47pt}|l|l}
    \hline
    \textbf{Algorithm} & \textbf{Construction} & \textbf{Initialization} & \textbf{Candidate} & \textbf{Neighbor Selection} & \textbf{Connectivity} & \textbf{Preprocessing} & \textbf{Seed} & \textbf{Routing} \\
    \hline
    \hline
    KGraph & refinement & random & expansion & distance & false & false & random & BFS \\
    \hline
    NGT & increment & VP-tree & search & distance \& distribution & false & true & VP-tree & RS \\
    \hline
    SPTAG1 & divide-and-conquer & TP-tree & subspace & distance \& distribution & false & true & KD-tree & BFS \\
    \hline
    SPTAG2 & divide-and-conquer & TP-tree & subspace & distance \& distribution & false & true & k-means tree & BFS \\
    \hline
    NSW & increment & random & search & distance & true & false & random & BFS \\
    \hline
    IEH & refinement & brute force & neighbors & distance & false & true & hashing & BFS \\
    \hline
    FANNG & refinement & brute force & neighbors & distance \& distribution & false & false & random & BFS \\
    \hline
    HNSW & increment & top layer & search & distance \& distribution & false & false & top layer & BFS \\
    \hline
    EFANNA & refinement & KD-tree & expansion & distance & false & true & KD-tree & BFS \\
    \hline
    DPG & refinement & NN-Descent & neighbors & distance \& distribution & false & false & random & BFS \\
    \hline
    NSG & refinement & NN-Descent & search & distance \& distribution & true & true & centroid & BFS \\
    \hline
    HCNNG & divide-and-conquer & clustering & subspace & distance & false & true & KD-tree & GS \\
    \hline
    Vamana & refinement & random & search & distance \& distribution & false & true & centroid & BFS \\
    \hline
    NSSG & refinement & NN-Descent & expansion & distance \& distribution & true & true & random & BFS \\
    \hline
    k-DR & refinement & brute force & neighbors & distance \& distribution & false & false & random & BFS or RS \\
    \hline
    \end{tabular}

\end{table*}

\subsection*{Appendix E. Characteristics of compared algorithms}
\label{Appendix_E}
We summarize some salient characteristics of compared algorithms in \autoref{tab: Characteristics_algorithms}. In the first line, \textbf{Construction} is the construction strategies of the algorithms, \textbf{Candidate} is the \textit{candidate neighbor acquisition}, \textbf{Preprocessing} is the \textit{seed preprocessing}, and \textbf{Seed} is the \textit{seed acquisition}. In the fourth column, ``search'' indicates that the algorithm obtains candidate neighbor by ANNS on graph, ``expansion'' indicates that the algorithm obtains candidate neighbor by neighbor propagation. For the fifth column, ``distance'' represents that the algorithm considers distance factor when neighbor selection to get as close as possible neighbors, ``distribution'' means that the algorithm considers distribution factor when neighbor selection so that the neighbors are evenly distributed. In the sixth and seventh columns, ``true'' indicates that the algorithm ensures corresponding process, ``false'' indicates that the algorithm does not ensure this process. As for the ninth column, BFS, GS, RS are the abbreviations of Best First Search, Guided Search, and Range Search respectively.

\subsection*{Appendix F. Algorithms description of best first search}
\label{Appendix_F}
We describe the execution process of BFS in \hyperref[Algorithm: BFS]{Algorithm 1}.

\begin{algorithm}[t]
  \label{Algorithm: BFS}
  \caption{BFS($G$, $q$, $c$, $\widehat{S}$)}
  \LinesNumbered
  \KwIn{graph $G$, query $q$, candidate set size $c$, seed set $\widehat{S}$}
  \KwOut{result set $\mathcal{R}$}
  
  candidate set $\mathcal{C} \gets \widehat{S}$, result set $\mathcal{R} \gets \widehat{S}$
  
  \While{$\mathcal{R}$ is updated}{
    $\hat{x} \gets \arg\, \min _{x\in \mathcal{C}}{\delta (x,q)}$
    
    $\mathcal{C} \setminus \left \{\hat{x} \right \}$
    
    $N(\hat{x}) \gets$ the neighbors of $\hat{x}$
    
    $\mathcal{ C} \cup N(\hat{x})$
    
    
    \While{$\vert \mathcal{C} \vert > c$}{
      $\hat{y}=\arg\, \max _{y\in \mathcal{C}}{\delta (y,q)}$

      $\mathcal{C} \setminus \left \{\hat{y} \right \}$
    }
    
    \ForAll{$n \in N(\hat{x})$}{	
            
      $\hat{z} \gets \arg \, \max _{z \in \mathcal{R}} {\delta (z,q)}$

      \If{$\delta (n,q) <\delta(\hat{z},q)$}{
        $\mathcal{R} \setminus \left \{\hat{z} \right \}$

        $\mathcal {R} \cup \left \{n \right \}$
      }
      
    }

  }
  return $\mathcal{R}$
\end{algorithm}

\subsection*{Appendix G. Characteristics of the synthetic datasets}
\label{Appendix_G}

\setlength{\textfloatsep}{0cm}
\setlength{\floatsep}{0cm}
\begin{table}[!tb]
  \centering
  \setlength{\abovecaptionskip}{0.05cm}
  \setstretch{0.9}
  \fontsize{8pt}{4mm}\selectfont
  \caption{Statistics of the synthetic datasets.}
  \label{tab: Synthetic Dataset}
  \setlength{\tabcolsep}{.0182\linewidth}{
  \begin{tabular}{l|l|l|l|l|l}
    \hline
    \textbf{\textbf{Dataset}} & \textbf{Dimension} & \textbf{cardinality} & \textbf{\# Cluster} & \textbf{SD} & \textbf{\# Query} \\
    \hline
    \hline
    d\_8 & 8 & 100,000 & 10 & 5 & 1,000\\
    \hline
    d\_32 & 32 & 100,000 & 10 & 5 & 1,000\\
    \hline
    d\_128 & 128 & 100,000 & 10 & 5 & 1,000\\
    \hline
    n\_10000 & 32 & 10,000 & 10 & 5 & 100\\
    \hline
    n\_100000 & 32 & 100,000 & 10 & 5 & 1,000\\
    \hline
    n\_1000000 & 32 & 1,000,000 & 10 & 5 & 10,000\\
    \hline
    c\_1 & 32 & 100,000 & 1 & 5 & 1,000\\
    \hline
    c\_10 & 32 & 100,000 & 10 & 5 & 1,000\\
    \hline
    c\_100 & 32 & 100,000 & 100 & 5 & 1,000\\
    \hline
    s\_1 & 32 & 100,000 & 10 & 1 & 1,000\\
    \hline
    s\_5 & 32 & 100,000 & 10 & 5 & 1,000\\
    \hline
    s\_10 & 32 & 100,000 & 10 & 10 & 1,000\\
    \hline
  \end{tabular}
  }
\end{table}

We summarize the characteristics of the nine synthetic datasets in this paper in \autoref{tab: Synthetic Dataset}, including dimension, cardinality, number of clusters (\textbf{\# Cluster}), and standard deviation of the distribution in each cluster (\textbf{SD}).

\subsection*{Appendix H. Parameters of the compared algorithms}
\label{Appendix_H}
The optimal parameters of all algorithms on our experimental datasets are available from our online github repository\textsuperscript{\ref{github_code}}.

\noindent\underline{\textbf{KGraph.}} When constructing the index, we search for the optimal values of KGraph's five sensitive parameters (\texttt{K}, \texttt{L}, \texttt{iter}, \texttt{S}, \texttt{R}), and other parameters adopt the default values recommended by the author~\cite{KGraph}. Increasing any of \texttt{K}, \texttt{L}, \texttt{S} and \texttt{R} has the effect of improving accuracy and slowing down speed at the same time. \texttt{iter} is the number of iterations of the NN-Descent, the larger its value, the higher the graph quality. For a more detailed analysis of the impact of these parameters on KGraph index construction and search performance, please see~\cite{KGraph}.

\setlength{\textfloatsep}{0cm}
\setlength{\floatsep}{0cm}
\begin{table*}[th!]
\setlength{\abovecaptionskip}{0.05cm}
\setstretch{0.9}
\fontsize{8pt}{4mm}\selectfont
    \centering
    \tabcaption{Maximum out-degree (D\_max) and minimum out-degree (D\_min) of the graph indexes of all compared algorithms}
    \label{tab: max_min out-degree}
    \setlength{\tabcolsep}{0.0042\linewidth}{
    \begin{tabular}{l|l|l|l|l|l|l|l|l|l|l|l|l|l|l|l|l}
    \hline
    \multirow{2}*{\textbf{Alg.}} & \multicolumn{2}{c|}{\textbf{UQ-V}} & \multicolumn{2}{c|}{\textbf{Msong}} & \multicolumn{2}{c|}{\textbf{Audio}} & \multicolumn{2}{c|}{\textbf{SIFT1M}} & \multicolumn{2}{c|}{\textbf{GIST1M}} & \multicolumn{2}{c|}{\textbf{Crawl}} & \multicolumn{2}{c|}{\textbf{GloVe}} & \multicolumn{2}{c}{\textbf{Enron}}\\
    \cline{2-17}
    ~ & D\_max & D\_min & D\_max & D\_min & D\_max & D\_min & D\_max & D\_min & D\_max & D\_min & D\_max & D\_min & D\_max & D\_min & D\_max & D\_min \\
    \hline
    \hline
    {KGraph} & 40 & 40 & 100 & 100 & 40 & 40 & 90 & 90 & 100 & 100 & 80 & 80 & 100 & 100 & 50 & 50 \\
    \hline
    {NGT-panng} & 2,379 & 5 & 1,108 & 6 & 320 & 14 & 738 & 10 & 5,181 & 4 & 58,677 & 4 & 29,999 & 11 & 869 & 5 \\
    \hline
    {NGT-onng} & 2,663 & 3 & 1,935 & 10 & 500 & 7 & 849 & 5 & 6,798 & 6 & 120,928 & 4 & 60,115 & 26 & 1,242 & 6 \\
    \hline
    {SPTAG-KDT} & 32 & 32 & 32 & 32 & 32 & 32 & 32 & 32 & \color{blue}\textbf{32} & 32 & 32 & 32 & 32 & 32 & 32 & 32 \\
    \hline
    {SPTAG-BKT} & 32 & 32 & 32 & 32 & 32 & 32 & 32 & 32 & \color{blue}\textbf{32} & 32 & 32 & 32 & 32 & 32 & 32 & 32 \\
    \hline
    {NSW} & 880 & 30 & 5,334 & 60 & 1,130 & 40 & 1,901 & 40 & 16,693 & 60 & 245,301 & 60 & 195,123 & 80 & 5,013 & 80 \\
    \hline
    {IEH} & 50 & 50 & 50 & 50 & 50 & 50 & 50 & 50 & 50 & 50 & 50 & 50 & 50 & 50 & 50 & 50 \\
    \hline
    {FANNG} & 90 & 90 & \color{blue}\textbf{10} & 10 & 50 & 50 & 70 & 70 & 50 & 50 & \color{blue}\textbf{30} & 30 & 70 & 70 & 110 & 110 \\
    \hline
    {HNSW} & 40 & \color{blue}\textbf{1} & 80 & \color{blue}\textbf{1} & 50 & \color{blue}\textbf{1} & 50 & \color{blue}\textbf{1} & 60 & \color{blue}\textbf{1} & 70 & \color{blue}\textbf{1} & 60 & \color{blue}\textbf{1} & 80 & \color{blue}\textbf{1} \\
    \hline
    {EFANNA} & 40 & 40 & 50 & 50 & \color{blue}\textbf{10} & 10 & 60 & 60 & 100 & 100 & 100 & 100 & 100 & 100 & 40 & 40 \\
    \hline
    {DPG} & 460 & 50 & 821 & 50 & 359 & 50 & 389 & 50 & 8,981 & 50 & 120,942 & 50 & 63,073 & 50 & 1,189 & 50 \\
    \hline
    {NSG} & 32 & \color{blue}\textbf{1} & 21 & \color{blue}\textbf{1} & 30 & \color{blue}\textbf{1} & 30 & \color{blue}\textbf{1} & 47 & \color{blue}\textbf{1} & 1,068 & \color{blue}\textbf{1} & 125 & \color{blue}\textbf{1} & 62 & \color{blue}\textbf{1} \\
    \hline
    {HCNNG} & 187 & 3 & 209 & 10 & 92 & 10 & 150 & 11 & 85 & 2 & 208 & 9 & 294 & 24 & 190 & 8 \\
    \hline
    {Vamana} & 30 & 30 & 30 & 30 & 50 & 50 & 50 & 50 & 50 & 50 & 50 & 50 & 110 & 110 & 110 & 110 \\
    \hline
    {NSSG} & \color{blue}\textbf{20} & \color{blue}\textbf{1} & 70 & \color{blue}\textbf{1} & 20 & \color{blue}\textbf{1} & \color{blue}\textbf{20} & 5 & 41 & \color{blue}\textbf{1} & 61 & \color{blue}\textbf{1} & \color{blue}\textbf{31} & \color{blue}\textbf{1} & \color{blue}\textbf{31} & \color{blue}\textbf{1} \\
    \hline
    {k-DR} & 128 & \color{blue}\textbf{1} & 250 & \color{blue}\textbf{1} & 91 & 2 & 202 & \color{blue}\textbf{1} & 2,685 & \color{blue}\textbf{1} & 27,107 & \color{blue}\textbf{1} & 7,580 & \color{blue}\textbf{1} & 399 & \color{blue}\textbf{1} \\
    \hline
    \end{tabular}
    }

\end{table*}

\noindent\underline{\textbf{NGT.}} It uses a method similar to NSW~\cite{NSW} to incrementally construct an approximate nearest neighbor graph (ANNG). The only difference from NSW is that the search algorithm used when NGT obtains candidate neighbors is range search. The parameter \texttt{$\epsilon$} defaults to 1.1. In ANNG, the upper bound of the number of connected bidirectional edges is \texttt{K} for each vertex. Since ANNG is an undirected graph, the number of edges connected to each vertex may actually exceed \texttt{K}. On the basis of ANNG, NGT-panng~\cite{NGT} performs path adjustment (an approximation to RNG, see \hyperref[Appendix_B]{Appendix B} for details) to cut redundant edges to ensure that the number of each vertex's neighbors is lower than the given parameter \texttt{R}. NGT-onng~\cite{NGT} first performs in-degree and out-degree adjustment on the basis of ANNG. The parameters involved are \texttt{out\_edge} and \texttt{in\_edge}, which respectively represent the number of outgoing and incoming edges (directed edges) of each vertex extracted from ANNG. Then, NGT-onng performs path adjustment like NGT-panng. For more details on these parameters, see \cite{yahoo2, yahoo3, NGT}.

\noindent\underline{\textbf{SPTAG.}} There are mainly two implementation versions of SPTAG, one is the original version~\cite{wang2012scalable, wang2013trinary,wang2012query} (SPTAG-KDT) and the other is an improved version~\cite{SPTAG} (SPTAG-BKT). The graph index of SPTAG-KDT is a KNNG, and it uses KD-Tree to get the entry point when searching. SPTAG-BKT's graph index adds the optimization of RNG on the basis of KNNG, and it uses k-means tree to get the entry point when searching. Detailed descriptions of relevant parameters can be found on the online web page\footnote{https://github.com/microsoft/SPTAG/blob/master/docs/Parameters.md}.

\noindent\underline{\textbf{NSW.}} \texttt{ef\_construction} controls the size of the candidate set, and it adjusts construction speed/index quality tradeoff. \texttt{max\_m0} is the maximum size of bidirectional edge of each vertex, and it controls the index size of the NSW.

\noindent\underline{\textbf{IEH.}} It contains three important parameters, i.e., \texttt{p}, \texttt{k}, \texttt{s}. \texttt{p} is the number of top nearest candidates, which are used for expansion in each iteration. \texttt{k} is the number of expansion. \texttt{s} is iteration number. According to the experiment evaluation in \cite{IEH}, \texttt{p} = 10, \text{k} = 50, \texttt{s} = 3 are reasonable for considering both high recall and low search time. However, our tests show that using the above recommended parameter values on most datasets does not receive the desired results. In order to get the specified recall rate, $p$ must be increased.

\noindent\underline{\textbf{FANNG.}} \texttt{L} controls the size of candidate neighbors, and \texttt{R} is the maximum number of neighbors.

\noindent\underline{\textbf{HNSW.}} \texttt{M0}, \texttt{M} and \texttt{ef\_construction} are used in the HNSW construction. \texttt{M0} is the maximum number of each vertex's neighbors in the bottom layer. \texttt{M} controls the maximum number of each vertex's neighbors in the high layer.  \texttt{ef\_construction} is the size of the candidate set when selecting neighbors.

\noindent\underline{\textbf{EFANNA.}} Different from the random initialization of KGraph, EFANNA initializes the neighbors of each vertex through KD-Tree. Therefore, two additional parameters are needed for EFANNA. \texttt{nTrees} controls the number of KD-Trees, and \texttt{mLevel} controls the maximum merged layers (see \cite{EFANNA} for details).

\noindent\underline{\textbf{DPG.}} DPG is acquired by diversifying KGraph's neighbors, thus, there are also the parameters \texttt{K}, \texttt{L}, \texttt{iter}, \texttt{S}, \texttt{R} of KGraph. In addition, the upper bound on the number of neighbors at each point is fixed at \texttt{K/2} during diversification. However, after adding the reverse edge operation, the number of neighbors at some points may surge back ($\gg$ \texttt{K/2}).

\noindent\underline{\textbf{NSG.}} Similar to DPG, NSG also reselects neighbors based on KGraph by appending an approximation to RNG. NSG has 3 additional parameters for the neighbor selection strategy, i.e., \texttt{L}, \texttt{R}, \texttt{C}. \texttt{L} is the size of the candidate set when acquiring candidate neighbors for each vertex with the greedy search. The larger the \texttt{L}, the closer the candidate neighbors to the target vertex, but the slower the acquisition operation. \texttt{C} controls the maximum size of the candidate neighbor set, \texttt{R} controls the index size of the graph, the best \texttt{R} is related to the intrinsic dimension of the dataset.

\noindent\underline{\textbf{HCNNG.}} Two parameters are used in the HCNNG construction: the number executions of hierarchical clustering procedures \texttt{m}, and the minimum size of clusters \texttt{n}. In addition, \texttt{nTrees} controls the number of KD-Tree for seed acquisition.

\noindent\underline{\textbf{Vamana.}} It first randomly initializes a graph index $G_{init}$, and then uses a heuristic edge selection strategy similar to HNSW~\cite{HNSW} to perform the neighbor update on $G_{init}$ to obtain the final graph index $G_{final}$, which is made two passes. We set the upper bound of the neighbors of $G_{init}$ and $G_{final}$ as the parameter \texttt{R}. During the neighbor update, the size of the candidate neighbor is set to \texttt{L}. According to the recommendation of the original paper~\cite{DiskANN}, in the first pass of the neighbor update, \texttt{$\alpha$} is set to 1, while in the second pass of the neighbor update, \texttt{$\alpha$} is set to 2.

\noindent\underline{\textbf{NSSG.}} NSSG is an optimization to NSG, it has additional parameters \texttt{L}, \texttt{R}, and \texttt{Angle} on the basis of KGraph. \texttt{L} controls the quality of the NSG, the larger the better, \texttt{L} > \texttt{R}. \texttt{R} controls the index size of the graph, the best \texttt{R} is related to the intrinsic dimension of the dataset. \texttt{Angle} controls the angle between two edges, generally, its optimal value is $60^{\circ}$~\cite{NSSG}.

\noindent\underline{\textbf{k-DR.}} There are two parameters of k-DR in the construction process, namely \texttt{k} and \texttt{R} (\texttt{R} $\le$ \texttt{k}). \texttt{k} is the neighbor upper bound of the initial KNNG, and it is also the number of candidate neighbors for subsequent trimming. \texttt{R} is the upper limit of neighbors reserved when performing edge pruning; the actual number of neighbors may exceed \texttt{R} due to the addition of reverse edges. The larger the value of the above two parameters, the higher the recall rate of the search results; but that will reduce the search efficiency, vice versa.

\setlength{\textfloatsep}{0cm}
\setlength{\floatsep}{0cm}
\begin{table*}[th!]
\setlength{\abovecaptionskip}{0.05cm}
\setstretch{0.9}
\fontsize{6.5pt}{4mm}\selectfont
    \centering
    \tabcaption{The construction time (CT) and queries per second (QPS) on synthetic datasets with different characteristics.}
    \label{tab: Scalability}
    \setlength{\tabcolsep}{0.00665\linewidth}{
    \begin{tabular}{l|l|l|l|l|l|l|l|l|l|l|l|l|l|l|l|l|l|l|l|l|l|l|l|l}
    \hline
    \multirow{2}*{\textbf{Alg.}} & \multicolumn{6}{c|}{\textbf{\#Dimensionality}} & \multicolumn{6}{c|}{\textbf{\#Cardinality}} & \multicolumn{6}{c|}{\textbf{\#Clusters}} & \multicolumn{6}{c}{\textbf{Standard deviation}} \\
    \cline{2-25}
    ~ &  \multicolumn{2}{c|}{{8}} & \multicolumn{2}{c|}{{32}} &\multicolumn{2}{c|}{{128}} & \multicolumn{2}{c|}{{$10^{4}$}} &\multicolumn{2}{c|}{{{$10^{5}$}}} & \multicolumn{2}{c|}{{$10^{6}$}} &  \multicolumn{2}{c|}{{1}} &  \multicolumn{2}{c|}{{10}} &  \multicolumn{2}{c|}{{100}} &  \multicolumn{2}{c|}{{1}} &  \multicolumn{2}{c|}{{5}} &  \multicolumn{2}{c}{{10}} \\
    \cline{2-25}
    ~ & CT & QPS & CT & QPS & CT & QPS & CT & QPS & CT & QPS & CT & QPS & CT & QPS & CT & QPS & CT & QPS & CT & QPS & CT & QPS & CT & QPS \\
    \hline
    \hline
    {KGraph} & \color{blue}\textbf{6} & 8,192 & \color{blue}\textbf{32} & 1,637 & 43 & \color{blue}\textbf{561} & 6 & 5,882 & \color{blue}\textbf{32} & 1,637 & 499 & 366 & 31 & 2,227 & \color{blue}\textbf{32} & 1,637 & 53 & 3,317 & \color{blue}\textbf{6} & 1,492 & \color{blue}\textbf{32} & 1,637 & \color{blue}\textbf{28} & 420 \\
    \hline
    {NGT-panng} & 377 & 1,643 & 145 & 343 & 134 & 119 & 9 & 1,677 & 145 & 343 & 1,422 & 115 & 88 & 148 & 145 & 343 & 251 & 471 & 155 & 464 & 145 & 343 & 155 & 196 \\
    \hline
    {NGT-onng} & 191 & 1,322 & 161 & 174 & 126 & 59 & 16 & 918 & 161 & 174 & 2439 & 42 & 123 & 46 & 161 & 174 & 185 & 397 & 176 & 184 & 161 & 174 & 107 & 77 \\
    \hline
    {SPTAG-KDT} & 138 & 70 & 243 & 39 & 178 & 3 & 43 & 48 & 243 & 39 & 1,327 & 33 & 108 & 4 & 243 & 39 & 267 & 67 & 228 & 34 & 243 & 39 & 1,110 & 10 \\
    \hline
    {SPTAG-BKT} & 244 & 51 & 284 & 46 & 342 & 2 & 17 & 65 & 284 & 46 & 3,107 & 31 & 173 & 1 & 284 & 46 & 137 & 4 & 256 & 64 & 284 & 46 & 637 & 4 \\
    \hline
    {NSW} & 77 & 4,250 & 45 & 1,150 & 59 & 335 & \color{blue}\textbf{1} & 4,624 & 45 & 1,150 & 2,962 & 243 & 79 & 763 & 45 & 1,150 & 84 & 2,405 & 38 & 1,301 & 45 & 1,150 & 51 & 508 \\
    \hline
    {IEH} & 70 & 623 & 87 & 53 & 177 & 511 & \color{blue}\textbf{1} & 588 & 87 & 53 & 8,393 & 14 & 88 & 512 & 87 & 53 & 89 & 2 & 88 & 91 & 87 & 53 & 90 & 1 \\
    \hline
    {FANNG} & 70 & 1,106 & 87 & 366 & 177 & 157 & \color{blue}\textbf{1} & 933 & 87 & 366 & 8,393 & 57 & 88 & 109 & 87 & 366 & 89 & 260 & 88 & 278 & 87 & 366 & 90 & 158 \\
    \hline
    {HNSW} & 34 & 8,577 & 461 & 676 & 2,535 & 528 & 4 & 5,073 & 461 & 676 & 52,782 & 252 & 1,931 & 2,877 & 461 & 676 & 227 & 4,009 & 458 & 7 & 461 & 676 & 1,075 & 299 \\
    \hline
    {EFANNA} & 28 & 13,464 & 33 & 1,639 & \color{blue}\textbf{27} & 84 & 9 & \color{blue}\textbf{10,871} & 33 & 1,639 & \color{blue}\textbf{313} & 386 & 34 & \color{blue}\textbf{6,131} & 33 & 1,639 & 118 & \color{blue}\textbf{5,141} & 104 & 1,466 & 33 & 1,639 & 53 & 426 \\
    \hline
    {DPG} & 25 & 11,131 & 71 & 1,531 & 63 & 374 & 4 & 4,275 & 71 & 1,531 & 405 & 605 & 27 & 4,252 & 71 & 1,531 & 72 & 3,789 & 13 & 1,482 & 71 & 1,531 & 33 & \color{blue}\textbf{1,259} \\
    \hline
    {NSG} & 12 & \color{blue}\textbf{14,081} & 39 & \color{blue}\textbf{2,580} & 37 & 401 & 5 & 7,064 & 39 & \color{blue}\textbf{2,580} & 818 & 657 & 77 & 2,709 & 39 & \color{blue}\textbf{2,580} & 100 & 4,315 & 36 & 1,451 & 39 & \color{blue}\textbf{2,580} & 100 & 1,077 \\
    \hline
    {HCNNG} & 43 & 11,848 & 48 & 1,996 & 111 & 504 & 3 & 7,544 & 48 & 1,996 & 3,073 & \color{blue}\textbf{658} & 66 & 1,074 & 48 & 1,996 & 52 & 4,035 & 38 & \color{blue}\textbf{2,025} & 48 & 1,996 & 53 & 868 \\
    \hline
    {Vamana} & 154 & 8,296 & 92 & 1,089 & 237 & 192 & 5 & 4,370 & 92 & 1,089 & 1,085 & 120 & 295 & 297 & 92 & 1,089 & \color{blue}\textbf{40} & 2,856 & 91 & 1,029 & 92 & 1,089 & 172 & 490 \\
    \hline
    {NSSG} & 45 & 10,524 & 84 & 918 & 105 & 23 & 7 & 3,985 & 84 & 918 & 1,526 & 369 & \color{blue}\textbf{17} & 1,833 & 84 & 918 & 51 & 4,361 & 10 & 20 & 84 & 918 & 83 & 1,177 \\
    \hline
    {k-DR} & 75 & 12,255 & 107 & 1,376 & 235 & 43 & 2 & 6,785 & 107 & 1,376 & 8,931 & 466 & 114 & 2,790 & 107 & 1,376 & 102 & 3,486 & 124 & 1,352 & 107 & 1,376 & 126 & 505 \\
    \hline
    \end{tabular}
    }
\end{table*}

\subsection*{Appendix I. Maximum and minimum out-degrees of the graph indexes of the compared algorithms}
\label{Appendix_I}

\autoref{tab: max_min out-degree} lists the maximum and minimum out-degrees of the graph index of each algorithm on the real-world dataset. In the search process, the algorithms can align the neighbor adjacent list to the same size (maximum out-degree), which will improve search efficiency by continuous memory access~\cite{NSSG}. However, some algorithms whose maximum out-degree is too large easily exceed the memory limit and cannot take advantage of the above memory optimization (e.g., NSW, DPG, k-DG).

\subsection*{Appendix J. Scalability of the graph-based ANNS algorithms}
\label{Appendix_J}

\noindent\underline{\textbf{Dimensionality.}} \autoref{tab: Scalability} reports that as the dimensionality increases, the CT of most algorithms increases. Interestingly, NSW and NGT are exactly the opposite of the above phenomenon. Note that they are both DG-based algorithms. Without exception, the QPS of all algorithms decreases as the dimensionality increases. Although the QPS of RNG-based algorithm (e.g., NSG) beats other categories of algorithms (e.g., KGraph, NSW) by a big margin in lower dimensionality, the QPS of KNNG- and MST-based algorithms (e.g., KGraph, HCNNG) surpasses some RNG-based algorithms (e.g., NSG) when the dimensionality is very high.

\noindent\underline{\textbf{Cardinality.}} As the cardinality increases, the CT of all algorithms increases, and the QPS decreases. When the cardinality is small, the QPS of KNNG- and DG-based algorithms (e.g., EFANNA, NSW) have a small gap with RNG- and MST-based algorithms (e.g., NSG, HCNNG). However, as the cardinality increases, the advantages of RNG- and MST-based algorithms (e.g., DPG, NSG, HCNNG) are gradually revealed (QPS is at least twice that of other categories).

\noindent\underline{\textbf{Clusters.}} When the number of clusters increases from 10 to 100, the CT of KNNG- and DG-based algorithms (e.g., KGraph, NSW) increases significantly, while the CT of RNG-based algorithm (e.g., NSSG, Vamana) decreases. It is worth noting that the CT of MST-based (e.g., HCNNG) and some KNNG-based (e.g., FANNG, k-DR) algorithms constructed by brute force is basically not affected by the number of clusters. RNG-based algorithms generally have better search performance on the datasets with more clusters, which is mainly due to the approximation to RNG, so that they can be better connectted on the clustered data.

\noindent\underline{\textbf{Standard deviation.}} The standard deviation of the distribution in each cluster reflects the difficulty of dataset~\cite{graph_survey}. As the standard deviation increases, the difficulty of the dataset increases, the CT of most algorithms increases, and the QPS decreases. Some exceptions occurred in some KNNG- and RNG-based algorithms (e.g., KGraph, NSG), and their QPS increase with the increase of the standard deviation at the beginning, and then drop. In particular, the QPS of NSSG increases as the standard deviation increases.

\noindent\underline{\textbf{Discussion.}} In general, as the difficulty of dataset increases (i.e., larger dimensionality, cardinality, cluster, and standard deviation), the index construction and search efficiency of each algorithm decline to varying degrees. Even so, RNG-based algorithms show the best scalability when performing searches on datasets with different characteristics. While for index construction on these datasets, the algorithms based on NN-Descent have the highest efficiency. On high-dimensional datasets, KNNG-, RNG- and MST-based algorithms have similar search performance. As the scale, cluster, and standard deviation increase, the advantages of RNG- and MST-based algorithms over KNNG- and DG-based algorithms become more and more obvious. In addition, algorithms that include additional tree structures are generally difficult to obtain high search performance.

\subsection*{Appendix K. Components settings for the benchmark algorithm used for components evaluation}
\label{Appendix_K}

\setlength{\textfloatsep}{0cm}
\setlength{\floatsep}{0cm}
\begin{table}[!tb]
  \centering
  \setlength{\abovecaptionskip}{0.05cm}
  \setstretch{0.9}
  \fontsize{6.5pt}{4mm}\selectfont
  \caption{The component settings of the benchmark algorithm.}
  \label{tab: component_settings}
  \setlength{\tabcolsep}{.017\linewidth}{
  \begin{tabular}{l|l|l|l|l|l|l}
    \hline
    \textbf{\textbf{C1}} & \textbf{C2} & \textbf{C3} & \textbf{C4} & \textbf{C5} & \textbf{C6} &\textbf{C7} \\
    \hline
    \hline
    \textit{C1\_NSG} & \textit{C2\_NSSG} & \textit{C3\_HNSW} & \textit{C4\_NSSG} & \textit{C5\_IEH} & \textit{C6\_NSSG} & \textit{C7\_NSW}\\
    \hline
  \end{tabular}
  }
\end{table}

\autoref{tab: component_settings} reports that the settings of each component of the benchmark algorithm. When evaluating a component, we make sure that the other components maintain the settings in \autoref{tab: component_settings}.

\subsection*{Appendix L. Evaluation of search performance under different NN-Descent iteration times}
\label{Appendix_L}

\begin{figure}
  \centering
  \subfigcapskip=-0.25cm
  \subfigure[Recall@10 (SIFT1M)]{  
    \captionsetup{skip=0pt}
    \vspace{-1mm}
    \includegraphics[scale=0.285]{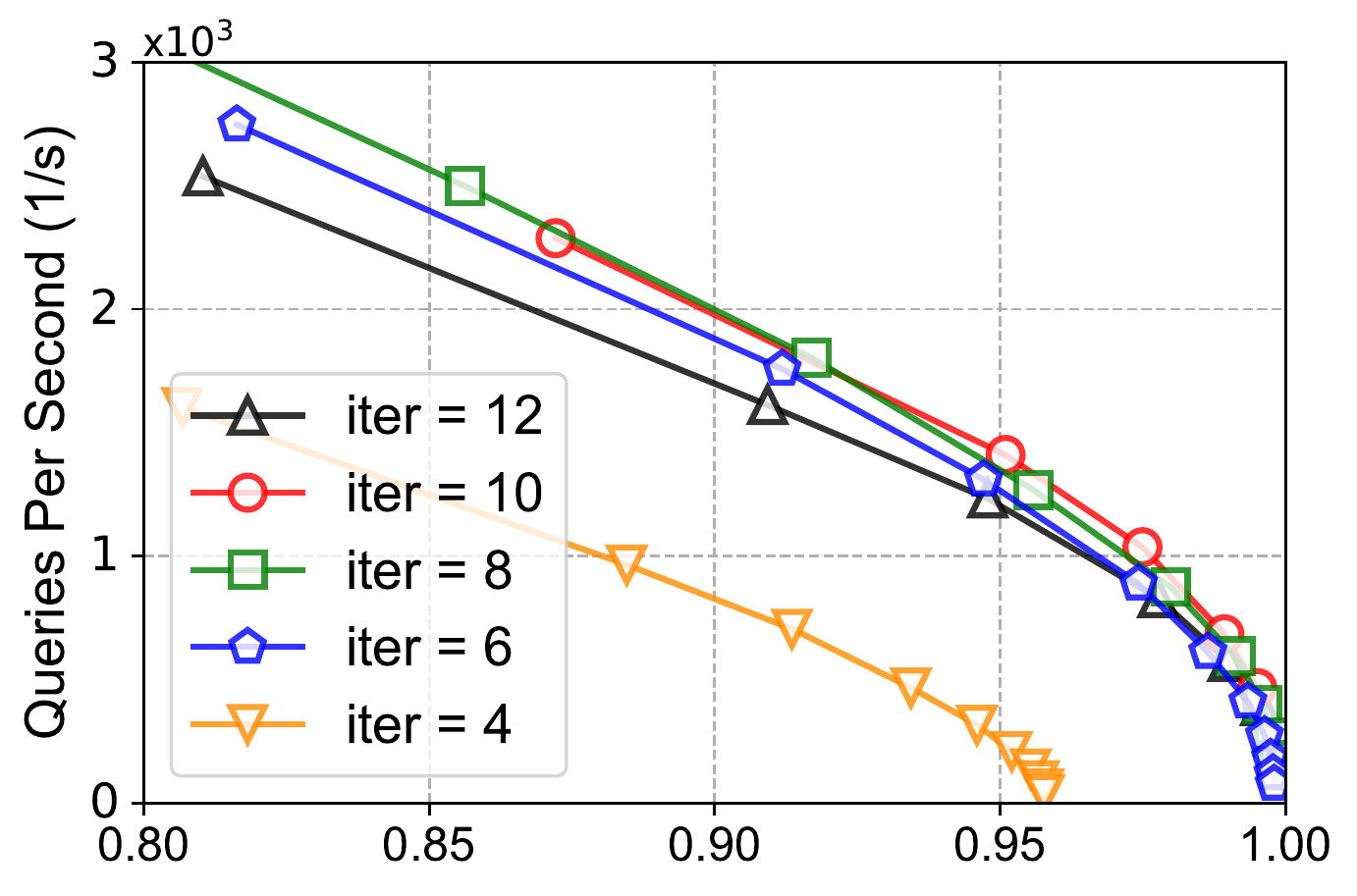}
    \label{fig: nndescent_iter_sift}
  }
  \subfigure[Recall@10 (GIST1M)]{ 
    \captionsetup{skip=0pt}
    \vspace{-1mm}
    \includegraphics[scale=0.285]{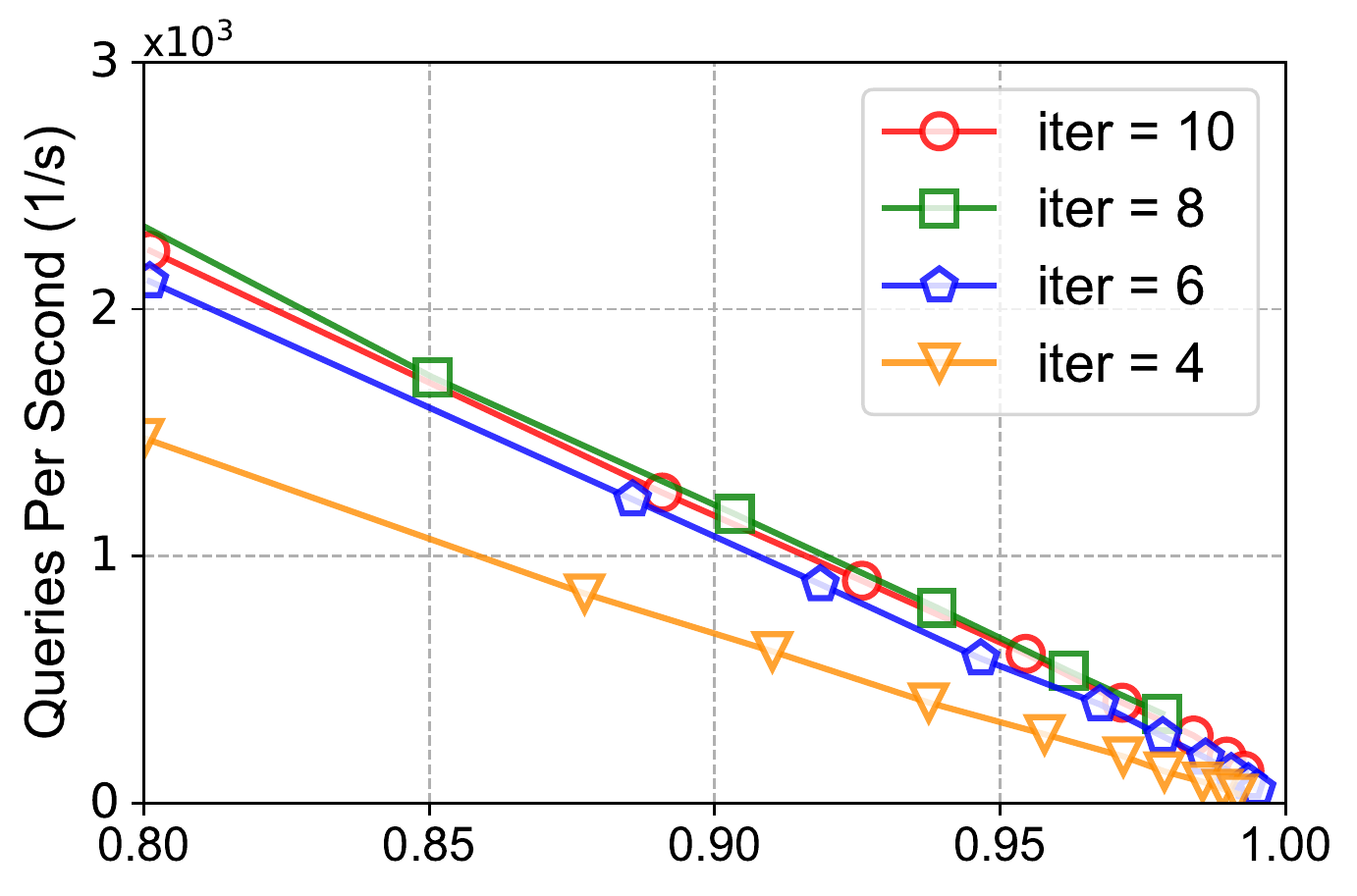}
    \label{fig: nndescent_iter_gist}
  }\vspace{-0.3cm}
  \vspace{-0.2cm}
  \caption{Search performance of the benchmark algorithm under different number of iterations.}\vspace{-0.4mm}
  \label{fig: nndescent_iter_search}
\end{figure}

\autoref{tab: Index Information} confirms that the highest graph quality does not necessarily achieve the best search performance. Therefore, we set different NN-Descent iteration times to obtain different initial graph quality, so as to get an optimal iteration value of the benchmark algorithm for component evaluation during initialization. As shown in \autoref{fig: nndescent_iter_search}, the search performance of the benchmark algorithm with different number of iterations shows the same trend on two real-world datasets. As the number of iterations increases, the search performance first increases and then decreases. This is consistent with the conclusion drawn in \autoref{tab: Index Information}. As far as search performance is concerned, the graph quality is not the higher the better. In addition, according to \autoref{tab: nndescent_iter_build_time}, the greater the number of iterations, the higher the index construction time. In summary, we set the number of iterations of NN-Descent to 8 for the benchmark algorithm during initialization.

\setlength{\textfloatsep}{0cm}
\setlength{\floatsep}{0cm}
\begin{table}[!tb]
  \centering
  \setlength{\abovecaptionskip}{0.05cm}
  \setstretch{0.9}
  \fontsize{8pt}{4mm}\selectfont
  \caption{Index construction time (s) of the benchmark algorithm under different number of iterations.}
  \label{tab: nndescent_iter_build_time}
  \setlength{\tabcolsep}{.0182\linewidth}{
  \begin{tabular}{l|l|l|l|l}
    \hline
    \textbf{\textbf{Dataset}} & \textbf{iter=4} & \textbf{iter=6} & \textbf{iter=8} & \textbf{iter=10} \\
    \hline
    \hline
    SIFT1M & \color{blue}\textbf{164} & 269 & 408 & 532 \\
    \hline
    GIST1M & \color{blue}\textbf{484} & 776 & 1,192 & 1,721 \\
    \hline
  \end{tabular}
  }
\end{table}

\subsection*{Appendix M. Index construction performance of components}
\label{Appendix_M}
\autoref{tab: components_buid_time} lists the index construction time of the benchmark algorithm using different components.

\setlength{\textfloatsep}{0cm}
\setlength{\floatsep}{0cm}
\begin{table}[!tb]
  \centering
  \setlength{\abovecaptionskip}{0.05cm}
  \setstretch{0.9}
  \fontsize{8pt}{4mm}\selectfont
  \caption{Index construction time (s) of algorithms with different components.}
  \label{tab: components_buid_time}
  \setlength{\tabcolsep}{.02\linewidth}{
  \begin{tabular}{l|l|l|l}
    \hline
    \multirow{2}*{\textbf{\textbf{Component}}} & \multirow{2}*{\textbf{Implementation Way}} & \multicolumn{2}{c}{\textbf{Dataset}} \\
    \cline{3-4}
    ~ & ~ & SIFT1M & GIST1M \\
    \hline
    \hline
    \multirow{3}*{C1} & \textit{C1\_NSG} & 408 & 1,192 \\
    \cline{2-4}
    ~ & \textit{C1\_KGraph} & \color{blue}\textbf{27} & \color{blue}\textbf{90} \\
    \cline{2-4}
    ~ & \textit{C1\_EFANNA} & 64 & 212 \\
    \hline
    \multirow{3}*{C2} & \textit{C2\_NSSG} & 408 & 1,192 \\
    \cline{2-4}
    ~ & \textit{C2\_DPG} & \color{blue}\textbf{392} & \color{blue}\textbf{1,179} \\
    \cline{2-4}
    ~ & \textit{C2\_NSW} & 547 & 1,719 \\
    \hline
    \multirow{5}*{C3} & \textit{C3\_HNSW} & \color{blue}\textbf{408} & \color{blue}\textbf{1,192} \\
    \cline{2-4}
    ~ & \textit{C3\_DPG} & 1,425 & 22,751 \\
    \cline{2-4}
    ~ & \textit{C3\_KGraph} & 1,065 & 22,175 \\
    \cline{2-4}
    ~ & \textit{C3\_NSSG} & 1,304 & 29,164 \\
    \cline{2-4}
    ~ & \textit{C3\_Vamana} & 1,236 & 24,932 \\
    \hline
    \multirow{6}*{C4 \& C6} & \textit{C4\_NSSG} & \color{blue}\textbf{408} & \color{blue}\textbf{1,192} \\
    \cline{2-4}
    ~ & \textit{C4\_HCNNG} & 493 & 1,465 \\
    \cline{2-4}
    ~ & \textit{C4\_IEH} & 458 & 1,299 \\
    \cline{2-4}
    ~ & \textit{C4\_NGT} & 432 & 1,305 \\
    \cline{2-4}
    ~ & \textit{C4\_NSG} & 490 & 1,472 \\
    \cline{2-4}
    ~ & \textit{C4\_SPTAG-BKT} & 559 & 1,774 \\
    \hline
    \multirow{2}*{C5} & \textit{C5\_NSG} & 408 & \color{blue}\textbf{1,192} \\
    \cline{2-4}
    ~ & \textit{C5\_Vamana} & \color{blue}\textbf{406} & 1,261 \\
    \hline
    \multirow{4}*{C7} & \textit{C7\_NSW} & 408 & 1,192 \\
    \cline{2-4}
    ~ & \textit{C7\_FANNG} & 449 & 1,354 \\
    \cline{2-4}
    ~ & \textit{C7\_HCNNG} & 428 & 1,276 \\
    \cline{2-4}
    ~ & \textit{C7\_NGT} & \color{blue}\textbf{399} & \color{blue}\textbf{1,173} \\
    \hline
  \end{tabular}
  }
\end{table}

\subsection*{Appendix N. Evaluation and analysis of k-DR algorithm}
\label{Appendix_N}

\noindent\underline{\textbf{Overview.}} \cite{aoyama2011fast} presents a fast approximate similarity search method that utilized a degree-reduced k-nearest neighbor graph (k-DR). k-DR first builds a simple KNNG $G(V,E)$, then for any vertex $x$ and its neighbor $y \in N(x)$, k-DR generates an undirected edge between $x$ and $y$ only if the BFS algorithm (\hyperref[Algorithm: BFS]{Algorithm 1}) starting from $y$ cannot find $x$ along the already existing edges on $G$; this is actually an approximation of RNG (similar to NGT, please refer to \hyperref[Appendix_B]{Appendix B} for proof). In addition, k-DR extends the similar range search to NGT~\cite{NGT} for routing.

We analyze the construction and search complexity of k-DR in \hyperref[Appendix_D]{Appendix D}, and summarize some important attributes of k-DR in \autoref{tab: attributes_kdr}. The characteristics of components within k-DR is depicted in \autoref{tab: Characteristics_algorithms}. \hyperref[Appendix_H]{Appendix H} discusses the parameters setting of k-DR. Index construction, search, and scalability performance evaluations are given in \autoref{tab: max_min out-degree}, \autoref{tab: Scalability}, \autoref{tab: index_time_size_kdr}, \autoref{tab: index_search_info_kdr}, \autoref{fig: qps-vs-recall}, and \autoref{fig: speedup-vs-recall}. In \autoref{tab: index_search_info_kdr}, GQ, AD, CC, CS, PL, and MO are the abbreviations of graph quality, average out-degree, connected components, candidate set size, query path length, and peak memory overhead respectively.

\setlength{\textfloatsep}{0cm}
\setlength{\floatsep}{0cm}
\begin{table}[!tb]
  \centering
  \setlength{\abovecaptionskip}{0.05cm}
  \setstretch{0.9}
  \fontsize{6.5pt}{4mm}\selectfont
  \caption{Index construction time (ICT, unit: s) and index size (IS, unit: MB) of k-DR and NGT on real-world datasets, and the bold values are optimal.}
  \label{tab: index_time_size_kdr}
  \setlength{\tabcolsep}{.0092\linewidth}{
  \begin{tabular}{l|l|l|l|l|l|l|l|l|l}
    \hline
    \multicolumn{2}{c|}{\textbf{Algorithm}} & \textbf{UQ-V} & \textbf{Msong} & \textbf{Audio} & \textbf{SIFT1M} & \textbf{GIST1M} & \textbf{Crawl} & \textbf{GloVe} & \textbf{Enron}\\
    \hline
    \hline
    \multirow{2}*{k-DR} & ICT &26,580 &36,415 &97 &15,814 &77,917 &131,943 &27,521 &1,416\\
    \cline{2-10}
    ~ & IS &\color{blue}\textbf{66} &\color{blue}\textbf{100} &\color{blue}\textbf{3.6} &\color{blue}\textbf{98} &\color{blue}\textbf{156} &\color{blue}\textbf{171} &\color{blue}\textbf{199} &\color{blue}\textbf{8.4}\\
    \hline
    \multirow{2}*{NGT-panng} & ICT &\color{blue}\textbf{2,094}& \color{blue}\textbf{2,224}& \color{blue}\textbf{51}& \color{blue}\textbf{1,142}& \color{blue}\textbf{5,938}& \color{blue}\textbf{64,507} & \color{blue}\textbf{15,014} & \color{blue}\textbf{236}\\
    \cline{2-10}
    ~ & IS &215& 225& 11& 229& 269& 470& 312& 225\\
    \hline
    \multirow{2}*{NGT-onng} & ICT &3,017& 4,814& 69& 1,989& 15,832& 174,996&138,348& 414\\
    \cline{2-10}
    ~ & IS &198& 224& 9.9& 214& 296& 527& 578& 21\\
    \hline
  \end{tabular}
  }
\end{table}

\setlength{\textfloatsep}{0cm}
\setlength{\floatsep}{0cm}
\begin{table}[!tb]
  \centering
  \setlength{\abovecaptionskip}{0.05cm}
  \setstretch{0.9}
  \fontsize{6.5pt}{4mm}\selectfont
  \caption{Index and search information of k-DR and NGT on real-world datasets, and the bold values are better.}
  \label{tab: index_search_info_kdr}
  \setlength{\tabcolsep}{.0092\linewidth}{
  \begin{tabular}{l|l|l|l|l|l|l|l|l|l}
    \hline
    \multicolumn{2}{c|}{\textbf{Algorithm}} & \textbf{UQ-V} & \textbf{Msong} & \textbf{Audio} & \textbf{SIFT1M} & \textbf{GIST1M} & \textbf{Crawl} & \textbf{GloVe} & \textbf{Enron}\\
    \hline
    \hline
    \multirow{6}*{k-DR} & GQ &0.574 &0.659 &0.613 &0.647 &0.754 &0.555 &0.747 &0.577\\
    \cline{2-10}
    ~ & AD &\color{blue}\textbf{16} &\color{blue}\textbf{25} &\color{blue}\textbf{17} &\color{blue}\textbf{25} &\color{blue}\textbf{40} &\color{blue}\textbf{21} &\color{blue}\textbf{43} &\color{blue}\textbf{22}\\
    \cline{2-10}
    ~ & CC &19,889& 1& 1& 2& 27& 10& 1& 6\\
    \cline{2-10}
    ~ & CS &1,440& 50& 30& 130& 130& 210& 210& 550\\
    \cline{2-10}
    ~ & PL &599& \color{blue}\textbf{61}& \color{blue}\textbf{29}& \color{blue}\textbf{166}& \color{blue}\textbf{305}& 2,713& 1,402& 383\\
    \cline{2-10}
    ~ & MO &\color{blue}\textbf{1,119}& \color{blue}\textbf{1,751}& \color{blue}\textbf{51}& \color{blue}\textbf{657}& \color{blue}\textbf{3,883}& \color{blue}\textbf{2,582}& \color{blue}\textbf{730}& \color{blue}\textbf{516}\\
    \hline
    \multirow{6}*{NGT-panng} & GQ &0.770& 0.681& 0.740& 0.762& 0.567& 0.628 & 0.589 & 0.646\\
    \cline{2-10}
    ~ & AD &52& 56& 49& 56& 67& 58& 66& 55\\
    \cline{2-10}
    ~ & CC &1& 1& 1& 1& 1& 1& 1& 1\\
    \cline{2-10}
    ~ & CS &\color{blue}\textbf{65}& \color{blue}\textbf{10}& \color{blue}\textbf{10}& \color{blue}\textbf{20}& \color{blue}\textbf{10}& \color{blue}\textbf{10}& \color{blue}\textbf{10}& \color{blue}\textbf{10}\\
    \cline{2-10}
    ~ & PL &\color{blue}\textbf{79}& 144& 33& 438& 1,172& 5,132& 2,281& \color{blue}\textbf{83}\\
    \cline{2-10}
    ~ & MO &1,432& 1,927& 63& 933& 4,111& 3,111& 928& 535\\
    \hline
    \multirow{6}*{NGT-onng} & GQ &0.431& 0.393& 0.412& 0.424& 0.266& 0.203&0.220& 0.331\\
    \cline{2-10}
    ~ & AD &47& 55& 45& 53& 75& 66& 124& 53\\
    \cline{2-10}
    ~ & CC &1& 1& 1& 1& 1& 1& 1& 1\\
    \cline{2-10}
    ~ & CS &1,002& 20& 15& 33& 33& 157& 74& 25\\
    \cline{2-10}
    ~ & PL &431& 227& 45& 392& 1,110& \color{blue}\textbf{244}& \color{blue}\textbf{388}& 131\\
    \cline{2-10}
    ~ & MO &1,411& 2,007& 63& 859& 4,088& 3,147& 1,331& 533\\
    \hline
  \end{tabular}
  }
\end{table}

\noindent\underline{\textbf{Analysis and discussion.}} Here we mainly discuss the performance difference between k-DR and NGT due to the similarity of the two. In \autoref{tab: Scalability}, k-DR exceeds NGT on simple dataset by a big margin, however, as the difficulty of the dataset increases, the performance gap between NGT and k-DR gradually narrows. Generally, the scalability of k-DR is better than NGT. As shown in \autoref{tab: index_time_size_kdr}, the index construction time of NGT is shorter than k-DR, which is mainly because the former initializes an exact KNNG while the initial graph of the latter is approximate. Although k-DR and NGT share the path adjustment strategy, k-DR implements a stricter constraint scheme, while NGT relaxes this constraint. Specifically, once there is an alternative path, k-DR directly deletes the corresponding edge, NGT has to consider the specific situation (\autoref{fig: NGT_path_adjustment}); this allows k-DR to have a smaller average out-degree, index size and memory overhead. As shown in \autoref{tab: index_search_info_kdr}, the query path length of k-DR is smaller on some simple datasets, but on some hard datasets, the query path length of NGT is smaller, and NGT-panng has a smaller candidate set size. In addition, the graph quality of k-DR is generally between NGT-onng and NGT-panng, k-DR achieves a better efficiency vs accuracy trade-off than NGT in \autoref{fig: qps-vs-recall}, and \autoref{fig: speedup-vs-recall}, which shows too high or too low graph quality does not produce better search performance. In summary, the overall performance of k-DR is better than NGT.

\setlength{\textfloatsep}{0cm}
\setlength{\floatsep}{0cm}
\begin{table}[!tb]
  \centering
  \setlength{\abovecaptionskip}{0.05cm}
  \setstretch{0.9}
  \fontsize{6.5pt}{4mm}\selectfont
  \caption{Important attributes of k-DR.}
  \label{tab: attributes_kdr}
  \setlength{\tabcolsep}{.0182\linewidth}{
  \begin{tabular}{l|l|l|l}
    \hline
    \textbf{Base Graph} & \textbf{Edge} & \textbf{Build Complexity} & \textbf{Search Complexity}\\
    \hline
    \hline
    KNNG+RNG & undirected & $O(|S|^2 +k \cdot |S|)$ & $O(\log^{5.51}(|S|))$ \\
    \hline
  \end{tabular}
  }
\end{table}

\subsection*{Appendix O. Trade-off for efficiency vs accuracy}
\label{Appendix_O}

In order to comprehensively evaluate the search performance of each algorithm, their trade-off curves of Queries Per Second (QPS) vs Recall@10 and Speedup vs Recall@10 ared measured on eight real-world datasets in \autoref{fig: qps-vs-recall} and \autoref{fig: speedup-vs-recall}. It is the most important part for the search performance evaluation of graph-based ANNS algorithms as the key of ANNS is seek a good trade-off between efficiency and accuracy. We mainly focus on the performance of each algorithm in the high-precision area due to actual needs~\cite{NSG, NSSG}. On GIST1M, Crawl, and GloVe datasets, SPTAG-BKT falls into the accuracy ``ceiling'' before reaching Recall@10=0.80, so it is not shown in \autoref{fig: qps-vs-recall} and \autoref{fig: speedup-vs-recall}.

\subsection*{Appendix P. Performance evaluation of the optimized algorithm}
\label{Appendix_P}
We evaluate the performance of the optimized algorithm (OA) and the state-of-the-art algorithms for index construction and search on two real world datasets of different difficulty in the same environment. According to our assessment, OA achieves the best overall performance.

\setlength{\textfloatsep}{0cm}
\setlength{\floatsep}{0cm}
\begin{table}[!tb]
  \centering
  \setlength{\abovecaptionskip}{0.05cm}
  \setstretch{0.9}
  \fontsize{6.5pt}{4mm}\selectfont
  \caption{Index construction time (s) of the optimized algorithm (OA) and the state-of-the-art algorithms.}
  \label{tab: oa_index_time}
  \setlength{\tabcolsep}{.017\linewidth}{
  \begin{tabular}{l|l||l|l|l|l|l|l}
    \hline
    \multicolumn{2}{c||}{\textbf{Algorithm}} & OA & NSG & NSSG & HCNNG & HNSW & DPG\\
    \hline
    \multirow{2}*{\textbf{Dataset}} & SIFT1M & 1,791 & 2,503 & 4,931 & 8,603 & 58,526 & \color{blue}\textbf{1,526}\\
    \cline{2-8}
    ~ & GIST1M & 12,440 & 14,965 & 13,157 & 9,934 & 104,484 & \color{blue}\textbf{6,188}\\
    \hline
  \end{tabular}
  }
\end{table}

\setlength{\textfloatsep}{0cm}
\setlength{\floatsep}{0cm}
\begin{table}[!tb]
  \centering
  \setlength{\abovecaptionskip}{0.05cm}
  \setstretch{0.9}
  \fontsize{6.5pt}{4mm}\selectfont
  \caption{Index size (MB) of the optimized algorithm (OA) and the state-of-the-art algorithms.}
  \label{tab: oa_index_size}
  \setlength{\tabcolsep}{.017\linewidth}{
  \begin{tabular}{l|l||l|l|l|l|l|l}
    \hline
    \multicolumn{2}{c||}{\textbf{Algorithm}} & OA & NSG & NSSG & HCNNG & HNSW & DPG\\
    \hline
    \multirow{2}*{\textbf{Dataset}} & SIFT1M & 88 & 97 & \color{blue}\textbf{80} & 394 & 202 & 293\\
    \cline{2-8}
    ~ & GIST1M & 79 & \color{blue}\textbf{53} & 102 & 326 & 234 & 362\\
    \hline
  \end{tabular}
  }
\end{table}

\setlength{\textfloatsep}{0cm}
\setlength{\floatsep}{0cm}
\begin{table}[!tb]
  \centering
  \setlength{\abovecaptionskip}{0.05cm}
  \setstretch{0.9}
  \fontsize{6.5pt}{4mm}\selectfont
  \caption{Graph quality (GQ), average out-degree (AD), and \# of connected components (CC) on graph indexes of the optimized algorithm (OA) and the state-of-the-art algorithms.}
  \label{tab: oa_index_info}
  \setlength{\tabcolsep}{.017\linewidth}{
  \begin{tabular}{l|l|l|l|l|l|l}
    \hline
    \multirow{2}*{\textbf{Algorithm}} & \multicolumn{3}{c|}{\textbf{SIFT1M}} & \multicolumn{3}{c}{\textbf{GIST1M}}\\
    \cline{2-7}
    ~ & GQ & AD & CC & GQ & AD & CC\\
    \hline
    \hline
    OA & 0.549 & \color{blue}\textbf{20} & \color{blue}\textbf{1} & 0.402 & 18 &\color{blue}\textbf{1}\\
    \hline
    NSG & 0.551 & 24 & \color{blue}\textbf{1} & 0.402 & \color{blue}\textbf{13} & \color{blue}\textbf{1}\\
    \hline
    NSSG & 0.579 & \color{blue}\textbf{20} & \color{blue}\textbf{1} & 0.399 & 26 & \color{blue}\textbf{1}\\
    \hline
    HCNNG & 0.887 & 61 & \color{blue}\textbf{1} & 0.354 & 42 & \color{blue}\textbf{1}\\
    \hline
    HNSW & 0.879 & 49 & 22 & 0.633 & 57 & 122\\
    \hline
    DPG & 0.998 & 76 & \color{blue}\textbf{1} & 0.992 & 94 & \color{blue}\textbf{1}\\
    \hline
  \end{tabular}
  }
\end{table}

\begin{figure*}[!t]
  \centering
  \setlength{\abovecaptionskip}{-0.3em}
  \setlength{\belowcaptionskip}{-0.45cm}
  \includegraphics[width=.5\linewidth]{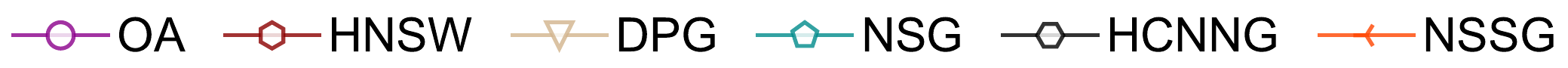}
\end{figure*}

\begin{figure*}
  \vspace{-4mm}
  \centering
  \subfigcapskip=-0.25cm
  \subfigure[Recall@10 (Msong)]{ 
    \captionsetup{skip=0pt}
    \vspace{-1.2mm}
    \includegraphics[scale=0.36]{figures/test_msong.pdf}
    \label{fig: test_msong}
  }
  \subfigure[Recall@10 (SIFT1M)]{ 
    \captionsetup{skip=0pt}
    \vspace{-1.2mm}
    \includegraphics[scale=0.36]{figures/test_sift.pdf}
    \label{fig: test_sift1M}
  }
  \subfigure[Recall@10 (GIST1M)]{ 
    \captionsetup{skip=0pt}
    \vspace{-1.2mm}
    \includegraphics[scale=0.36]{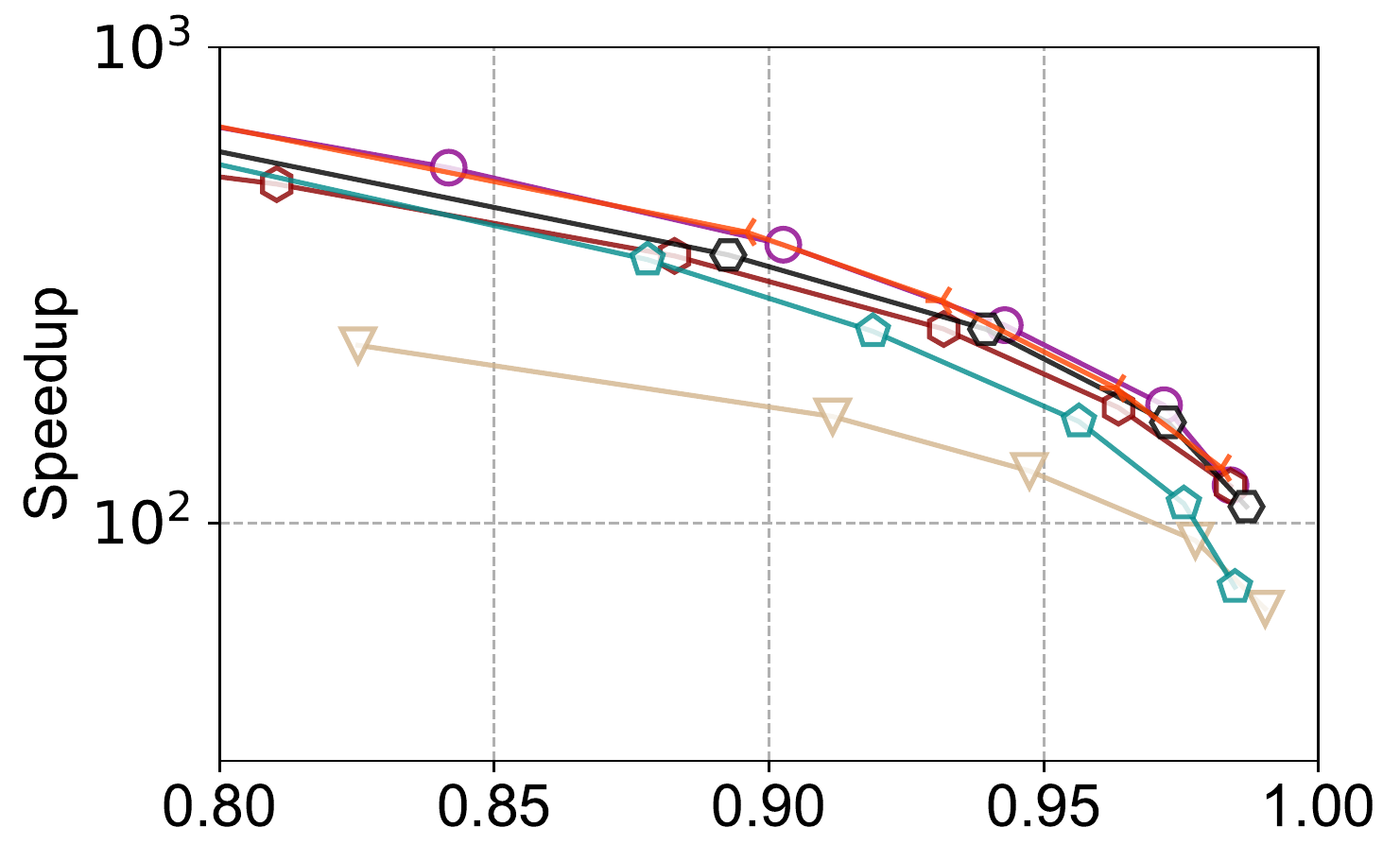}
    \label{fig: test_gist}
  }

  \vspace{-0.5cm}
  \caption{The Speedup vs Recall@10 of the optimized algorithm (OA) and the state-of-the-art algorithms. (top right is better).}\vspace{-0.4mm}
  \label{fig: test_oa}
\end{figure*}

\setlength{\textfloatsep}{0cm}
\setlength{\floatsep}{0cm}
\begin{table}[!tb]
  \centering
  \setlength{\abovecaptionskip}{0.05cm}
  \setstretch{0.9}
  \fontsize{6.5pt}{4mm}\selectfont
  \caption{Candidate set size (CS), query path length (PL), and peak memory overhead (MO) of the optimized algorithm (OA) and the state-of-the-art algorithms.}
  \label{tab: oa_search_info}
  \setlength{\tabcolsep}{.017\linewidth}{
  \begin{tabular}{l|l|l|l|l|l|l}
    \hline
    \multirow{2}*{\textbf{Algorithm}} & \multicolumn{3}{c|}{\textbf{SIFT1M}} & \multicolumn{3}{c}{\textbf{GIST1M}}\\
    \cline{2-7}
    ~ & CS & PL & MO & CS & PL & MO\\
    \hline
    \hline
    OA & 99 & 95 & 682 & 266 & 380 &3,846\\
    \hline
    NSG & 101 & 85 & 653 & 867 & 826 & \color{blue}\textbf{3,781}\\
    \hline
    NSSG & 255 & 157 & \color{blue}\textbf{640} & 280 & 270 & 3,829\\
    \hline
    HCNNG & 97 & 37 & 1,056 & 371 & 179 & 4,159\\
    \hline
    HNSW & 66 & 47 & 1,206 & 181 & 130 & 4,372\\
    \hline
    DPG & \color{blue}\textbf{37} & \color{blue}\textbf{30} & 851 & \color{blue}\textbf{55} & \color{blue}\textbf{124} & 4,091\\
    \hline
  \end{tabular}
  }
\end{table}

\noindent\underline{\textbf{Index construction performance.}} As shown in \autoref{tab: oa_index_time}, \autoref{tab: oa_index_size} and \autoref{tab: oa_index_info}, compared with the state-of-the-art algorithms, the index construction efficiency of the optimized algorithm (OA) ranks very high (second only to DPG, but OA performs better than DPG in other aspects), which is mainly because OA is not committed to achieving high graph quality at an expensive time cost (\autoref{tab: oa_index_info}), and its neighbor acquisition does not involve distance calculation. There is no additional structure attached to the graph index for OA, which makes it obtain a smaller index size; and that is also due to its smaller average out-degree. In addition, OA ensures the accessibility from the entries to any other point, which is backed up by the number of connected components.

\noindent\underline{\textbf{Search performance.}} As shown in \autoref{fig: test_oa}, OA obtains the optimal speedup vs recall trade-off on SIFT1M and GIST1M. At the same time, its candidate set size, query path length, and peak memory overhead are all close to the optimal values in \autoref{tab: oa_search_info}.

\subsection*{Appendix Q. Multiple trials for randomized parts of the algorithms}
\label{Appendix_Q}

\setlength{\textfloatsep}{0cm}
\setlength{\floatsep}{0cm}
\begin{table}[!tb]
  \centering
  \setlength{\abovecaptionskip}{0.05cm}
  \setstretch{0.9}
  \fontsize{6.5pt}{4mm}\selectfont
  \caption{Index construction time (ICT, unit: s) and index size (IS, unit: MB) of Vamana under different trails (a, b, and c).}
  \label{tab: index_build_vamana_trail}
  \setlength{\tabcolsep}{.017\linewidth}{
  \begin{tabular}{l|l|l|l|l|l}
    \hline
    \multicolumn{2}{c|}{\textbf{Trial}} & {\textbf{UQ-V}} & {\textbf{Msong}} & {\textbf{Audio}} & {\textbf{SIFT1M}}\\
    \hline
    \hline
    \multirow{4}*{\textbf{ICT}}& a & 1,451 & 1,786 & 158 & 2,657\\
    \cline{2-6}
    ~ & b & 1,575 & 1,736 & 145 & 2,756\\
    \cline{2-6}
    ~ & c & 1,378 & 1,695 & 139 & 2,608\\
    \cline{2-6}
    ~ & Average & 1,468 & 1,739 & 147 & 2,674\\
    \hline
    \hline
    \multirow{4}*{\textbf{IS}}& a & 119 & 118 & 11 & 195\\
    \cline{2-6}
    ~ & b & 119 & 118 & 11 & 195\\
    \cline{2-6}
    ~ & c & 119 & 118 & 11 & 195\\
    \cline{2-6}
    ~ & Average & 119 & 118 & 11 & 195\\
    \hline
  \end{tabular}
  }
\end{table}

\begin{figure*}
  \vspace{-4mm}
  \centering
  \subfigcapskip=-0.25cm
  \subfigure[Recall@10 (UQ-V)]{  
    \captionsetup{skip=0pt}
    \vspace{-1mm}
    \includegraphics[scale=0.29]{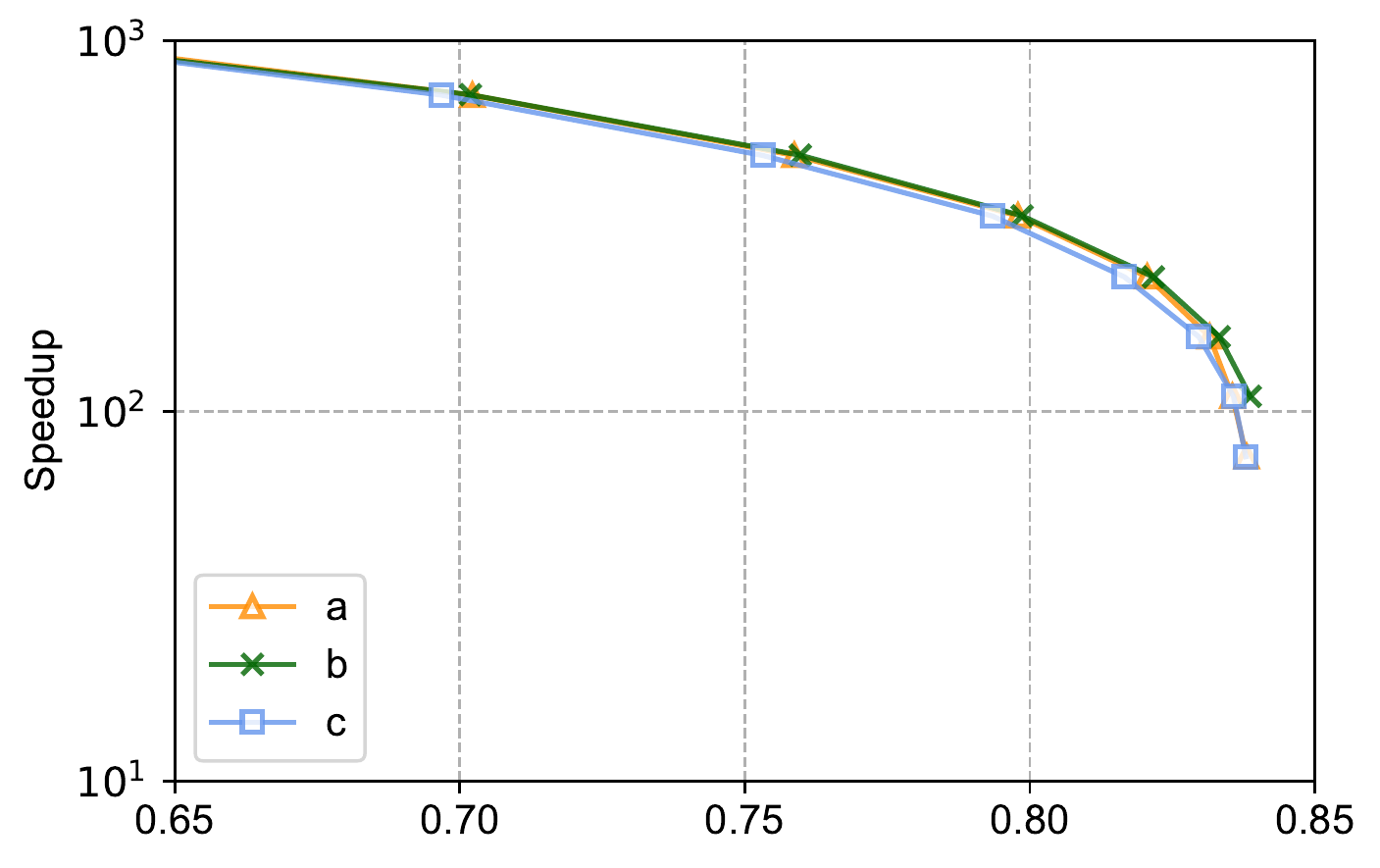}
    \label{fig: multiple_trail_vamana_uqv}
  }
  \subfigure[Recall@10 (Msong)]{ 
    \captionsetup{skip=0pt}
    \vspace{-1mm}
    \includegraphics[scale=0.29]{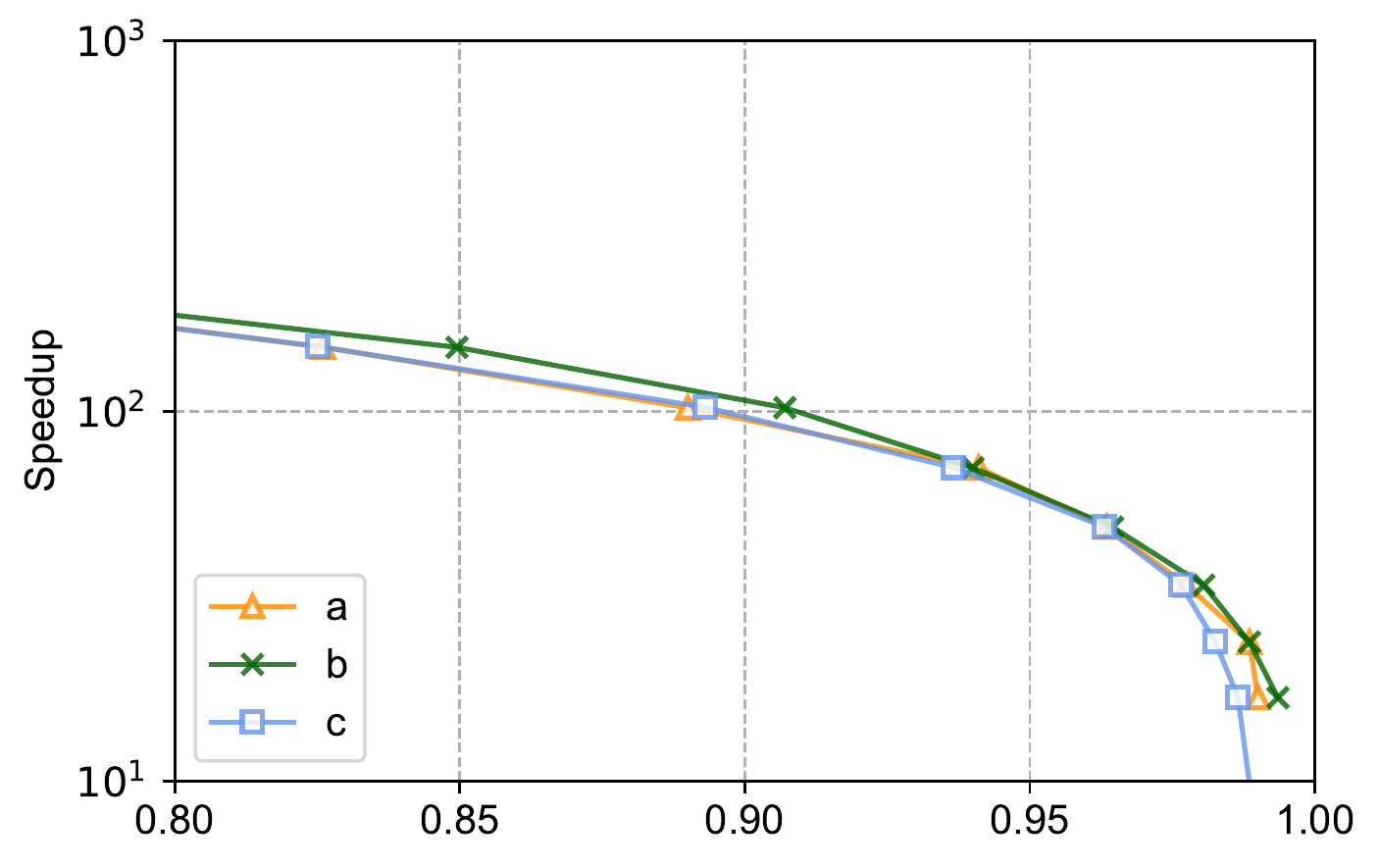}
    \label{fig: multiple_trail_vamana_msong}
  }
  \subfigure[Recall@10 (Audio)]{ 
    \captionsetup{skip=0pt}
    \vspace{-1mm}
    \includegraphics[scale=0.29]{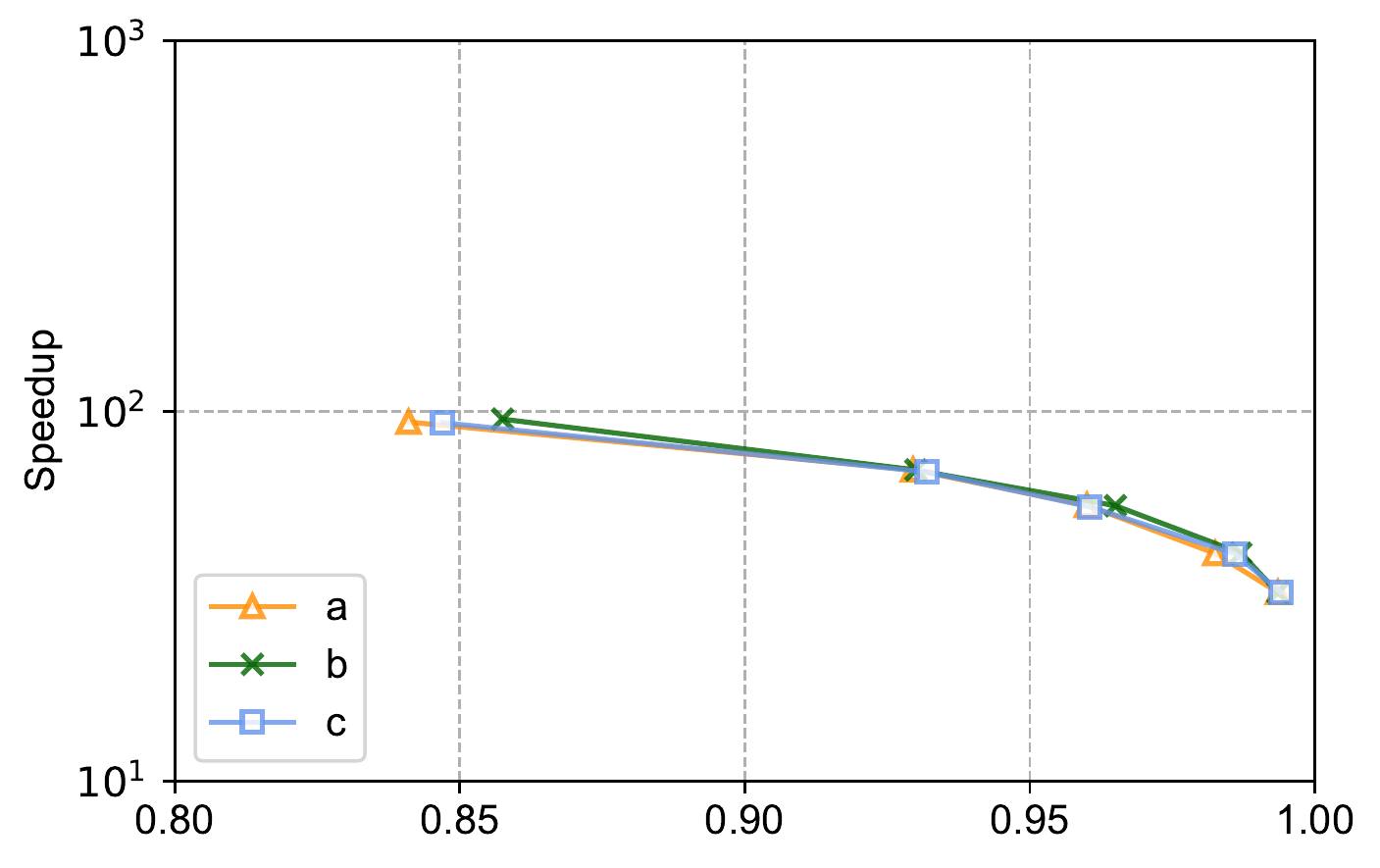}
    \label{fig: multiple_trail_vamana_audio}
  }
  \subfigure[Recall@10 (SIFT1M)]{ 
    \captionsetup{skip=0pt}
    \vspace{-1mm}
    \includegraphics[scale=0.29]{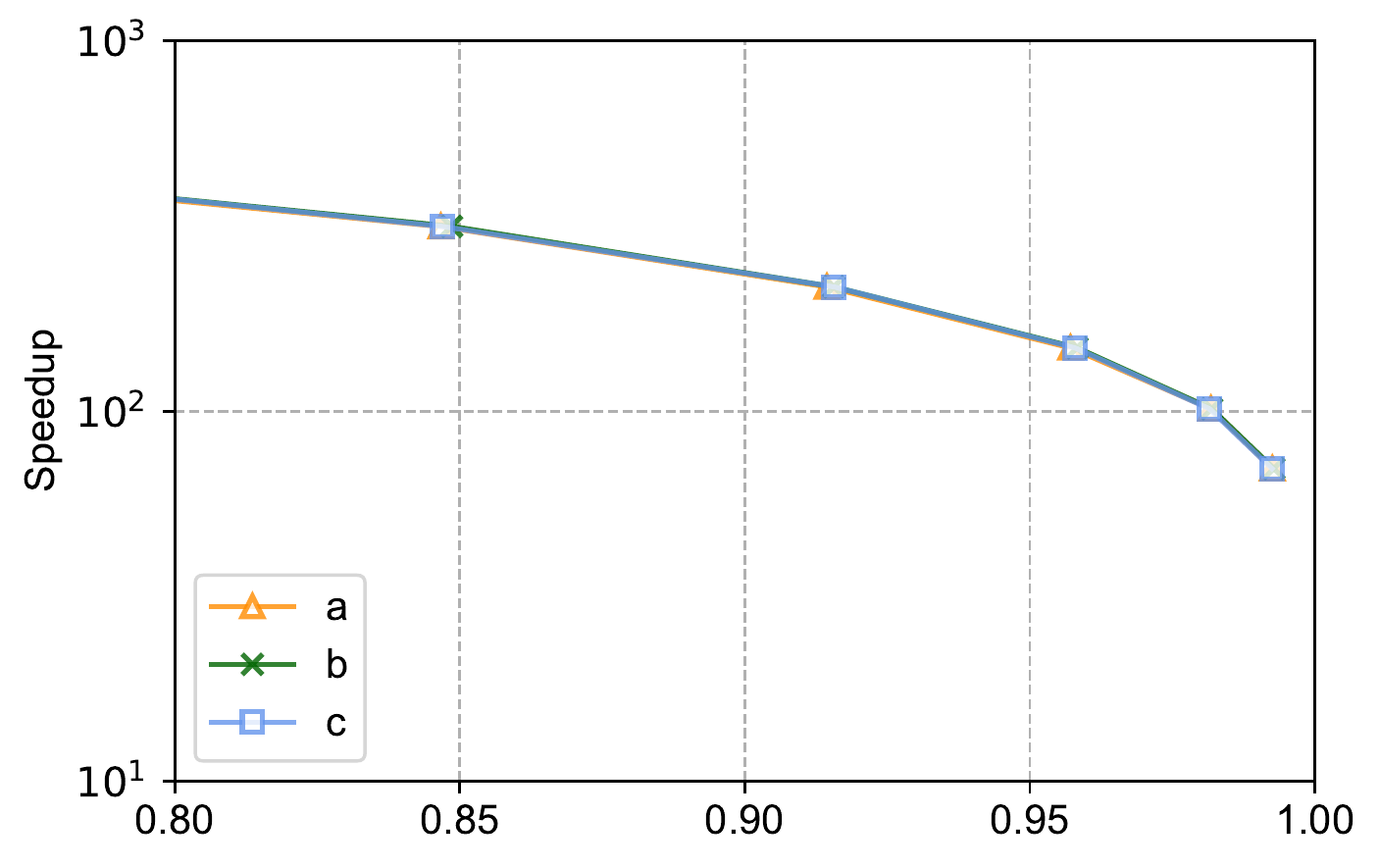}
    \label{fig: multiple_trail_vamana_sift}
  }

  \vspace{-0.5cm}
  \caption{Speedup vs Recall@10 of Vamana under different trails (a, b, and c).}\vspace{-0.4mm}
  \label{fig: search_vamana_trail}
\end{figure*}

\begin{figure*}
  \vspace{-4mm}
  \centering
  \subfigcapskip=-0.25cm
  \subfigure[Recall@10 (UQ-V)]{  
    \captionsetup{skip=0pt}
    \vspace{-1mm}
    \includegraphics[scale=0.29]{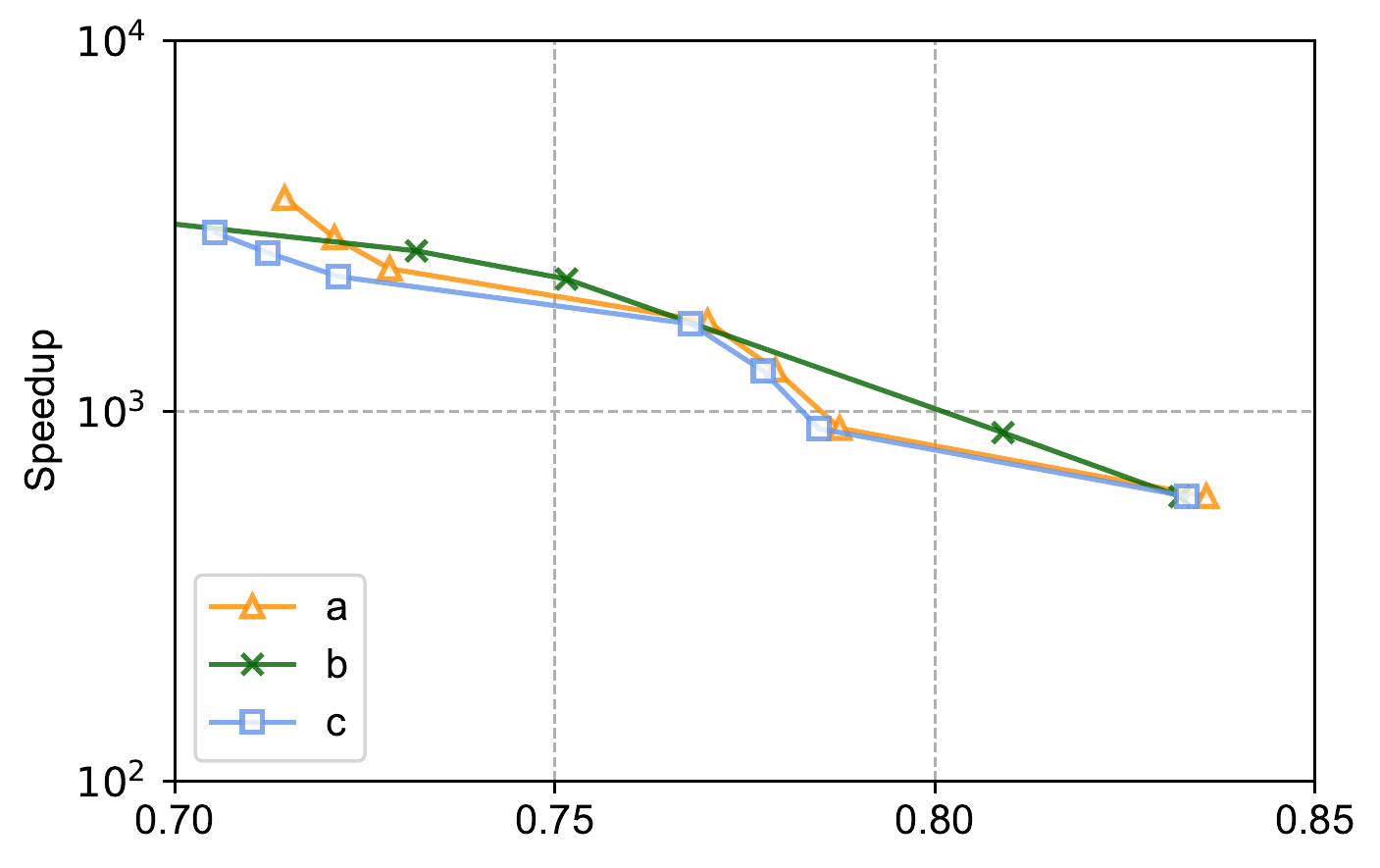}
    \label{fig: multiple_trail_nssg_uqv}
  }
  \subfigure[Recall@10 (Msong)]{ 
    \captionsetup{skip=0pt}
    \vspace{-1mm}
    \includegraphics[scale=0.29]{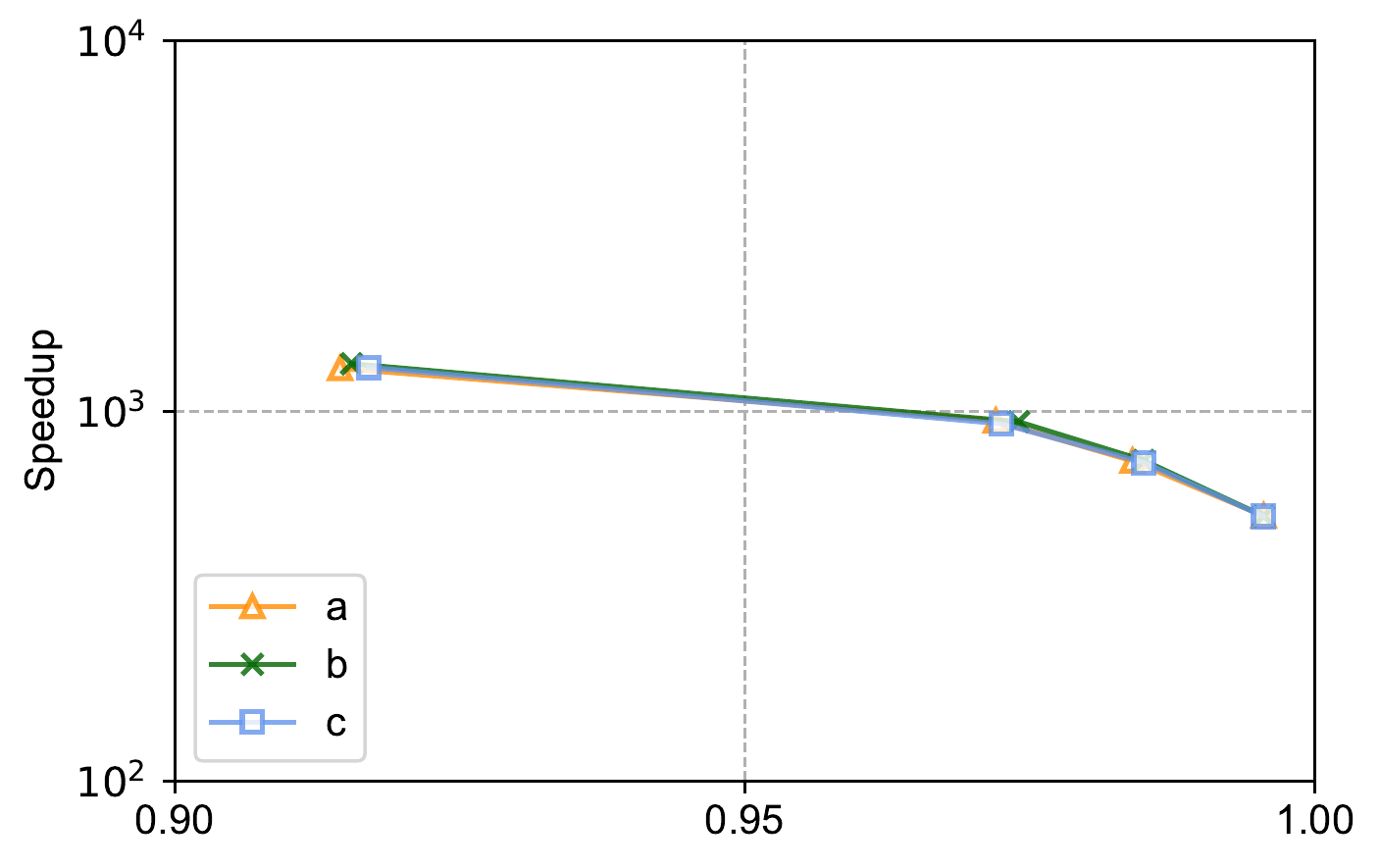}
    \label{fig: multiple_trail_nssg_msong}
  }
  \subfigure[Recall@10 (Audio)]{ 
    \captionsetup{skip=0pt}
    \vspace{-1mm}
    \includegraphics[scale=0.29]{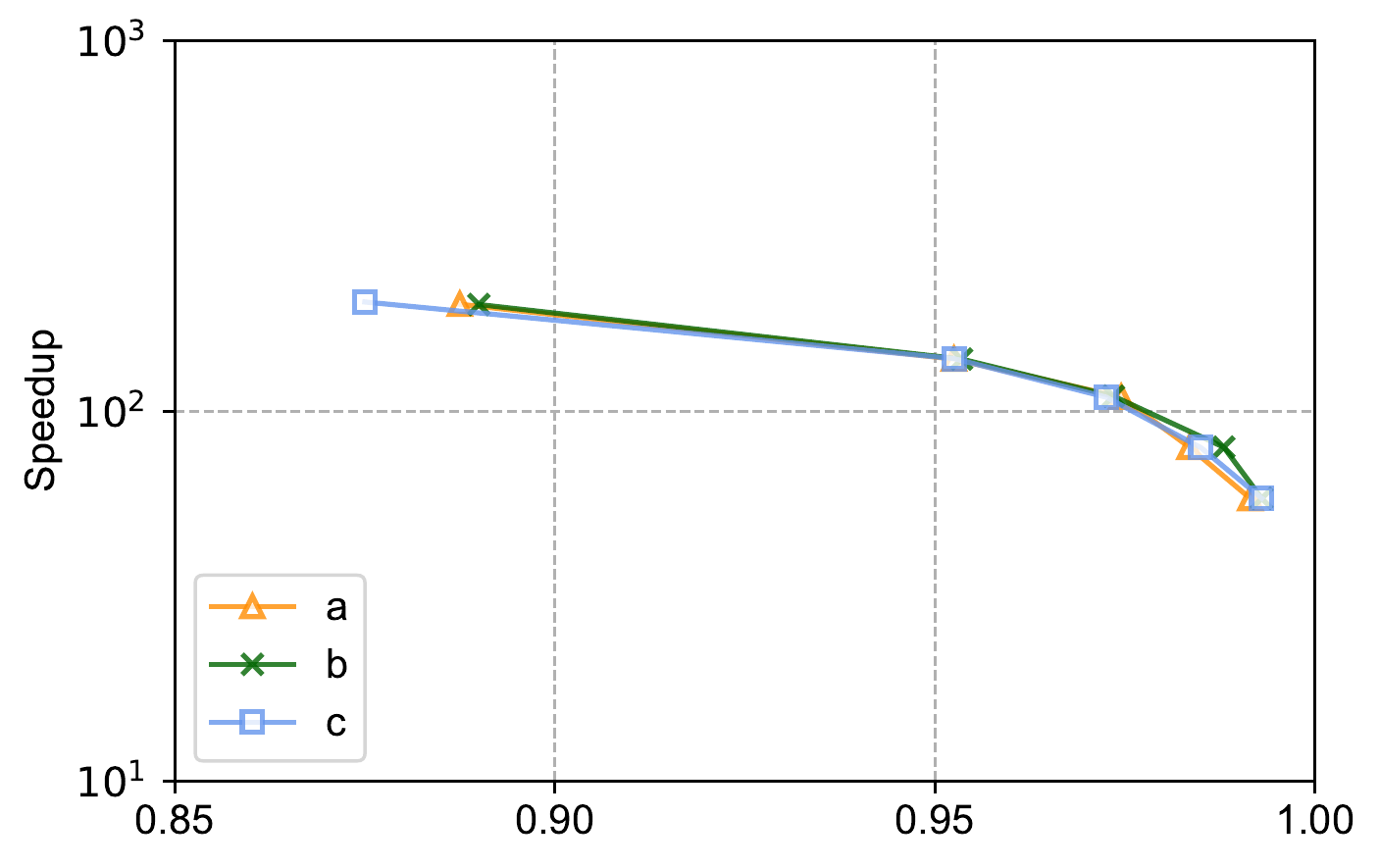}
    \label{fig: multiple_trail_nssg_audio}
  }
  \subfigure[Recall@10 (SIFT1M)]{ 
    \captionsetup{skip=0pt}
    \vspace{-1mm}
    \includegraphics[scale=0.29]{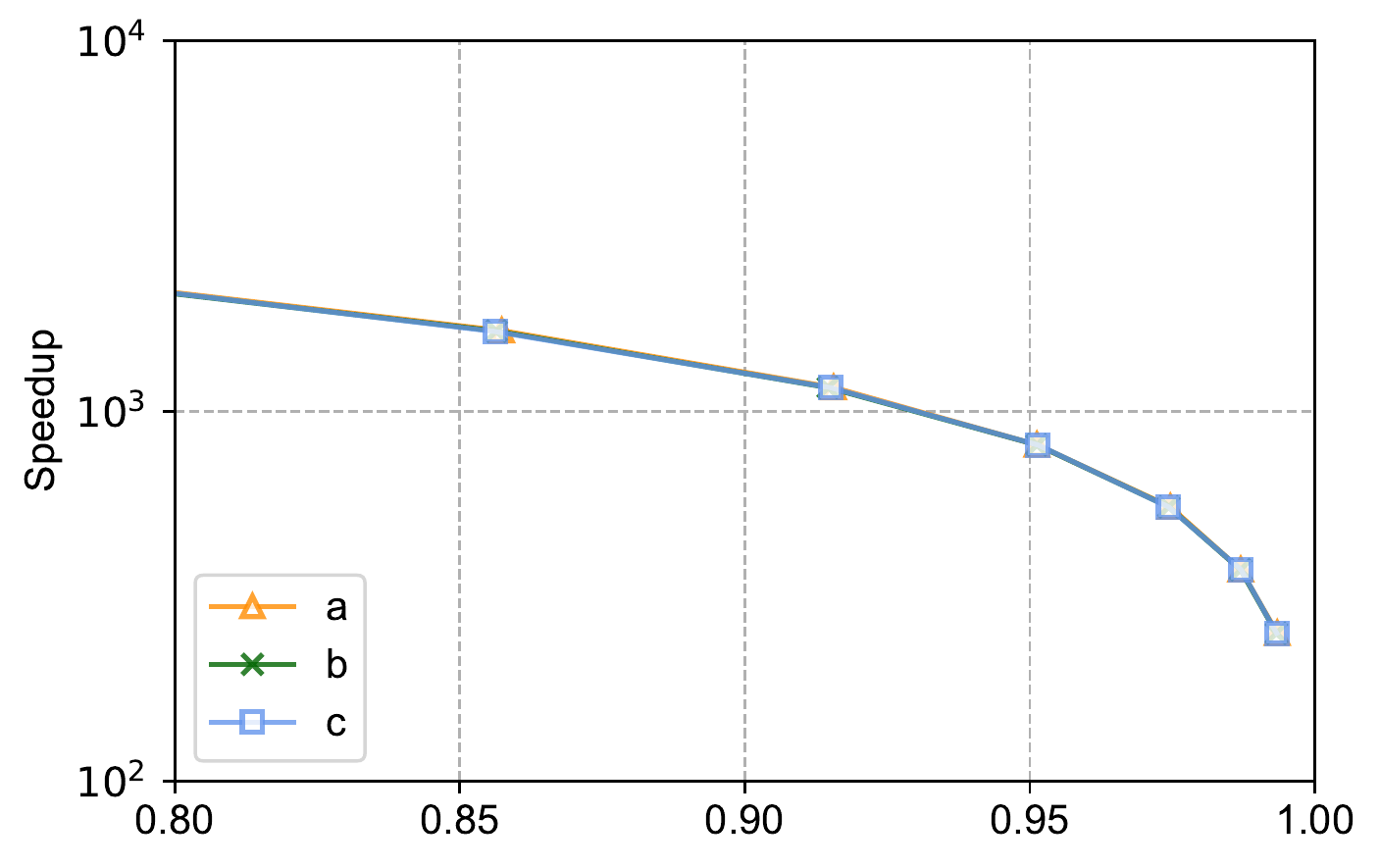}
    \label{fig: multiple_trail_nssg_sift}
  }

  \vspace{-0.5cm}
  \caption{Speedup vs Recall@10 of NSSG under different trails (a, b, and c).}\vspace{-0.4mm}
  \label{fig: search_nssg_trail}
\end{figure*}

For some algorithms that include the randomization, we perform multiple experiments under the same environment and report the average value. According to our experimental results in \autoref{tab: index_build_vamana_trail}, \autoref{fig: search_vamana_trail} and \autoref{fig: search_nssg_trail}, we conclude that a single value is very close to the average value. Below we take Vamana and NSSG as examples to explain the reasons for the above phenomenon.

\noindent\underline{\textbf{Vamana.}} The initialization of the graph index is to randomly select a given number of neighbors for each element on the dataset. For the convenience of description, we divide any vertex's neighbors to ``good neighbors'' (GN) and ``bad neighbors'' (BN). If we initialize GN for each vertex, Vamana’s index construction efficiency can reach the optimal value, as GN enable Vamana to perform ANNS more efficiently when acquiring candidate neighbors. In contrast, if BN are initialized for each element, it will lead to the worst index construction efficiency. However, the probability of the above two situations happening is extremely low (close to zero). Actually, the ratio of GN and BN of each element is relatively stable. In general, Vamana's ANNS performance with different initial graphs is close (see \autoref{fig: search_vamana_trail}). When ignoring the cost of neighbor selection (it does not affected by the initial graph), its index construction efficiency mainly depends on ANNS on the initial graph for obtaining candidate neighbors; so it is almost the same under different trails.

\noindent\underline{\textbf{NSSG.}} The seed acquisition component of NSSG is randomized. As shown in \autoref{fig: search_nssg_trail}, the search performance curves of NSSG under different experiments almost overlap. For the convenience of description, we also divide the seeds into ``good seeds'' (GS) and ``bad seeds'' (BS). Performing search from GS can get the best search performance, while starting search from BS will result in the worst search performance. Due to the sufficient amount of query data (> 200, \autoref{tab: Dataset}), the probability that the randomly obtained seeds is GS or BS for all queries is almost zero; for a given batch of queries, the ratio of the two is stable. Therefore, random seeds with multiple repetitions produce similar search performance.

\subsection*{Appendix R. Evaluation and analysis of machine learn (ML) based methods}
\label{Appendix_R}

\noindent\underline{\textbf{Overview.}} In general, ML-based approaches append additional optimizations on existing graph-based algorithms. For example, \cite{learn_to_route} (ML1) learns vertex representation on graph-based algorithms (e.g., NSW) to provide a better routing; \cite{li2020improving} (ML2) performs ANNS on HNSW through learned adaptive early termination, it builds and trains gradient boosting decision tree models to learn and predict when to stop searching for a certain query; \cite{prokhorenkova2020graph} (ML3) maps the dataset into a space of lower dimension while trying to preserve local geometry by ML, then it can be combined with any graph-based algorithms like HNSW or NSG.

\noindent\underline{\textbf{Setting.}} We implement ML1 and ML3 optimization on NSG, and implement ML2 optimization on HNSW (HNSW is selected in the original paper \cite{li2020improving}); considering that when ML2 is applied to NSG, some additional optimization is required (we will add ML2 to NSG soon for evaluation after solving these problems). We focus on the index processing time, memory consumption during index construction, and speedup vs recall trade-off of each method. Note that ML1's index preprocessing training is very time-consuming and memory-consuming (more than 125G on SIFT1M), so we use GPU to accelerate the process and use the smaller SIFT100K and GIST100K datasets. In addition, the number of nearest neighbors recalled is uniformly set to 1 for each query due to the limitation of ML1~\cite{learn_to_route}, and Recall@1 represents the corresponding recall rate.

\begin{figure*}
  \vspace{-4mm}
  \centering
  \subfigcapskip=-0.25cm
  \subfigure[Recall@1 (SIFT100K)]{  
    \captionsetup{skip=0pt}
    \vspace{-1mm}
    \includegraphics[scale=0.36]{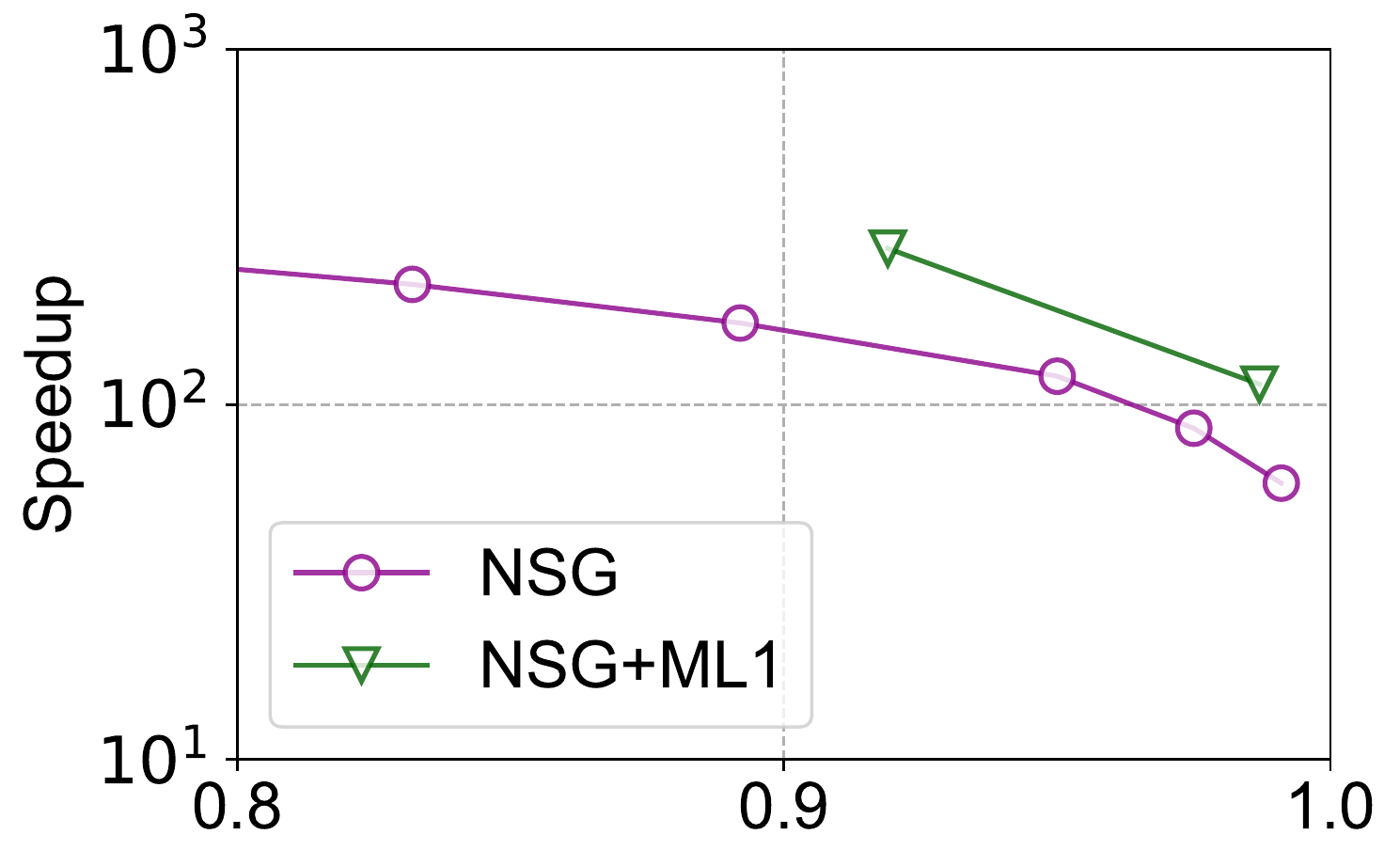}
    \label{fig: tml1_sift}
  }\hspace{-1mm}
  \subfigure[Recall@1 (GIST100K)]{ 
    \captionsetup{skip=0pt}
    \vspace{-1mm}
    \includegraphics[scale=0.36]{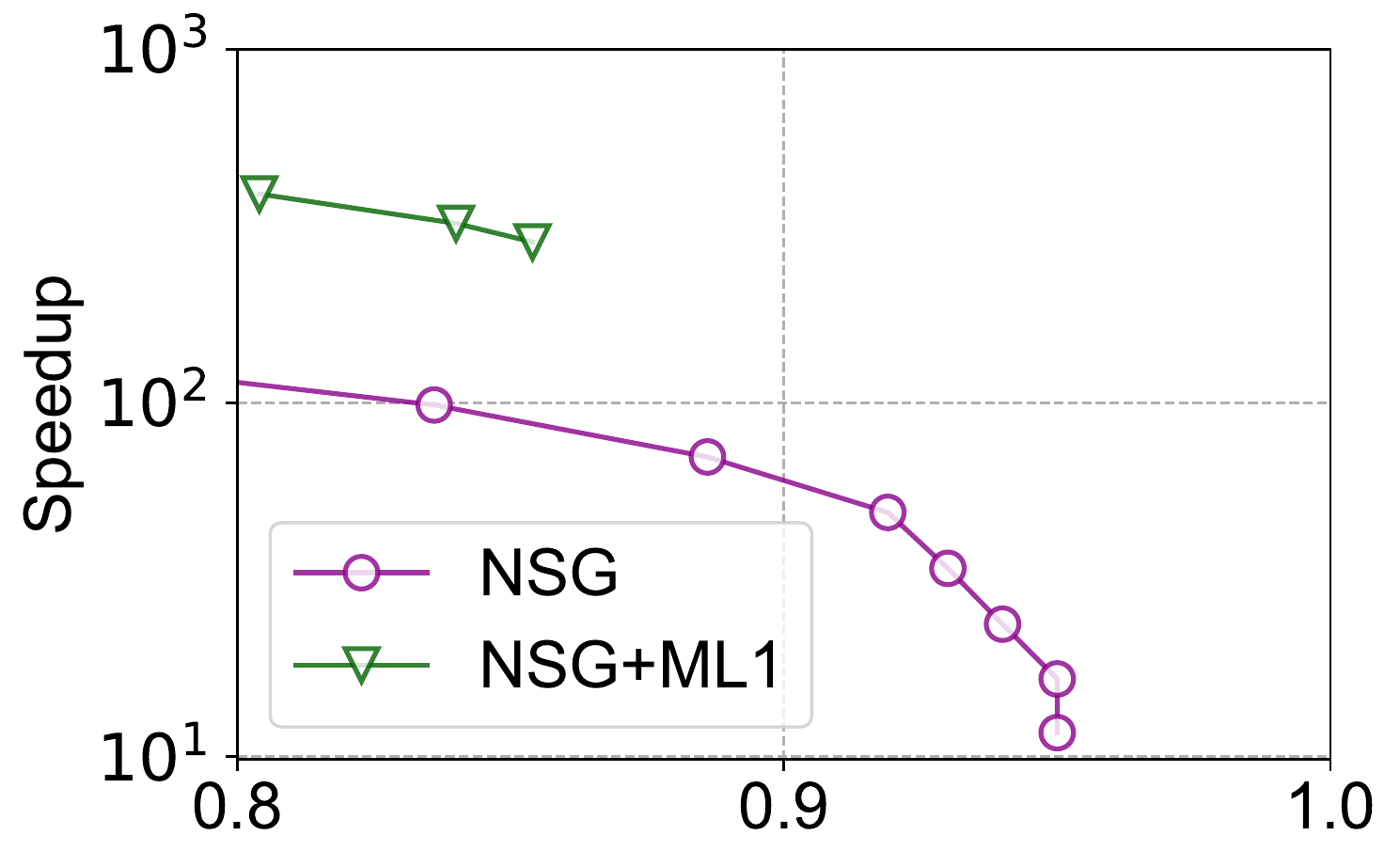}
    \label{fig: ml1_gist}
  }\hspace{-1mm}
  \subfigure[Recall@1 (SIFT100K)]{ 
    \captionsetup{skip=0pt}
    \vspace{-1mm}
    \includegraphics[scale=0.36]{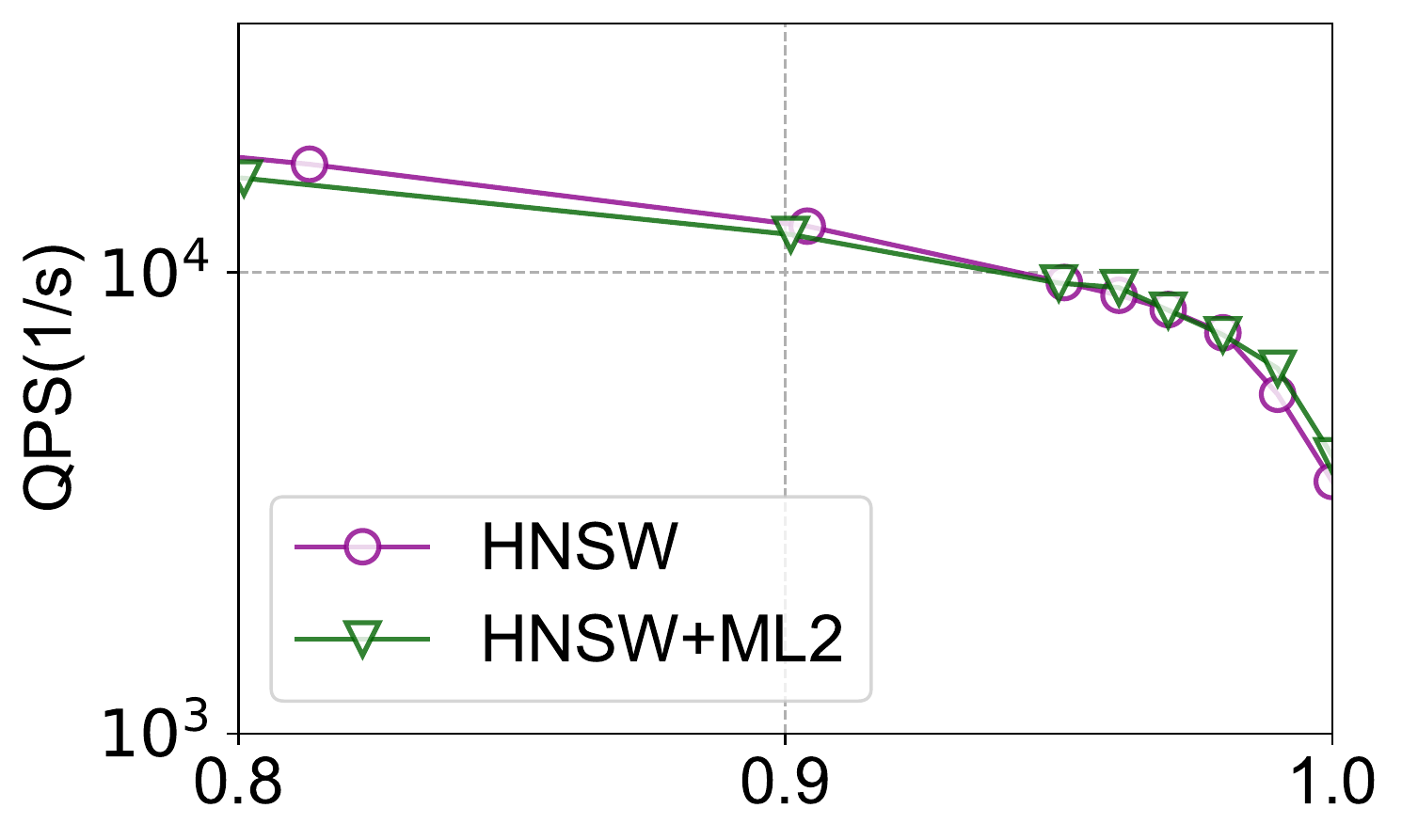}
    \label{fig: ml2_sift}
  }\hspace{-1mm}
  \subfigure[Recall@1 (GIST100K)]{ 
    \captionsetup{skip=0pt}
    \vspace{-1mm}
    \includegraphics[scale=0.36]{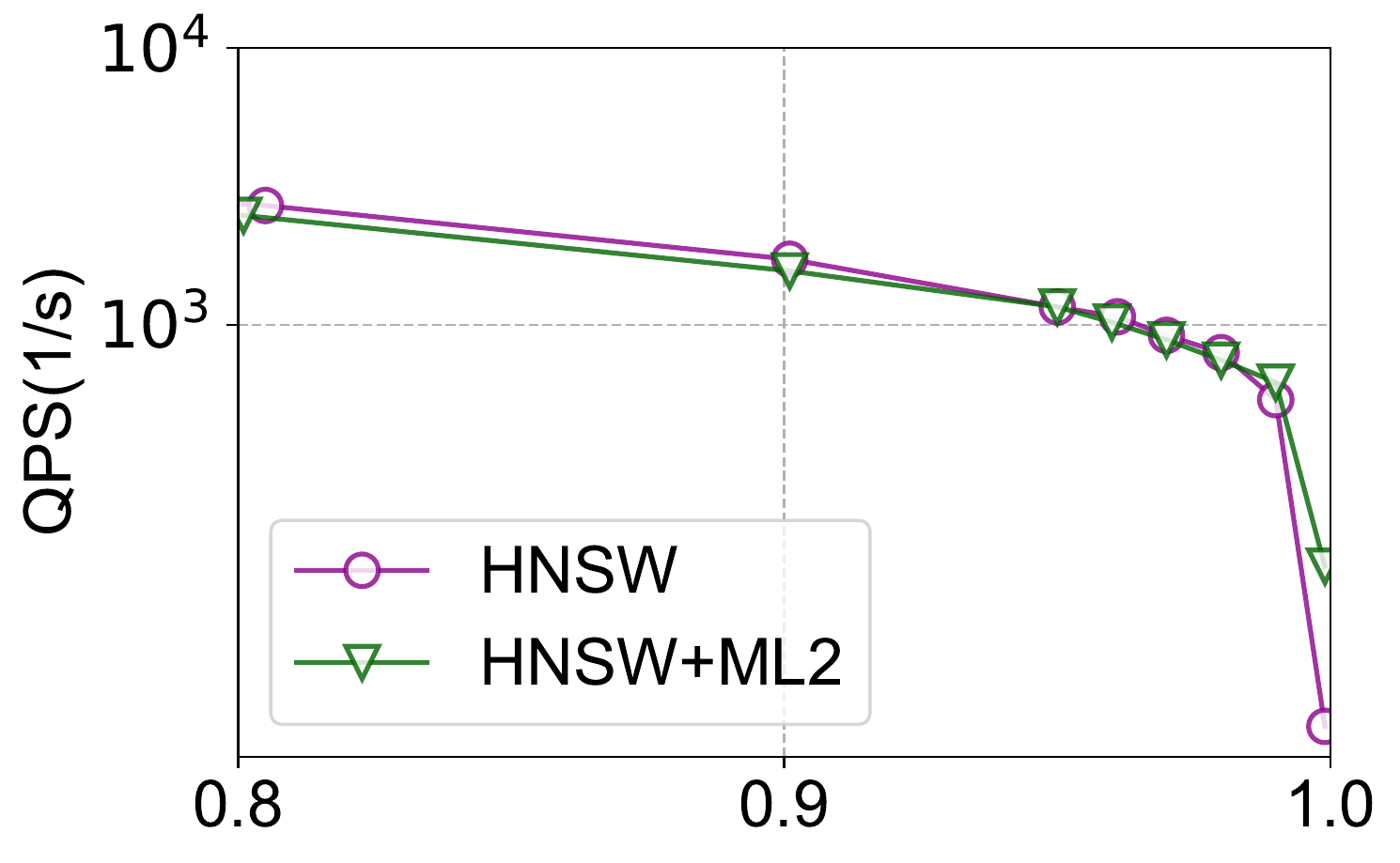}
    \label{fig: ml2_gist}
  }\hspace{-1mm}
  \subfigure[Recall@1 (SIFT100K)]{ 
  \captionsetup{skip=0pt}
  \vspace{-1mm}
  \includegraphics[scale=0.36]{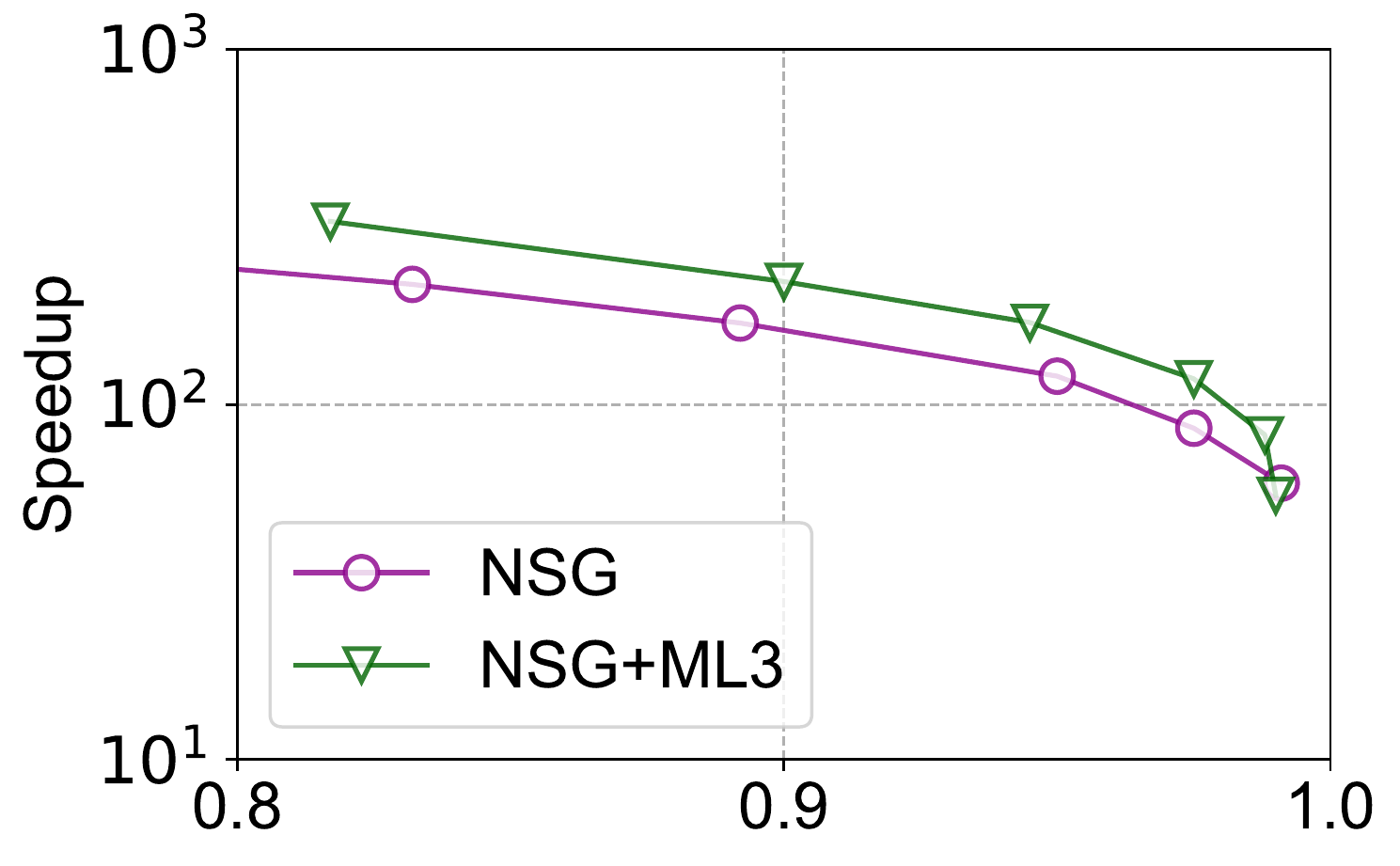}
  \label{fig: ml3_sift}
  }\hspace{-1mm}
  \subfigure[Recall@1 (GIST100K)]{ 
  \captionsetup{skip=0pt}
  \vspace{-1mm}
  \includegraphics[scale=0.36]{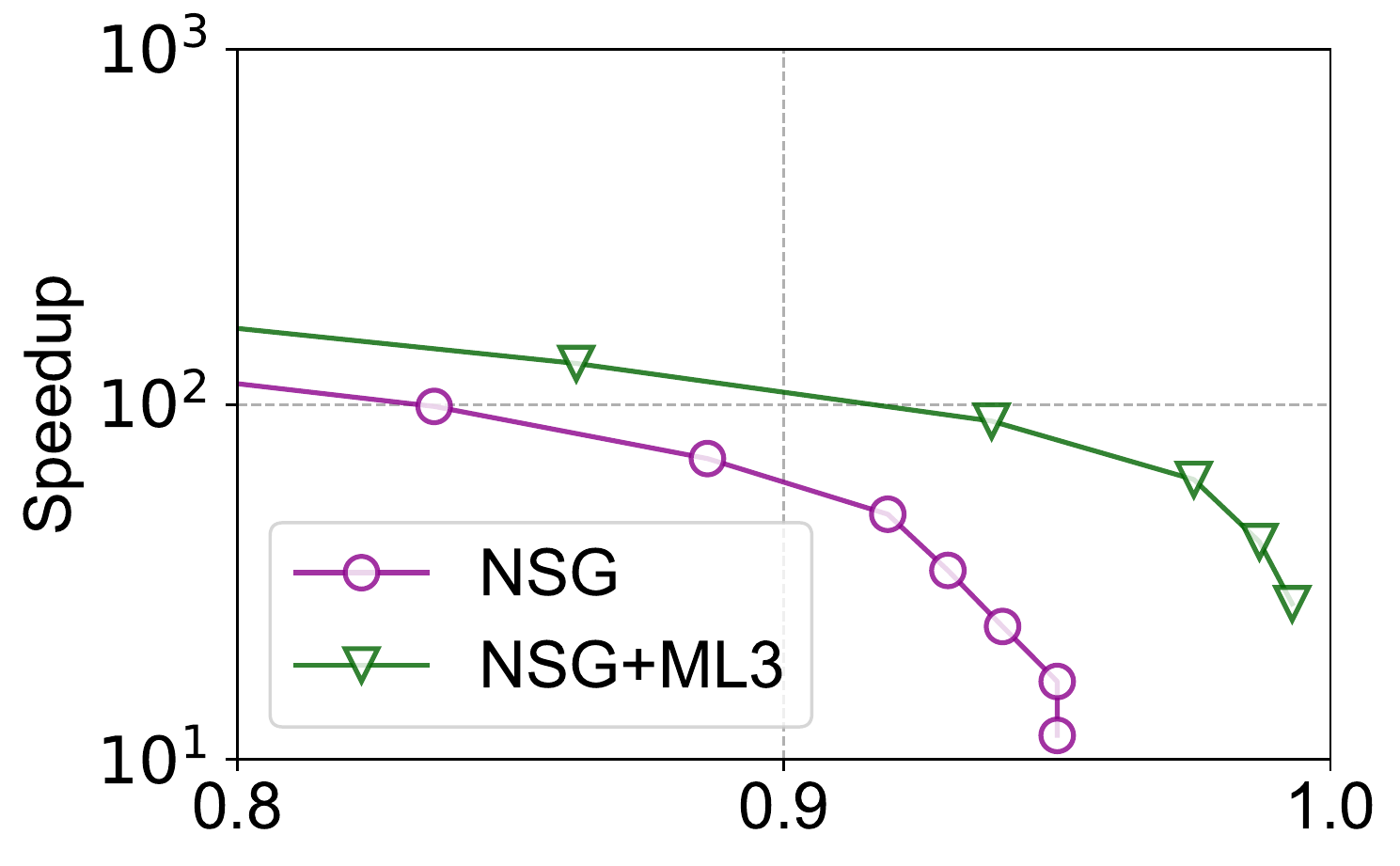}
  \label{fig: ml3_gist}
  }

  \vspace{-0.5cm}
  \caption{Speedup vs Recall@10 of ML-based methods.}\vspace{-0.4mm}
  \label{fig: search_ml}
\end{figure*}

\setlength{\textfloatsep}{0cm}
\setlength{\floatsep}{0cm}
\begin{table}[!tb]
  \centering
  \setlength{\abovecaptionskip}{0.05cm}
  \setstretch{0.9}
  \fontsize{6.5pt}{4mm}\selectfont
  \caption{Index processing time (IPT) and memory consumption (MC) of ML-based methods.}
  \label{tab: index_build_ml}
  \setlength{\tabcolsep}{.017\linewidth}{
  \begin{tabular}{l|l|l|l}
    \hline
    \multicolumn{2}{c|}{\textbf{Method}} & {\textbf{SIFT100K}} & {\textbf{GIST100K}}\\
    \hline
    \hline
    \multirow{4}*{\textbf{IPT(s)}}& NSG & \color{blue}\textbf{55} & \color{blue}\textbf{142}\\
    \cline{2-4}
    ~ & NSG+ML1 & 67,315 & 45,742 \\
    \cline{2-4}
    ~ & HNSW+ML2 & 2,018 & 2,626 \\
    \cline{2-4}
    ~ & NSG+ML3 & 1,260 & 1,287 \\
    \hline
    \hline
    \multirow{4}*{\textbf{MC(GB)}}& NSG & {\color{blue}\textbf{0.37}} & {\color{blue}\textbf{0.68}} \\
    \cline{2-4}
    ~ & NSG+ML1 & 23.8 & 58.7 \\
    \cline{2-4}
    ~ & HNSW+ML2 & 3 & 5.7 \\
    \cline{2-4}
    ~ & NSG+ML3 & 23 & 25.5 \\
    \hline
  \end{tabular}
  }
\end{table}

\noindent\underline{\textbf{Discussion.}} As shown in \autoref{tab: index_build_ml} and \autoref{fig: search_ml}, there ML-based optimizations generally obtain better speedup (or QPS) vs recall tradeoff than the original algorithms (NSG or HNSW) at the expense of more time and memory. ML2 only provides slight latency reduction in the high-precision area. It is worth noting that ML2 have a smaller impact on IPT and IS (compared to HNSW). The IPT of ML1 is significantly higher than the original NSG, and it also requires additional index overhead. Although ML3 improves the speedup vs recall tradeoff by a large margin, it also significantly increases memory consumption.

\begin{figure*}[!t]
  \centering
  \vspace{-1.4cm}
  \setlength{\abovecaptionskip}{-0.3em}
  \setlength{\belowcaptionskip}{-0.45cm}
  \includegraphics[width=\linewidth]{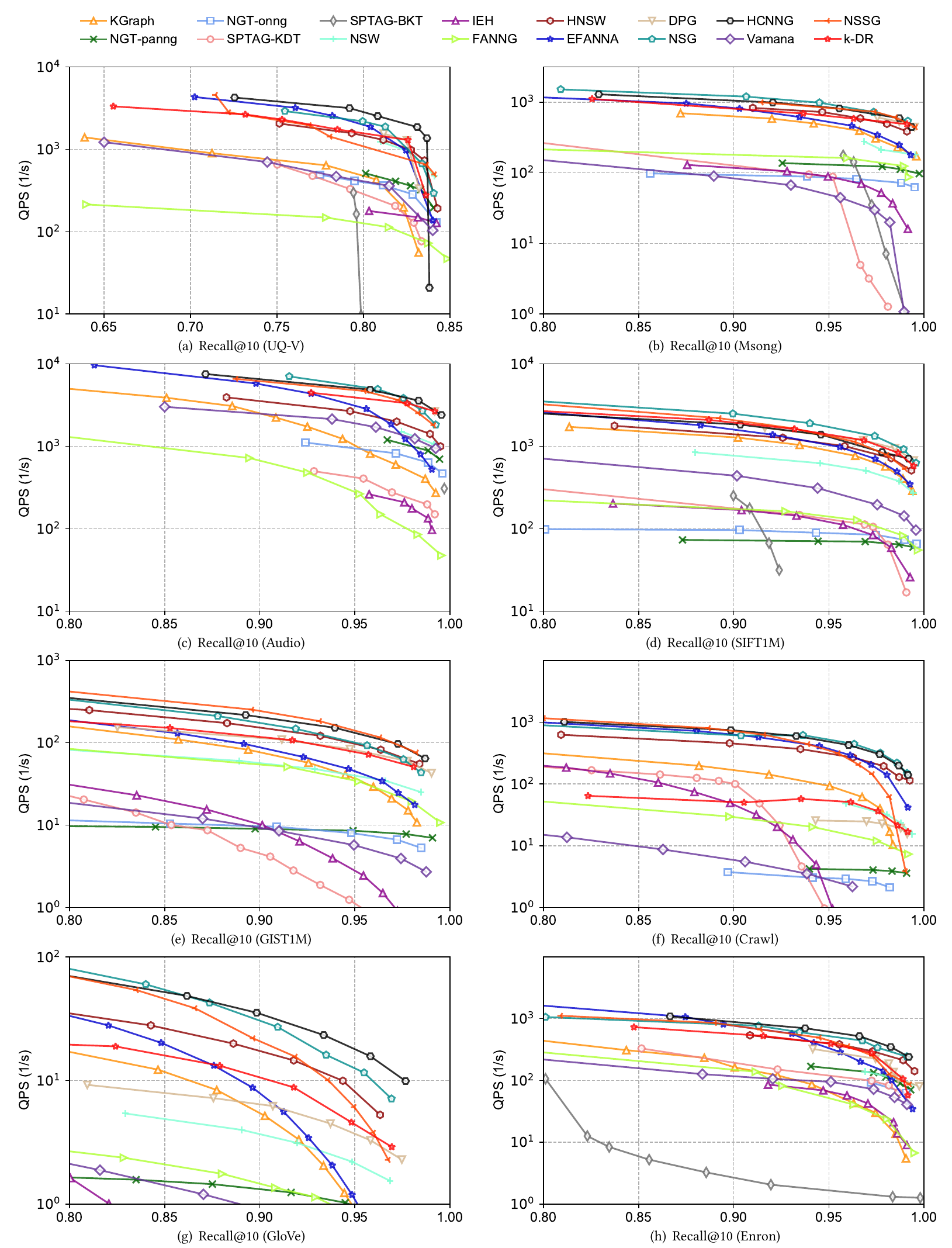}
  \vspace{-0.3cm}
  \caption{The Queries Per Second (QPS) vs Recall@10 of graph-based ANNS algorithms with their optimal indices in high-precision region on the eight real world datasets (top right is better).}\vspace{-0.4mm}
  \label{fig: qps-vs-recall}
\end{figure*}

\begin{figure*}[!t]
  \centering
  \vspace{-1.4cm}
  \setlength{\abovecaptionskip}{-0.3em}
  \setlength{\belowcaptionskip}{-0.45cm}
  \includegraphics[width=\linewidth]{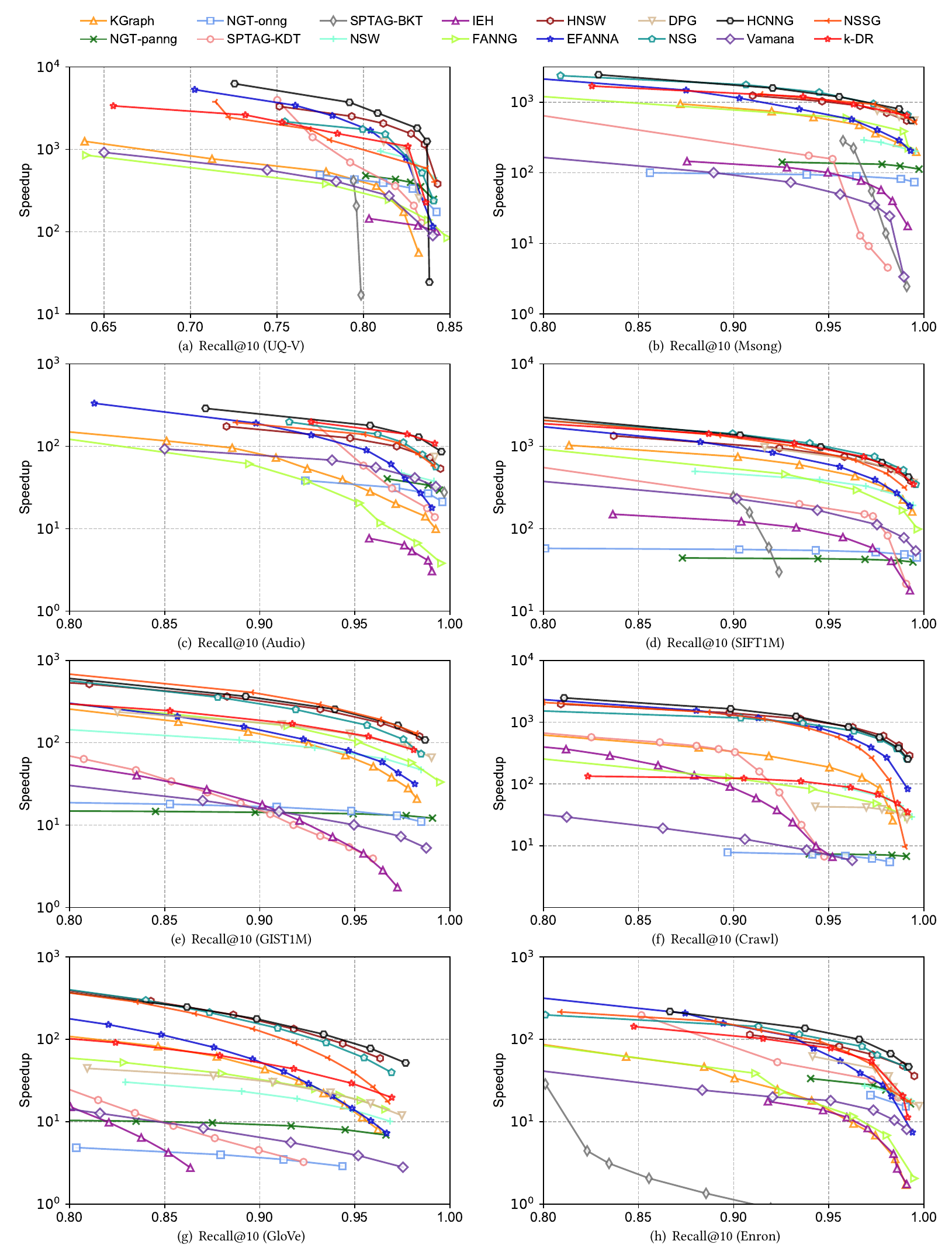}
  \vspace{-0.3cm}
  \caption{The Speedup vs Recall@10 of graph-based ANNS algorithms with their optimal indices in high-precision region on the eight real world datasets (top right is better).}\vspace{-0.4mm}
  \label{fig: speedup-vs-recall}
\end{figure*}

\end{document}